\documentclass[journal,comsoc,numbers,sort&compress]{IEEEtran}
\usepackage{cite}
\usepackage{graphicx}
\usepackage{epstopdf}
\usepackage{enumitem}[topsep=3pt,itemsep=3pt]
\usepackage{amssymb, amsmath}
\usepackage{amsthm}
\DeclareMathOperator*{\argmax}{argmax}
\usepackage[switch,pagewise,columnwise]{lineno}
\newenvironment{sequation}{\begin{equation}\small}{\end{equation}}

\usepackage{makecell,multirow,diagbox}
\usepackage{stfloats}
\usepackage{array}
\newtheorem{myDef}{Definition}

\usepackage{comment}
\usepackage{stfloats}
\usepackage{colortbl} 
\usepackage{xtab,booktabs}
\usepackage{rotating}
\usepackage{xcolor}
\usepackage[normalem]{ulem}
\usepackage{longtable}
\usepackage{bm}
\usepackage{multicol}
\usepackage{longtable} 
\usepackage[switch,columnwise]{lineno}
\definecolor{color1}{rgb}{0.0, 0, 0}

\begin{document}	
\title{Applications of Game Theory in Vehicular Networks: A Survey}
	\author{
		Zemin~Sun,~\IEEEmembership{Student Member, IEEE},
		Yanheng~Liu,\IEEEmembership{}
		Jian~Wang,~\IEEEmembership{Member, IEEE},
		Guofa~Li,~\IEEEmembership{Member, IEEE},
		Carie~Anil,\IEEEmembership{}
    	Keqiang~Li,\IEEEmembership{}
    	Xinyu~Guo,\IEEEmembership{}
    	Geng~Sun,~\IEEEmembership{Member, IEEE},
    	Daxin~Tian,~\IEEEmembership{Senior Member, IEEE}, and
		Dongpu~Cao,~\IEEEmembership{Senior Member, IEEE}
			
		\thanks{Manuscript received July 22, 2020, revised.  This study is supported in part by the National Natural Science Foundation of China (62172186, 62002133, 61872158, 61806083), in part by the National Key Research and Development Program of China (2018YFC0831706), in part by the Science and Technology Development Plan Project of Jilin Province (20190701019GH, 20190701002GH, 20190103051JH, 20200201166JC), and in part by the Central government funds for guiding local scientific and Technological Development (2021Szvup047) (\textit{Corresponding authors: Yanheng Liu, Geng Sun, and Daxin Tian.)}}	
		\thanks{ Zemin Sun, Yanheng Liu,  Jian Wang, Xinyu Guo, and Geng Sun are with the College of Computer Science and Technology, Jilin University, Changchun 130012, China, and Key Laboratory of Symbolic Computation and Knowledge Engineering of Ministry of Education, Jilin University, Changchun 130012, China (e-mail: laurasun166@gmail.com, yhliu@jlu.edu.cn, wangjian591@jlu.edu.cn, xinyug19@mails.jlu.edu.cn, sungeng@jlu.edu.cn).}
		\thanks{ Guofa Li is with the Institute of Human Factors and Ergonomics, College of Mechatronics and Control Engineering, Shenzhen University, Shenzhen, Guangdong, 518060, China (e-mail: hanshan198@gmail.com). }
		\thanks{ Carie Anil is with the School of Computer Science and	Engineering, VIT-AP, Amaravati-522237, Andhra Pradesh, India, and the College of Engineering, Nanjing Agricultural University, Nanjing 210031, China.(email:carieanil@gmail.com). }
		\thanks{ Keqiang Li is with the State Key Laboratory of Automotive Safety and Energy, Department of Automotive Engineering, Tsinghua University, Beijing 100084, China (e-mail: likq@tsinghua.edu.cn).}
		\thanks{ Daxin Tian is with the Beijing Advanced Innovation Center for Big Data and Brain Computing, Beijing Key Laboratory for Cooperative Vehicle Infrastructure Systems and Safety Control, School of Transportation Science and Engineering, Beihang University, Beijing 100191, China (e-mail:dtian@buaa.edu.cn; ).}
		\thanks{ Dongpu Cao is with Waterloo Cognitive Autonomous Driving Lab, University of Waterloo, N2L 3G1, Canada (e-mail: dongpu.cao@uwaterloo.ca). }

}
	\markboth{Journal of \LaTeX\ Class Files,~Vol.~, No.~, ~}
	{Shell \MakeLowercase{\textit{et al.}}: Bare Demo of IEEEtran.cls for Computer Society Journals}
	
	\maketitle
	\begin{abstract}

In the Internet of things (IoT) era, vehicles and other intelligent components in an intelligent transportation system (ITS) are connected, forming vehicular networks (VNs) that provide efficient and safe traffic and ubiquitous access to various applications. However, as the number of nodes in  an ITS increases, it is challenging to satisfy a varied and large number of service requests with different quality of service (QoS) and security requirements in highly dynamic VNs. Intelligent nodes in VNs can compete or cooperate for limited network resources to achieve the objective for either an individual or a group. Game theory (GT), a theoretical framework designed for strategic interactions among rational decision makers sharing scarce resources, can be used to model and analyze individual or group behaviors of communicating entities in VNs. This paper primarily surveys the recent developments of GT in solving various challenges of VNs. This survey starts with an introduction to the background of VNs. A review of GT models studied in the VNs is then introduced, including the basic concepts, classifications, and applicable vehicular issues. After discussing the requirements of VNs and the motivation of using GT, a comprehensive literature review on GT applications in dealing with the challenges of current VNs is provided. Furthermore, recent contributions of GT to VNs that are integrated with diverse emerging 5G technologies are surveyed. Finally, the lessons learned are given, and several key research challenges and possible solutions of applying GT in VNs are outlined.

\end{abstract}
	
	\maketitle

	\IEEEpeerreviewmaketitle

\section{Introduction}
\label{sec:introduction}
	
Internet of things (IoT) is seen as the most promising technology to realize the vision of connecting several things at any time, from any place, and to any network, and continues to be a hot research topic. Integrating IoT technologies to vehicles and infrastructures, intelligent transportation Systems (ITSs) aim to improve traffic safety, relieve traffic congestion, and increase energy efficiency by providing real-time information for road users and transportation system operators. Vehicular networks (VNs) have been regarded as an important component in ITS for intelligent decision making through vehicle-to-vehicle (V2V),  vehicle-to-infrastructure (V2I), vehicle-to-network (V2N), or vehicle-to-pedestrian (V2P) communication \cite{gasmi2019vehicular}. With the emergence of the cloud control system concept, vehicles are further connected with more intelligent nodes in ITS by integrating the physical layer,  the cyber layer, and the application layer of these nodes. However, as the number of connected nodes in ITS continues to increase, it is challenging to satisfy these varied and large number of service requests with different quality of service (QoS) and security requirements in highly dynamic VNs.
 
 The salient characteristics of VNs, such as the high mobility of vehicles,  the unstable network topology, and  the constrained resources (e.g., bandwidth), impose challenges on decision making process of various intelligent devices (e.g., vehicles and road-side units (RSUs)). As a result, trade-off decisions may be made between the contradictory or conflicting goals such as throughput and delay, resource utility and energy consumption,  and QoS and security. The devices have to make rational (often compromised) decisions in a highly dynamic and resource-restricted environment. For example, RSUs should make balanced decisions on resource allocation to improve resource utilization while minimizing computational overheads and energy consumption. Besides, resource consumers such as vehicles should decide to either selfishly compete for limited resources or cooperate with others to maximize the overall network performance. Moreover, to defend against selfish or malicious behaviors, an intelligent node should make a trade-off between the communication performance and the security level because these two contradictory goals consume the node's limited resources.  The game theory (GT) technology has shown its potential by mobile users to be always best connected in the highly dynamic and unstable environment because it provides a mathematical model for optimizing  complex issues  \cite{friedman1986game}, where multiple players with contradictory objectives compete for limited resources or cooperate for maximizing common interests. This study systematically surveys the application of GT in modeling the strategic behaviors (i.e., competition or cooperation) of intelligent nodes in VNs.

  \begin{figure*}[!hbt]
  	\centering
  	\includegraphics[width =7in]{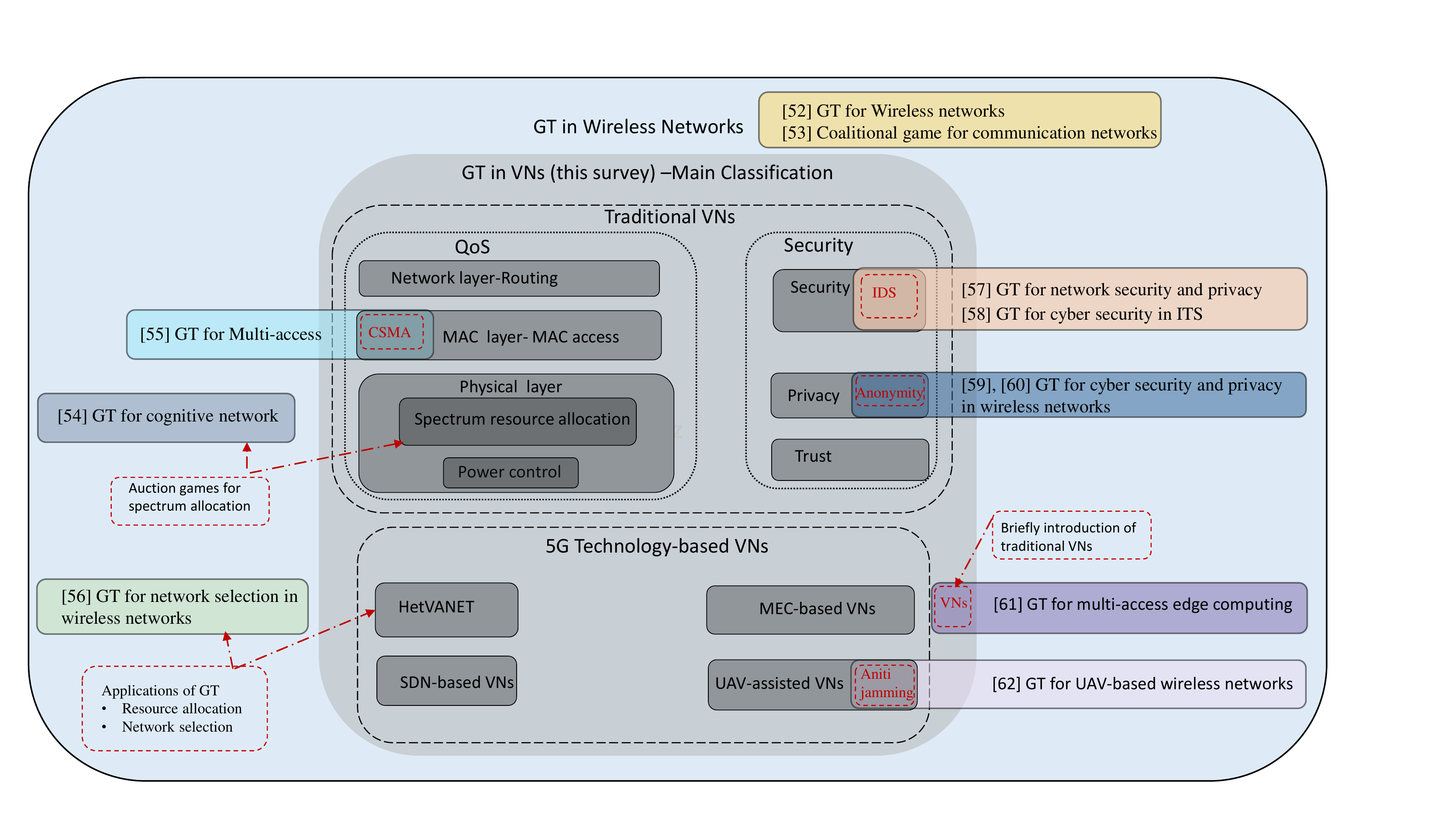}
  	\caption{Comparison between this survey and the existing surveys on GT application in wireless networks.}
  	\label{fig_ConnecGTinWNs}
  \end{figure*}
 
 VNs have been extensively studied in different contexts \cite{karagiannis2011vehicular,gerla2011vehicular,al2014comprehensive,liu2016survey,liang2015vehicular,nanda2019internet}, and  recently published surveys cover many aspects of VNs. Several surveys focus on the QoS of VNs involving the medium access control (MAC) protocols \cite{booysen2011survey,hadded2015tdma,jayaraj2016survey,johari2020tdma,rebei2020mac},  the routing or message dissemination solutions \cite{chen2010broadcasting,panichpapiboon2011review,suthaputchakun2011routing,bilal2013position,chaqfeh2014survey,bitam2014bio,silva2016vehicular,awang2017routing,peng2018vehicular}, {\color{color1} the} energy harvesting techniques \cite{alsabaan2012vehicular,atallah2016energy}, and {\color{color1} the} mobility management \cite{aljeri2020mobility}. {\color{color1} Furthermore, considering the vulnerability of VNs to various attacks, {\color{color1} many} studies investigate the topics on the security of VNs, such as the cyber security protection \cite{engoulou2014vanet,bariah2015recent,hasrouny2017vanet}, the privacy preserving \cite{petit2014pseudonym,boualouache2017survey,ali2019authentication,babaghayou2020pseudonym}, the trust management \cite{kerrache2016trust}, or all these three aspects \cite{lu2018survey,manivannan2020secure}.} Besides, some survey papers focus on the standard protocols of VNs, such as the dedicated short-range communication (DSRC) employing the IEEE 802.11p standard \cite{li2010overview, miao2012evaluation}, {\color{color1} the} long term evolution (LTE) standard \cite{araniti2013lte}. With the development of the fifth-generation (5G) communication technology \cite{shah20185g}, recent surveys investigate the {next-generation VNs that are integrated with  \color{color1} the  5G technologies or applications}, including {\color{color1} the} vehicular cloud computing-enabled VNs \cite{whaiduzzaman2014survey,mekki2017vehicular}, {\color{color1} the} vehicular edge computing (VEC)-enabled VNs \cite{liu2019vehicular,dziyauddin2019computation}, {\color{color1} the} big-data-driven VNs \cite{cheng2018big}, {\color{color1} the software defined networks (SDN)-based VNs \cite{jaballah2020security}}, {\color{color1} the} unmanned aerial vehicles (UAV)-assisted VNs \cite{shi2018drone}, and {\color{color1} the} heterogeneous VNs (HetVNs) \cite{zheng2016soft,zheng2015heterogeneous}. However, these studies mainly focus on characteristics, requirements, attacks, and the corresponding solutions for VNs, but with little coverage on the contradiction or trade-off problems of {\color{color1} the decision making} in {\color{color1} resource-limited} VNs.

Recently, the paradigm {\color{color1}of} cooperative communication \cite{mansourkiaie2014cooperative,sami2016survey} has been proposed to study the cooperative behaviors in wireless networks. {\color{color1} Issues specific to vehicular nodes in a cooperative vehicular network (CVN) are studied in \cite{ahmed2018cooperative}.} However, most of the solutions for CVN aim at the single-objective optimization, which {\color{color1} cannot be} applied for the VNs with varying aspects--{\color{color1} a problem} mentioned as a challenge and future trend in \cite{ahmed2018cooperative}. Furthermore, game theoretical approaches are considered as {\color{color1} the} impressive {\color{color1} tools} that provide fair and distributed optimization schemes for future works of cooperative communications \cite{mansourkiaie2014cooperative}.

GT approaches have been widely applied to {\color{color1} analyze} the interactions of competitive and cooperative behaviors in wireless networks. The applications of GT in wireless networks have been generally investigated in \cite{felegyhazi2006game,saad2009coalitional, wang2010game}. {\color{color1} In addition, a number of surveys focus on the applications of GT on the specific aspects of wireless networks such as the multiple access \cite{akkarajitsakul2011game}, the network selection \cite{trestian2012game}, and the security or privacy solutions \cite{manshaei2013game,sedjelmaci2019cyber,do2017game,pawlick2019game}.} Besides, with the development of 5G, several recent surveys are mainly concerned with the applications of GT in {\color{color1} the} emerging wireless communications such as {\color{color1} the} multi-access edge computing wireless networks \cite{moura2018game} and {\color{color1} the} UAV-assisted wireless networks \cite{mkiramweni2019survey}.

\begin{figure*}[!hbt]
	\centering
	\includegraphics[width =7in]{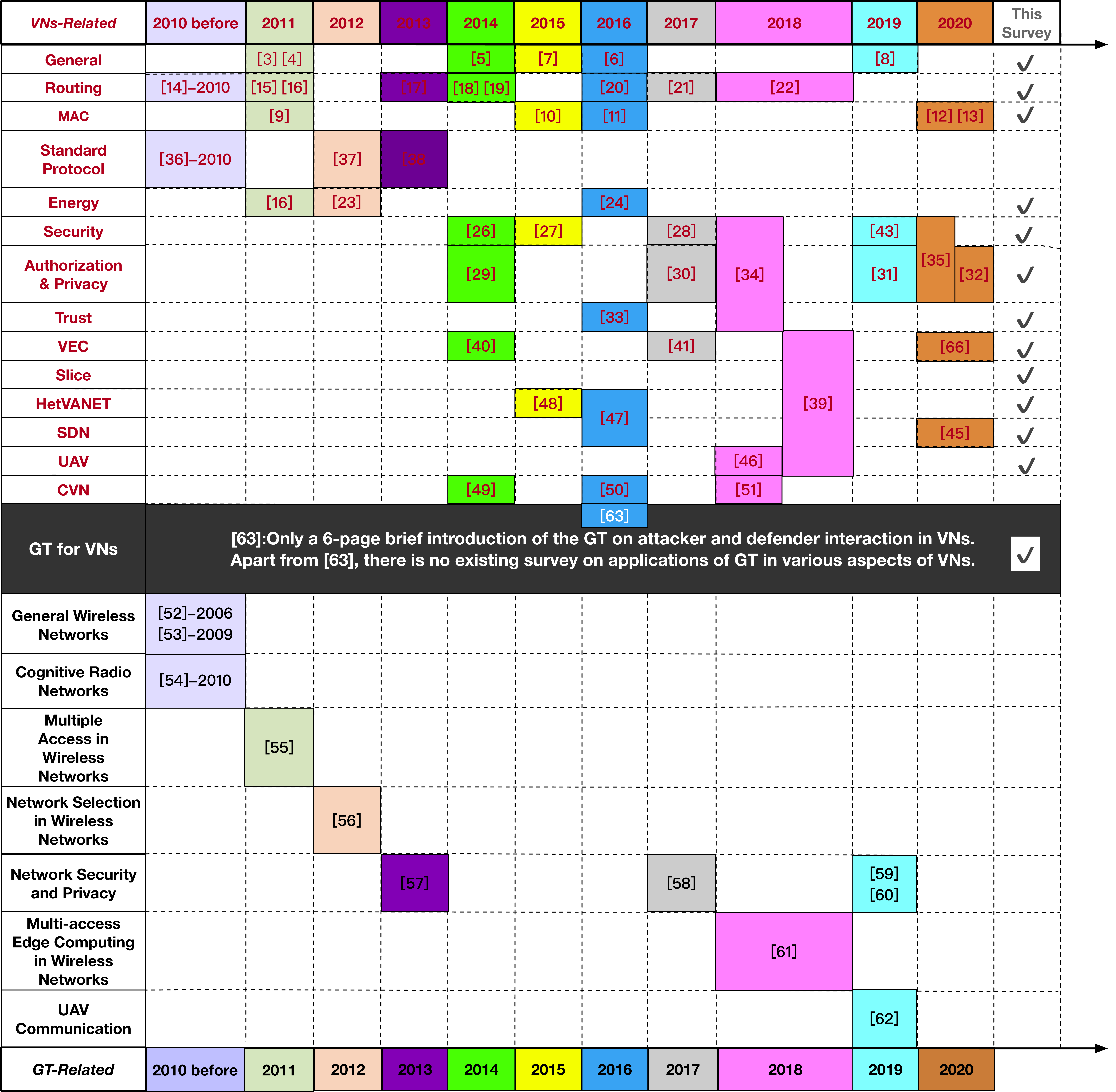}
	\caption{\textcolor{color1}{Comparison between this survey with the existing VN-related and GT-related surveys.}}
	\label{fig_relatedwork}
\end{figure*}

 {\color{color1}
 	We compare this survey with the existing surveys on VNs and GT applications from the following aspects. First, we compare the contents of this survey with that of the existing surveys on GT application in wireless networks, which is shown in Fig. \ref{fig_ConnecGTinWNs}. Then, we further compare this survey with the existing surveys on GT application in wireless networks in detail from the aspects of ``area", ``taxonomy", ``main classification", ``similarities", and ``differences",  as shown in Table \ref{fig_relatedwork}. Besides, an application-oriented comparison of this survey with the existing surveys on GT application in wireless networks and the existing surveys on VNs is shown in Fig. \ref{fig_relatedwork}. It can be concluded from the above comparisons that to our best knowledge, the only survey on GT application in VNs is \cite{bahamou2016game}. However, this paper mainly gives a brief introduction to the GT application in modeling the interaction between attackers and defenders in VNs. There is no existing survey that gives a comprehensive investigation on GT approaches to solving problems on various aspects of VNs. }

\begin{table*}
	\scriptsize
	\caption{\centering{\color{color1} Comparison between this work and the existing surveys on GT application in wireless networks in terms of ``area",``taxonomy", ``main classification", ``similarities", and ``differences"}}
	\label{tab_CompGTinWNs}
	\renewcommand*{\arraystretch}{.5}
	\begin{center}
		\begin{tabular}{|p{.02\textwidth}|p{.05\textwidth}|p{.08\textwidth}|p{.26\textwidth}|p{.16\textwidth}|p{.33\textwidth}|}
			\hline
			\textbf{Ref}&\textbf{Area}&\textbf{Taxonomy}&\textbf{Main Classification}&\textbf{Similarities}&\textbf{Differences}\\
			\hline
			\cite{felegyhazi2006game}&Wireless networks&\begin{itemize} [leftmargin=3pt] \item Protocol layer \item Non-cooperative GT \end{itemize} &\begin{itemize} [topsep=0pt,leftmargin=5pt,itemsep=0pt]\item Protocol layer\begin{itemize} [leftmargin=10pt,topsep=0pt,itemsep=0 pt]\item {\color{color1} Network layer }\item MAC layer\item Physical layer\end{itemize}\item Non-cooperative GT \begin{itemize} [leftmargin=10pt,topsep=0pt,itemsep=0 pt]\item Complete information \item Incomplete information \end{itemize}\end{itemize}& \begin{enumerate}[leftmargin=6pt] \item Similar classification {based on protocol \color{color1}layers}\end{enumerate}& \begin{enumerate}[leftmargin=5pt] \item Games:\begin{itemize}[leftmargin=5pt]  \item Non-cooperative games are mainly surveyed in \cite{felegyhazi2006game}\item  Both cooperative and non-cooperative models are considered in this survey\end{itemize} \item Applications of GT to security are not covered by \cite{felegyhazi2006game}. \end{enumerate}  \\
			\hline
			\cite{saad2009coalitional}&Wireless and communication networks& Cooperative GT&\begin{itemize}[itemsep=0pt,topsep=0pt,leftmargin=3pt]\item Canonical coalitional {\color{color1}games}\begin{itemize}[itemsep=0pt,topsep=0pt,leftmargin=3pt]\item Rate allocation in MAC channel \item Receiver and transmitter cooperation\end{itemize} 
			\item Coalitional formation {\color{color1}games} \begin{itemize}\setlength{\itemsep}{0pt}\item Transmitter and cost in a TDMA system\item Spectrum sensing in cognitive radio networks\end{itemize}  \item Coalitional graph {\color{color1}games} \begin{itemize}\setlength{\itemsep}{0pt}\item Distributed uplink in WiMAX IEEE 802.16j \end{itemize} \end{itemize}&Applications of GT to: \begin{enumerate}[leftmargin=5pt] \item MAC channel access \item Modeling cooperation between transmitters and receivers \item  Spectrum sensing \end{enumerate}&\begin{enumerate}[leftmargin=4pt] \item Game: \begin{itemize}[leftmargin=5pt]  \item The cooperative (coalitional) game is the focus {\color{color1} of} \cite{saad2009coalitional}\item Both cooperative and non-cooperative games (modeling selfishness in VNs) are considered in this survey\end{itemize}\item MAC techniques: \begin{itemize}[leftmargin=5pt]  \item Research in \cite{saad2009coalitional} is mainly based on protocols {\color{color1} used in general} wireless communications such as TDMA and IEEE 802.16j \item Research in this survey is mainly based on CSMA/CA and IEEE 802.11p techniques for VNs\end{itemize}\end{enumerate}\\
			\hline 
			\cite{wang2010game}&Cognitive radio networks&GT&\begin{itemize}[leftmargin=5pt] \item Non-cooperative games \item Economic games \item Cooperative games \item Stochastic games\end{itemize}&\begin{enumerate}[leftmargin=5pt] \item Applications of GT to spectrum resources\item Similar types of games and related concepts are considered \end{enumerate}&\begin{enumerate}[leftmargin=5pt] \item There is little research on applications of the cooperative game to spectrum allocation in VNs due to the difficulty of constructing stable coalitions among mobile and selfish competitors (vehicles) \end{enumerate} \\
			\hline
		 \cite{akkarajitsakul2011game}&Multiple access in wireless networks&Access scheme&\begin{itemize}[leftmargin=5pt] \item Contention free \begin{itemize}[leftmargin=5pt] \item Time-division multiple access (TDMA) \item Frequency-division multiple access (FDMA)\item Code-division multiple access (CDMA)\end{itemize} \item Contention-based  \begin{itemize}[leftmargin=5pt] \item ALOHA \item Carrier sense multiple access (CSMA)\end{itemize}\end{itemize}&\begin{enumerate}[leftmargin=5pt]\item Applications of GT to {\color{color1} modeling nodes} behaviors (e.g., selfishness and contention) of {\color{color1}MAC} \end{enumerate}&\begin{enumerate}[leftmargin=5pt] \item This paper mainly discusses CSMA/CA (which is applied to VNs according to  {\color{color1}IEEE 802.11p}) in Section \ref{sec_GTforMAC} \end{enumerate}\\
			 \hline
			  \cite{trestian2012game}&Network selection in wireless networks&Players&\begin{itemize}[leftmargin=5pt, itemsep=0 pt, topsep=0pt]  \item {\color{color1}Player interactions}\begin{itemize}[leftmargin=5pt]\item Users vs. Users  \item Networks  vs. Users \item Networks vs. Networks \end{itemize}\end{itemize}&\begin{enumerate}[leftmargin=5pt]\item Applications of GT to: \begin{enumerate}[leftmargin=5pt]\item Network selection/access
			  \item  Network resource allocation \end{enumerate}\end{enumerate}&\begin{enumerate}[leftmargin=5pt,topsep=0pt,itemsep=0 pt]\item Network selection: \begin{itemize}[leftmargin=5pt,topsep=0pt,itemsep=0 pt]  \item \color{color1} 4G- wireless network  selection {is the main focus in \cite{trestian2012game}} \item This paper considers HetVNs selection \end{itemize}
			  \item GT applied to security protection in wireless networks is not discussed in \cite{trestian2012game}\end{enumerate} \\ 
			  \hline 
			  \cite{manshaei2013game}&Network security and privacy&Applications and objectives&\begin{itemize}[topsep=0pt,leftmargin=5pt] \item Security of physical and MAC {\color{color1}layers}\item Security of self-organizing {\color{color1}networks}\item Security of IDS \item Privacy\item Economics of network security \item Cryptography \end{itemize}&\begin{enumerate}[leftmargin=5pt]\item Applications of GT to IDS
			  \end{enumerate}&\begin{enumerate}[leftmargin=5pt,topsep=0pt,itemsep=0 pt,] \item   Content:\begin{itemize}[leftmargin=5pt,topsep=0pt,itemsep=0 pt]\item General security and privacy of networks (including the economic perspective) are surveyed in \cite{manshaei2013game}\item Security, privacy and trust of VNs are surveyed {\color{color1} in this paper} \end{itemize} \item New schemes: \begin{itemize}[leftmargin=5pt]\item The research (from 2003-2011) in \cite{manshaei2013game} do not cover the new schemes \item GT applied for security protection in 5G-based VNs are also {\color{color1}surveyed} in Section \ref{sec_GTin5G}\end{itemize} \end{enumerate}\\
		  \hline
			  \cite{sedjelmaci2019cyber}&Cyber security for ITS&GT&\begin{itemize}[leftmargin=5pt] \item Non-cooperative games \item Cooperative games\end{itemize}&\begin{enumerate}[leftmargin=5pt]\item Applications of GT to cyber security \end{enumerate}&\begin{enumerate}[leftmargin=5pt]\item Schemes:\begin{itemize}[leftmargin=5pt] \item IDS for ITS is the major focus in \cite{sedjelmaci2019cyber}
			  		\item IDS, IPS, and IRS in VNs are covered in {\color{color1} this survey} \end{itemize}\end{enumerate}\\
		  	\hline
		 	\cite{do2017game,pawlick2019game}&General Cybersecurity and Privacy&\begin{itemize}[topsep=0pt,itemsep=0 pt,leftmargin=3pt]\item Application \cite{do2017game}\item Deception types \cite{pawlick2019game}\end{itemize}&	{\color{color1}\item\cite{do2017game}}\begin{itemize}[leftmargin=5pt,topsep=3pt,itemsep=0 pt]\item Cyber-physical security \item Communication security \item Privacy  \end{itemize}\item	{\color{color1}\cite{pawlick2019game}}\begin{itemize}[leftmargin=5pt,topsep=3pt,itemsep=0 pt] \item Perturbation \item Moving target defense \item Obfuscation \item Mixing \item Honey-x \item Attacker engagement \end{itemize}&\begin{enumerate}[leftmargin=5pt,topsep=0pt,itemsep=0 pt] \item {\color{color1}Applications of GT to} the location privacy of VNs are {\color{color1}surveyed} Section 3.4 in \cite{pawlick2019game}, in Section 5.2 in \cite{do2017game}, and in Section \ref{sec_GTprivacy} in this survey\end{enumerate}&\begin{enumerate}[leftmargin=5pt] \item \textcolor{color1}{The studies of} \cite{do2017game,pawlick2019game} mainly focus on general network security and privacy, where only ``anonymity" {\color{color1}\cite{do2017game}} and  ``mixing" {\color{color1}\cite{pawlick2019game}} are applied to VNs \end{enumerate}\\
		  	\hline
		\end{tabular}
	\end{center}
\end{table*}

\begin{table*}
	\scriptsize
	\label{tab_GTinWNs1}
	\renewcommand*{\arraystretch}{.05}
	\begin{center}
		\begin{tabular}{|p{.02\textwidth}|p{.05\textwidth}|p{.08\textwidth}|p{.26\textwidth}|p{.16\textwidth}|p{.33\textwidth}|}
			\hline
			\textbf{Ref}&\textbf{Area}&\textbf{Taxonomy}&\textbf{Main Classification}&\textbf{Similarities}&\textbf{Differences}\\
			\hline

			  \cite{moura2018game}& Multiaccess edge computing&\begin{itemize}[leftmargin=5pt, topsep=0pt,itemsep=5pt]\item Application\item Game\end{itemize}& \begin{itemize}[leftmargin=3pt] \item Applications:\begin{itemize}[leftmargin=3pt] \item Small cell networks \item D2D networks \item VNs \item Wireless sensor networks \item Cellular networks \end{itemize}   \item Game: \begin{itemize}[leftmargin=3pt] \item  Non-cooperative GT \item Cooperative GT\end{itemize}\end{itemize}&\begin{enumerate}[leftmargin=5pt,topsep=0pt,itemsep=0 pt]\item Similar taxonomy \item Applications of GT to VNs in Section III-A-6) in \cite{moura2018game},  		 \cite{mkiramweni2019survey}, Section IV-F and Section \ref{sec_GTVEC} in this survey \end{enumerate}&\begin{enumerate} [leftmargin=5pt]\item  \begin{itemize}[leftmargin=6pt] \item \textcolor{color1}{The study of \cite{moura2018game}} generally surveys the applications of GT to multiple access edge computing in wireless networks;
			  \cite{moura2018game}  briefly discusses the GT in traditional VNs in Section III.A.6), where VEC-enabled VNs are not covered. 
			  \item This paper discusses GT applications in \textcolor{color1}{VEC-enabled VNs} {\color{color1}(e.g., MEC, FC, cloudlet, and VC)} in detail in Section \ref{sec_GTVEC} \end{itemize} \end{enumerate}\\
			\hline	
			 \cite{mkiramweni2019survey}&UAV communication&\begin{itemize}[leftmargin=3pt]\item Application\item Game\end{itemize}&
			\begin{itemize}[leftmargin=3pt]\item Applications \begin{itemize}[leftmargin=3pt]\item Optimal height and coverage\item UAV coordination and mobility control\item Performance optimization\end{itemize}\item Game: \begin{itemize}[leftmargin=3pt] \item  Non-cooperative GT \item Cooperative GT\end{itemize}\end{itemize} &\begin{enumerate}[leftmargin=5pt,topsep=0pt,itemsep=0 pt] \item Similar taxonomy \item Applications of GT to anti-jamming for UAV-VNs in Section IV-E in \cite{moura2018game} and in Section \ref{sec_UAV} in this survey \end{enumerate}&\begin{enumerate} [leftmargin=6pt]\item \textcolor{color1}{The study of \cite{mkiramweni2019survey}} surveys the applications of GT to the UAV communications in wireless networks\end{enumerate}\\
			\hline
		\end{tabular}
\end{center}
\end{table*}

This paper aims to provide a comprehensive survey of the current development on GT approaches in solving the challenges of VNs. The main contributions of this paper can be summarized as follows:

\begin{itemize}[itemsep=3pt, topsep=3pt]
	\item We provide a comprehensive overview of GT applications for current and next-generation VNs, their advantages, existing challenges, and possible solutions.
	
	\item We prepare an initial background of VNs for non-specialized readers and present the unique features that may result in the challenges in VNs.
	
	\item We present an overview of GT models used in VNs, giving the basics and classifications from several aspects, proposing a taxonomy (Table \ref{tab_game}) on the models applied in VNs, and discussing the key characteristics or advantages of each model that are appropriate to solve specific problems in VNs.
	
	\item We highlight the requirements of issues in VNs and discuss the motivation of using GT, analyzing the key characteristics of GT in fulfilling the specific requirements in VNs.

	\item We conclude the lessons learned, {\color{color1} and we analyze} how and which GT models can be employed to efficiently solve the corresponding problems in VNs.
	
	\item We identify {\color{color1} various} remaining challenges and future research directions to advance GT development in VNs.
	
\end{itemize}

\begin{figure*}[!hbt]
	\centering
	\includegraphics[width =7in]{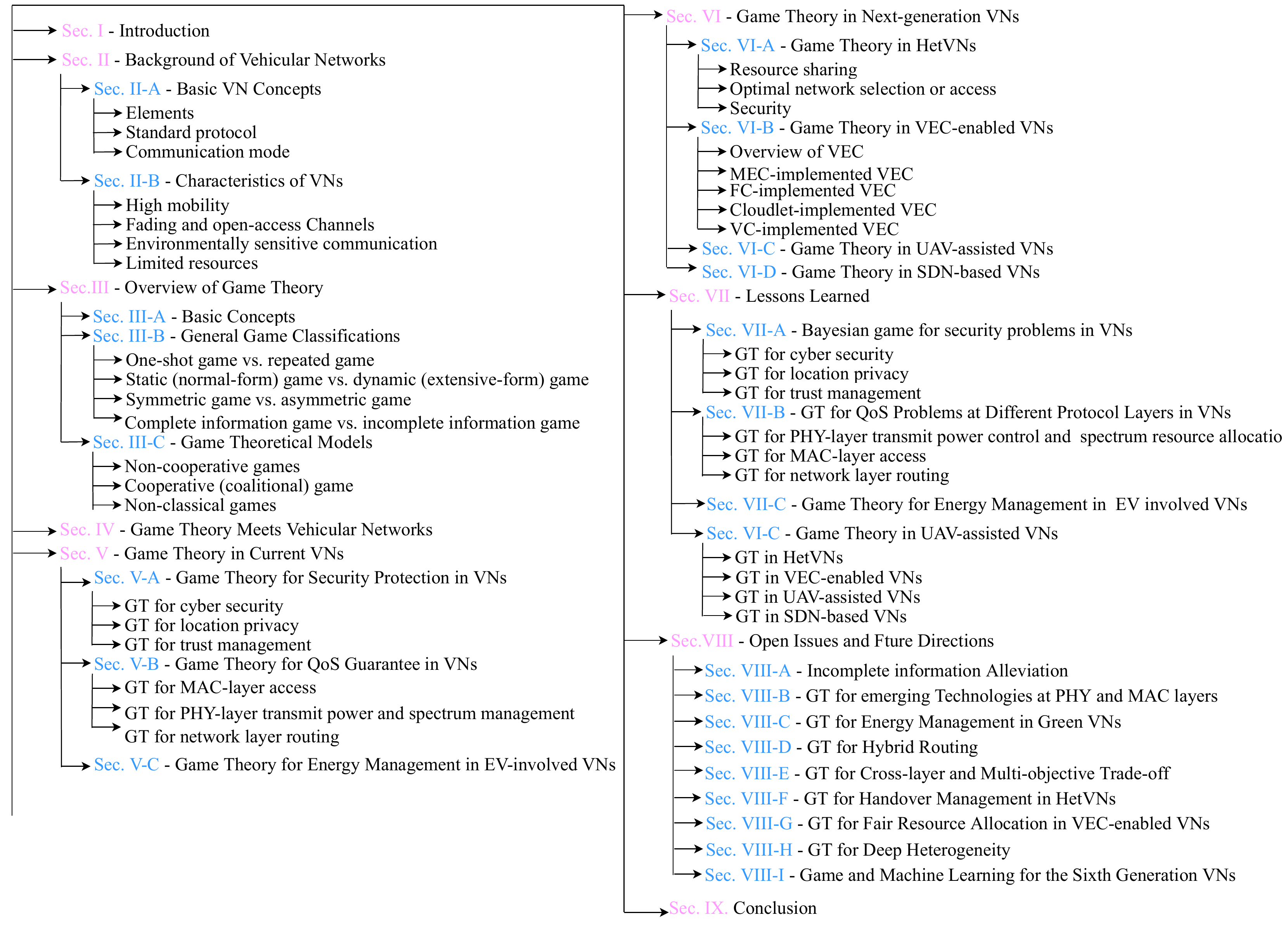}
	\caption{\textcolor{color1}{Structure overview of this survey.}}
	\label{fig_outline}
\end{figure*}

The rest of the paper is structured as portrayed in Fig. \ref{fig_outline}. Section \ref {sec_OverviewVNs} introduces the background of the VNs. Section \ref{sec_BgGT} provides a comprehensive overview of {\color{color1} the} GT models used in VNs, including the basics and classifications from several aspects, the taxonomy  on the models applied in VNs, and the key characteristics of each model to solve the specific problems in VNs. {\color{color1} Section \ref{sec_GTMeetsVNs} discusses the motivation of applying GT in VNs.} Section \ref{sec_GTinVNs} reviews the GT solutions {\color{color1} to address} the challenges of current VNs. Section \ref{sec_GTin5G} further investigates the applications of GT in VNs with 5G-based technologies.  Section \ref{sec_lesson} provides a description of lessons learned. Section \ref{sec_challenge} highlights the key challenges and future works. The paper is concluded in Section \ref{sec_conclusion}. The abbreviations in this paper are listed in Table \ref{tab_abb}.

	\begin{table}
	\scriptsize
	\caption{ABBREVIATIONS}
	\label{tab_abb}
	\renewcommand*{\arraystretch}{1}
	\begin{center}
		\begin{tabular}{p{.04\textwidth} p{.16\textwidth}p{.04\textwidth} p{.17\textwidth}}
			\hline
			\textbf{Acronyms} &\textbf{Description}&\textbf{Acronyms} &\textbf{Description}\\
			\hline
			4G& The Fourth Generation &	5G&	The Fifth Generation \\
			\textcolor{color1}{6G}&\textcolor{color1}{The Sixth Generation}& AI&Artificial intelligence\\
			BNE&Bayesian Nash Equilibrium&	BSM&Basic Safety Message\\
			
			CBR&Channel Busy Ratio&CC&Cloud Computing\\
			CH&Cluster Head&
			CH-IDS&Coalition Head with Intrusion Detection Head\\
			CR&  Cognitive Radio&	CS& Carrier Sense \\			CSMA&Carrier Sense Multiple Access&
			CSMA/CA&Carrier Sense Multiple Access/Collision Avoidance\\
				
			D2D&Device to Device&
			DoS&Denial of Service\\
			DSRC&Dedicated short-range communication &
			ECD&Edge Computing Device\\
			ESS&Evolutionary Stable Strategy&
			FC&Fog Computing\\
			F-SBS&Fog-Small-BS&
			GDS&Global Decision System\\
			GIDS&Global Intrusion Detection System&
			GT&Game Theory\\
			HetVN&Heterogeneous Vehicular Network&
			HP & High Priority\\
			IDA&Intrusion Detection Agent&
			IDS&Intrusion Detection System\\
			IoT&Internet of Things&
			IPS& Intrusion Prediction System\\
			IRS& Intrusion Reaction System &
			ISPs& Internet SPs \\
			ITS &Intelligent Transportation Systems&
			LA&Learning Automata\\
			LP&Low Priority&
			LIDS&Local Intrusion System Detection\\
			LSA& Licensed Shared Access&
			LTE&Long Term Evolution\\
			MAC&Medium Access Control &
			MANETs& Mobile Ad-hoc Networks\\
			MEC&Mobile Edge Computing&
			\textcolor{color1}{MIMO}&\textcolor{color1}{Multi-Input Multi-Output}\\
			mmWave& Millimeter-wave (mmWave)&
			MNO&Multiple Network Operators\\
			NBS&Nash Bargaining Solution	&        
			NE&Nash Equilibrium\\
			OBUs &On Board Units &
			PCN&Post Crash Notification\\
			\textcolor{color1}{PC5}&\textcolor{color1}{Direct interface of LTE-V2X}&
			\textcolor{color1}{PE}&\textcolor{color1}{Pareto Efficiency}\\
			PKI&Public Key Infrastructure&
		    \textcolor{color1}{PO}& \textcolor{color1}{Pareto Optimality}\\
			PU&Primary User&
			QoE&Quality of Experience\\
			QoS&Quality of Service&
			RHCN&Road Hazard Condition Notification\\
			RL&Reinforcement Learning&
			ROS&Requirement of Service\\
			RSU&Road-Side Unit&
			\textcolor{color1}{SAE}&\textcolor{color1}{Society of automotive engineers}\\
			SDN&\textcolor{color1}{Software Defined Network}&
			SINR&Signal-to-Interference-plus-Noise-Ratio\\
			SNR& Signal-to-Noise Ratio&
			SP&Service Provider\\
			SU&Secondary User&
			SVA&Stopped/Slow Vehicle Advisor\\
			TDMA&Time Division Multiple Access&
			TA&Trusted Authority\\
			UAV&Umanned Assisted Vehicles&
			\textcolor{color1}{Uu}&\textcolor{color1}{Cellular interface of LTE-V2X}\\
			\textcolor{color1}{UE}&\textcolor{color1}{User equipment}&
			\textcolor{color1}{V2C}&\textcolor{color1}{Vehicle-to-cloud}\\
			\textcolor{color1}{V2E}&\textcolor{color1}{Vehicle-to-edge}&
			V2I &Vehicle-to-Infrastructure\\
			\textcolor{color1}{V2N} &Vehicle-to-Network&
			\textcolor{color1}{V2P} &Vehicle-to-Pedestrian\\
			V2R&  Vehicle-to-Roadside&
			V2U&Vehicle-to-UAV\\
			V2V &Vehicle-to-Vehicle&
			\textcolor{color1} {V2X} &\textcolor{color1}{Vehicle-to-Everything}\\
			VC&Vehicular Cloud&
			VEN&Vehicular Edge Node\\
			VFC&Vehicular Fog Computing&

			VM&Virtual Machine\\
			VN&Vehicular Networks&
			WAVE&Wireless Access in Vehicular Environments\\
			\hline
		\end{tabular}
	\end{center}
\end{table}

\section{Background of Vehicular Networks}
\label{sec_OverviewVNs}
In this section, the basic concepts, communication types, and characteristics of VNs are introduced.

\subsection{{\color{color1} Basic VN Concepts}}
\label{sec_ConceptsVNs} 
This subsection mainly introduces the basic VN concepts, including the elements in VNs, the basic standard protocols that are applied to VNs, and the communication mode of VNs.

 \vspace{6 pt}

\subsubsection{Elements}
\label{sec_VNelements}
VNs are a specialized form of mobile ad-hoc networks (MANETs). The elements in VNs can be classified into traditional elements and next-generation elements.

\vspace{3pt}
 \textbf{\textit{a) Traditional elements in traditional VNs}}
\vspace{3pt}

As it is shown in Fig. \ref{fig_fit}(a), traditional VNs mainly consist of three types of intelligent elements, i.e., vehicles with on-board units (OBUs), RSUs, and trusted authorities (TAs).

\begin{itemize}[itemsep=3pt, topsep=3pt,]
	\item\textbf{OBU}  
	
	Each vehicle is equipped with an OBU to communicate with other vehicles and with the transport infrastructures.  An OBU comprises sensors (e.g., GPS, Lidars, and cameras), computing units, storage devices, and network devices. The vehicle sends its own information (speed, direction, and angle) or the perception information (location of other vehicles) via the OBU.
	
	\item\textbf{RSU}
	
	RSUs are usually deployed on both sides of roads and intersections. The RSU periodically sends road information to vehicles within its communication range. The RSU extends the communication distance of vehicles and uploads safety information (emergency or traffic accident) to the core network.
	
	\item \textbf{TA}
	
	The TA, {\color{color1} which must have a} high computational performance and adequate storage capacity is responsible for the trust and security management of the entire VNs, including verifying the authenticity of vehicles and revoking nodes in the case {\color{color1} vehicles broadcast fake messages or behave maliciously}. 
\end{itemize}	

 \textbf{\textit{b) New elements in next-generation VNs}}

\vspace{3pt}
		
	With the ever-growing number of connected vehicles, the emergence of various vehicular applications, the volume and variety of data would proliferate tremendously. The traditional elements, which have the disadvantages of limited computing abilities,  insufficient flexibility, and low scalability, would not be able to meet the  diverse and stringent requirements on communication performances, security, and quality-of-experience (QoE). In this context, the emerging 5G elements are integrated into the next-generation VNs to fulfill these requirements efficiently, securely, and consistently.
	The emerging elements in next-generation VNs include the CC center, VEC servers, UAVs, and SDN controllers, as shown in Fig. \ref{fig_fit}(a). 
	
	\begin{itemize}[itemsep=3pt, topsep=3pt]

			\item \textbf{CC center}
			
			The CC center provides cloud storage and computing power for vehicular communications. The intelligent nodes in VNs can access to the shared pool
			of resources ubiquitously. However, these resources are located in the remote core network and could cause high latency of the real-time vehicular communication.

			\item\textbf{VEC server}
			
			By extending the CC capabilities to the edge of VNs, the VEC server provides storage, computing, and control resources to process large amounts of data or compute-intensive tasks quickly and locally.
			
			\item \textbf{UAV}
			
			By cooperating with the on-ground vehicles or RSUs, UAV-assisted VNs can enhance the communication connectivity among vehicles, extend the coverage range of RSUs, provide the flexibility of deployment, and improve the ability of information collection.
			
			\item\textbf{ SDN }

			SDN is an emerging architecture that aims to overcome the complexity, staticity, and inflexibility of the traditional distributed VNs. By decoupling the control plane from the data plane, the SDN provides directly programmability, logically centralized control, and  global knowledge for the VN. Accordingly, the integration of the SDN architecture enables the next-generation VNs to handle massive vehicular requirements, heterogeneous vehicular communications, and dynamic VN conditions more efficiently.		
	\end{itemize}

 \vspace{6 pt}
\subsubsection{Standard protocol}
\label{sec_protocol}
\begin{figure*}[!hbt]
	\centering
	\includegraphics[width =7in]{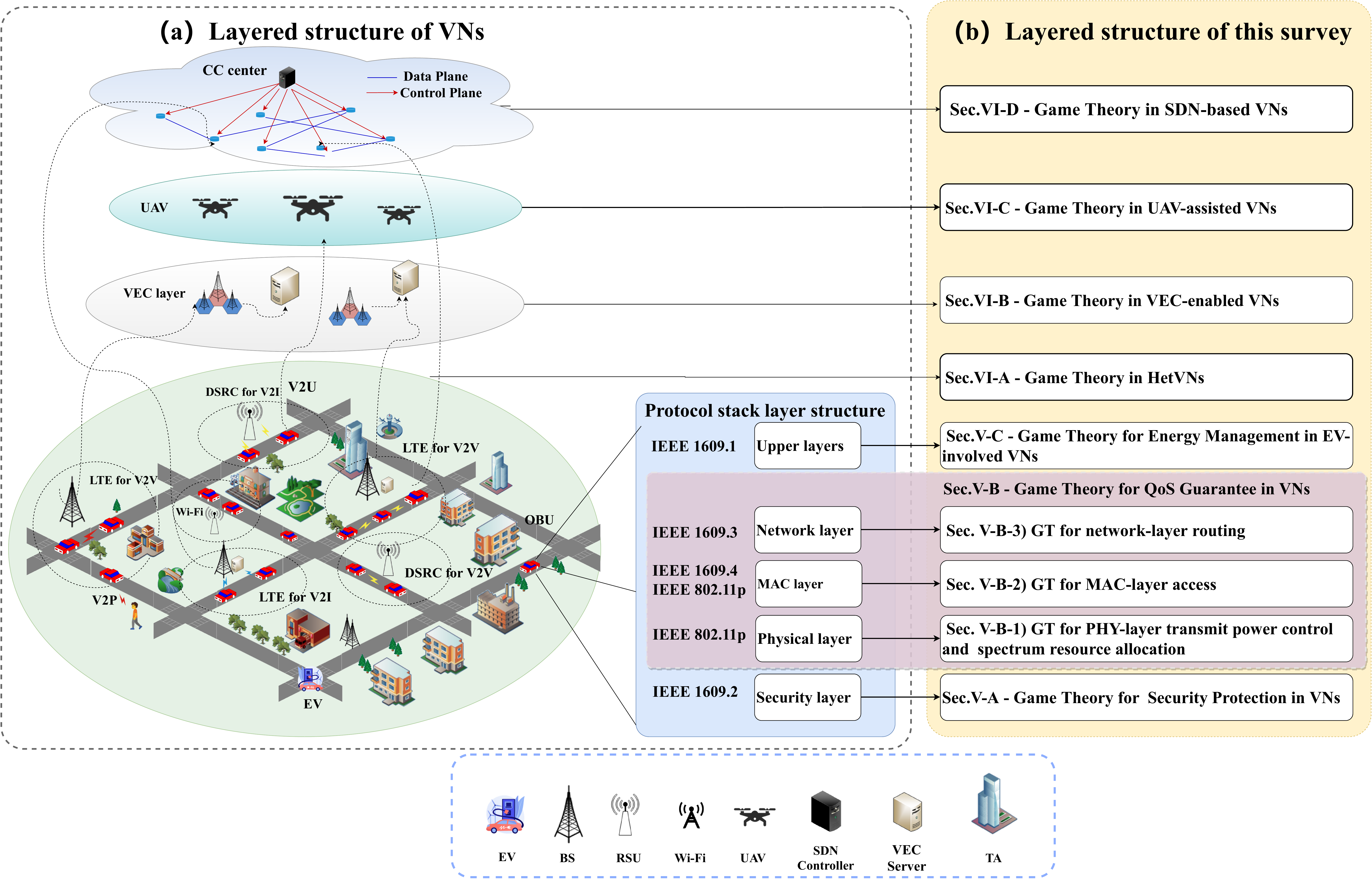}
	\caption{\textcolor{color1}{(a) The layered structure of VNs. (b) The layered structure of this survey.}}
	\label{fig_fit}
\end{figure*}

{\color{color1} The two major technologies that support vehicular communications are DSRC released by IEEE 802.11p \cite{ieee2010ieee} and LTE-V2X released by 3GPP LTE \cite{3gppevolved}. }
\begin{itemize}[itemsep=3 pt, topsep=0 pt]
	\item\textbf{DSRC}	
	
{\color{color1} DSRC standards suit, which is based on several IEEE standards, is developed for direct communication between vehicles, or between vehicles and infrastructures. DSRC employs IEEE 802.11p standard as the protocols for VN's lower layers, including the orthogonal frequency division multiplexing (OFDM) protocol for the physical (PHY) layer and {\color{color1} the} carrier sense multiple access with collision avoidance (CSMA/CA) protocol for the MAC layer. IEEE 802.11p standard, an amendment to the IEEE 802.11 standard (the basis of products marketed as Wi-Fi), includes the wireless access in vehicular environments (WAVE). WAVE \footnote{WAVE is the core of DSRC; however the two terms of DSRC and WAVE are commonly used arbitrarily \cite{5462975}.} is a term used to address IEEE 1609 family.  Specifically, IEEE 1609 family of standards consist of IEEE 1609.1 for the upper layer, IEEE 1609.2 for security, IEEE 1609.3 for the network layer, and IEEE 1609.4 for the MAC layer. Besides, the message format used by the IEEE 1609 protocols is standardized with the cooperation of Society of Automotive Engineers (SAE), which is a global professional association and standards developing organization. The WAVE protocol stack for VNs is depicted in \textcolor{color1}{Fig. \ref{fig_fit}(a)}}.

\item \textbf{LTE-V2X}
	
	The third-generation partnership project (3GPP) proposes {\color{color1}the use of the} LTE-V2X. Compared to DSRC, {\color{color1}LTE-V2X} introduces two interfaces for radio access, i.e., {\color{color1}the direct interface PC5 and the cellular interface Uu}.  The PC5 interface enables vehicles to directly communicate with other vehicles without the help of base stations. It has been proved that the performance of the LTE-V2X is superior to that of {\color{color1} the} DSRC in terms of throughput, latency, and {\color{color1} spectrum utilization}. {\color{color1} However,  the communication performance of the LTE-V2X drops more rapidly than that of the DSRC in the highly dense VNs because LTE-V2X lacks the mechanism like CSMA/CA of DSRC to prevent message collisions.
} 

\end{itemize}

\vspace{6pt}

\subsubsection{Communication mode}
\label{sec_communication mode}
The VNs have multiple types of communication modes, including V2V communication, V2I communication, V2N communication, and V2P communication, which are collectively referred to V2X.

\begin{itemize}[itemsep=3pt, topsep=3pt]
	\item\textbf{V2V communication}
	
	Vehicles can directly communicate with other vehicles through V2V communication to transmit or receive safety and non-safety application information (e.g., intersection collision warnings and emergency vehicle  warnings).
	
	\item \textbf{V2I communication}
	
	Road infrastructures and vehicles share valuable information via the V2I communication. Specifically, the vehicles obtain transport information such as traffic light conditions from the infrastructures; the roadside infrastructures obtain vehicle state information such as overtaking vehicle warning and cooperative merging assistance from vehicles.	
	
	\item \textbf{V2N communication}	
	V2N communication enables LTE-assisted V2V communications over cellular networks.  V2N is a unique use case of the LTE-V2X but is not supported by the IEEE 802.11p because the IEEE 802.11p  is mainly designed for the direct communications.
	
	\item \textbf{V2P communication} 
	
	{\color{color1} V2P communication enables the communication between vehicles and road pedestrians} {\color{color1} when people are walking or riding bicycles, and children are in strolers}. {\color{color1}  V2P communication can provide real-time information (e.g., blind spot warnings) for both pedestrians (through  handheld devices or wearables) and  drivers (through OBUs).}
	
		
	\item \textbf {5G-based communication modes}
	
	{\color{color1} The emerging heterogeneous networks support both V2V and V2I communications with DSRC, LTE, or Bluetooth technology. These Besides, with the emergence of the UAV technology, the 5G-vehicular communication also incorporates V2U communication.}
	
\end{itemize}

\subsection{Characteristics of VNs}
\label{sec_VNchara}
Compared with MANETs, VNs are characterized by the following unique features:

\begin{itemize}[itemsep= 3 pt,topsep = 3 pt]
	\item \textbf{High mobility}
	
	The {\color{color1}high} mobility of vehicles is the unique characteristic of VNs which distinguishes them from the traditional MANETs. 
	The high mobility of vehicles directly induces {\color{color1} highly dynamic network topology.} The network topology of VNs changes rapidly and frequently, especially in the lack-of-infrastructure places where the temporal V2V connections among vehicles are  autonomously formed and distributed. The temporal variability in VNs make the vehicular communication unreliable and uncertain. Furthermore, the {\color{color1} vehicular} communications are vulnerable since it is difficult to identify the attackers {\color{color1}in real time}.

	\item \textbf{Fading and open-access channels}
	
	The vehicular channels may experience various fading (e.g., Relay, Rician, or Nakagami) due to the multipath propagation, {\color{color1} Doppler effects}, weather {\color{color1}conditions}, or shadowing of obstacles. The signal also suffers more severe {\color{color1}losses} with the communication distance. Similarly, this characteristic may result in temporary failure of the communication.  Furthermore, the open-access vehicular channel is the main reason that causes communication vulnerability to attacks such as eavesdropping and false message injection.

	\item \textbf{Environmentally sensitive communication}
	
	Vehicular communications are sensitive to the environment, which in essence results from the dynamic nature of the VNs. The above-mentioned channel fading varies with the different situated vehicular environments (such as urban, rural areas, {\color{color1} and} central city.) and different weather conditions (particularly rain). Moreover, the frequent communication {\color{color1} induces} congestion, especially in {\color{color1} dense} traffic {\color{color1}scenarios} because most of the information in VNs is broadcast. At the same time, the broadcast communication could be exposed to more attackers such as the eavesdroppers in {\color{color1}dense environment}.
 
   \item \textbf{Limited resources}
   
   {The VN environment} is also characterized by limited resources such as congested spectrum or access media, limited bandwidth, computational capacity, space, {\color{color1} and} energy.  With the growth of vehicles and services, limited resources could lead to conflicts among various requirements due to {\color{color1} this} competition. For example, adopting strong security algorithms {\color{color1} may cause considerable} overheads on communication delay, energy, or space. On the other hand, although a weak security mechanism {\color{color1}could save} computational resources, it {\color{color1} would provide} little protection.

\end{itemize}

\section{Overview of Game Theory}
\label{sec_BgGT}
In this section, basic concepts of GT are briefly introduced, and game models used for solving the {\color{color1} VN issues} are discussed.

\subsection{Basic Concepts}
\label {sec_BgGTBasic}
	GT is a branch of applied mathematics for analyzing the strategic interactions among multiple decision makers (players) \cite{friedman1986game}. These decision makers cooperatively or competitively take rational actions that have conflicting consequences. 
	
	A normal form game is a triplet $ \mathbf{G}=(\mathcal{N}, \mathcal{S}, \mathcal{U})$, where:
\begin{itemize}[itemsep=3pt,topsep=3pt]
	\item  $\mathcal{N}=\{1, \ldots, n\}$  denotes a finite set of $n$ players, which are the intelligent nodes.
	\item  $\mathcal{S}=S_1 \ldots \ldots \times S_n$ denotes players' strategies,  where $S_i=\{s_{i}^1,\ldots, s_i^{N_s^i}\}, {i\in\mathcal{N}}$ is a set of $N_s^i$ possible strategies for player $i$. The strategy $s_i^j\in S_i$  that is selected deterministically is called the pure strategy. Besides, probabilistically is called the \textit{mixed strategy}, where $\Delta S_i=\{\sigma_{i}:S_i\rightarrow [0,1]:\sum_{s_i^j\in S_i}\sigma_{i}(s_i^j)=1\}$ is the set of probability distributions over $S_i$.
	\item  $\mathcal{U}=\{u_1, \ldots, u_n\}$, where $ u_i:\mathcal{S} \rightarrow \mathbb{R}$ denotes the utility (or payoff) function \footnote{The notions of utility and payoff functions are generally used interchangeably.} of player $i$.  Each player in the game aims to select the strategy to maximize its utility function given the other players' strategies.
\end{itemize}

 Each player $i\in \mathcal{N}$ in the game aims to select the ``best" strategy to obtain the best utility. Therefore, the player would compare its strategies before making decisions. The term of \textit{ preference} is used to quantify the comparison between the strategies, which is defined as Definition \ref{def_preference}.

\begin{myDef}
	\label{def_preference}
	Given two strategies $s_i^j,s_i^k\in S_i$ of player $i$, a preference relation  $s_i^j \succeq s_i^k$ denotes that the strategy $s_i^j $ is preferred to the strategy $s_i^k$. The preference relation between strategies can be represented by the relation between the corresponding utilities: $s_i^j \succeq s_i^k \Leftrightarrow  u_i(s_i^j)\geq u_i(s_i^k)$.
\end{myDef}

GT solutions vary based on the nature of the different games. However, almost all of them depend on the concepts of \textit{Nash equilibrium} (NE) \cite{games1953john}.
As the central idea of the NE, the concept of the  \textit{best response} will be introduced firstly. The best response of the player $i$ is the optimal strategy that generates the maximum utility given the known strategies of the other players.

\begin{myDef}
	\label{def_BS}
	The player $i$'s strategy $s_{i}^* \in S_i$ is a \textbf{best response } if
	\begin{equation}
	\label{eq_BS}
	br_i(s_{i}^*,s_{-i})=\arg \max_{s_{i}^j\in S_{i}}u_i(s_{i}^j,s_{-i}),
	\end{equation}
	where $s_{-i}$ is the other players' strategies, namely the strategy vector of all players except the player $ i$.
\end{myDef}

A strategy profile is an NE if each player in the game adopts the best response strategy.  In this case,  no player  can increase his utility by unilaterally changing his strategies. Therefore, no player has an incentive to deviate from the NE.
\begin{myDef}
	\label{def_NE}
	A strategy profile $s^*=\{s_{i}^*,s_{-i}^*\}$  is an NE, if for each player $i \in \mathcal{N}$
	\begin{equation}
	\label{eq_NE}
	u_{i}\left(s_{i}^{*}, s_{-i}^{*}\right) \geq u_{i}\left(s_{i}^j, s_{-i}^{*}\right), \forall s_{i}^j \in S_i,
	\end{equation}
	where $s_{-i}^*$ denotes the other players' best response strategies and	{\color{color1} $u_i(s_i^*, s_{-i}^*)$ is the maximum utility function of the player $i$ over the set of all its strategies given {\color{color1} the} other players' best response strategies.}
	
\end{myDef}

A game could have more than one NEs \cite{bier2012game}.  An NE is less efficient if there is another strategy profile leading to a higher utility value than  the NE strategy.  Several measures are proposed to quantify the efficiency or inefficiency of the game. One of the most  basic and rudimentary notions is the  Pareto optimality (PO) or the Pareto efficiency (PE) \cite{aumann2010pareto}. The PO or PE captures the idea that a game solution is inefficient if it is possible to achieve an improvement for the players simultaneously \cite{aumann2010pareto}. That is to say, a PO or a PE can be reached if it is impossible to make any player better off without making at least one player worse off.

\begin{myDef}
	\label{def_PO}
	A strategy profile $S^p=\{s_1^p,\ldots,s_n^p\}\in \mathcal{S}$ is Pareto optimal, if and only if there does not exist another strategy profile $S^{'}=\{s_1^{'},\ldots,s_n^{'}\} \in \mathcal{S}$ 

	\begin{equation}
	\label{eq_Pareto}
	\forall i\in \mathcal{N},u_{i}(s^{'}) > u_{i}(s^{p}) \ \&\& \ \ 
	\exists j\in\mathcal{N}, j\neq i,	u_{j}\left(s^{'}\right) > u_{j}(s^{p}).	 
	\end{equation} 	
\end{myDef}
Although the PO provides efficiency,  Dubey \cite{dubey1978finiteness} has shown that the NE may generally be Pareto inefficient.  The inefficiency of  an NE  is typically quantified by the term \textit{price of anarchy} (PoA) \cite{papadimitriou2001algorithms}, which is defined by using a worst-case approach.

\begin{myDef}
	\label{def_poa}
	The PoA  is the ratio between the quality of the worst-case NE strategy profile and that of the optimal strategy profile.
	\begin{equation}
	\label{eq_poa}
	PoA=\frac{\min_{s_{w}\subseteq S_{NE}}SW(s_{w})}{SW(s_{o})},
	\end{equation}
	where $s_{w}$ denotes the worst-case NE strategy profile that is the subset of the NE set $S_{NE}$, $SW(\cdot)$ denotes the social welfare that is defined by Definition \ref{def_SW}, and $s_{o}$ denotes the social maximum that is defined by  Definition \ref{def_SO}.
\end{myDef}

\begin{myDef}
	\label{def_SW}
	The social welfare is used to quantify the quality of the strategy. It is defined as the sum of the players' utilities given the strategy profile $s\in \mathcal{S}$:
	\begin{equation}
	\label{eq_SW}
	SW(s)=\sum_{i\in \mathcal{N}}u_i(s).
	\end{equation}
\end{myDef}

\begin{myDef}
	\label{def_SO}
	The social optimum  is used to quantify the optimal strategy profile. It is defined as the strategy profile that maximizes the social welfare:
	\begin{equation}
	\label{eq_SO}
	s_{o}=\arg \max_{s\in\mathcal{S}} SW(s),
	\end{equation}
\end{myDef}where $s_o$ is also called the social optimal strategy.

\begin{figure*}[!hbt] 
	\centering
	\includegraphics[width=6.6in]{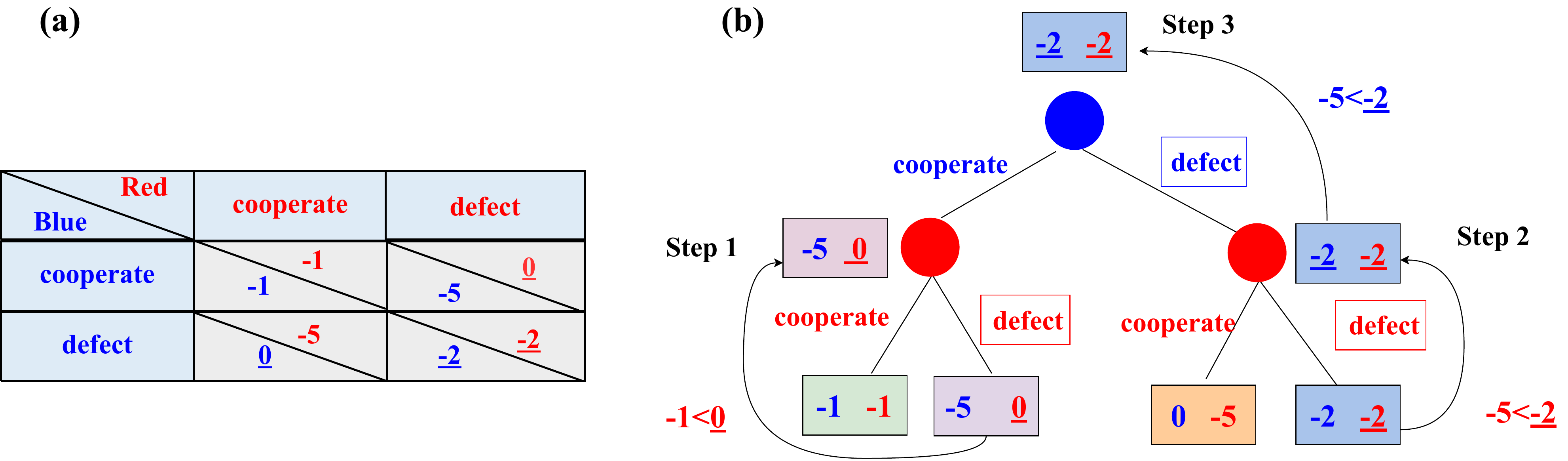}
	\caption{\textcolor{color1}{Representations of games. (a) Normal-form representation of a static game. (b) Extensive-form representation of a dynamic game.}}
	\label{fig_extensive}
\end{figure*}

\subsection{General Game Classifications}
\label {sec_GTclass}
Games can be classified as one-shot game vs. repeated game,  static game  vs. dynamic game, or complete information game vs. incomplete information game according to the characteristics of the game.

\vspace{6pt}

\subsubsection{One-shot game {\color{color1} vs.} repeated game}
\label {sec_oneshot}

A one-shot game can be played only once where players often behave selfishly because each player has little knowledge of the other players' strategies. Oppositely, a repeated game can be played more than once where players can interact and cooperate to play social optimal strategies. On the one hand, a player's reputation will affect {\color{color1} the} other payers' future strategies. On the other hand, a player learns {\color{color1} the} other players' reputations from their history behaviors and decides his best action. In the repeated game, a critical approach to inducing interaction is {\color{color1} by} punishing the players who deviate from the cooperation. {\color{color1} The players may obtain better utilities by adjusting the optimal strategies each time in the repeated game. However, repeating a game a number of times could, in general, results in very different NE outcomes as well as different optimal strategies because repeating the game could lead to the emergence of new NEs \cite{Damme1989Renegotiation}.}

\vspace{6pt}
\subsubsection{Static (normal-form) game {\color{color1} vs.} dynamic (extensive-form) game}
\label {sec_static}

In static games, {\color{color1} the} players make decisions simultaneously, {\color{color1} so} one player {\color{color1} cannot observe the} other players' historical actions. {\color{color1} In dynamic games,} players move sequentially or repeatedly over time.  The static game is a one-shot game, and the dynamic game is a repeated game. {\color{color1} The} normal form and {\color{color1} the} extensive form are two types of representation used to describe the static game and {\color{color1} the}  dynamic game, respectively.

A normal-form representation generally uses a matrix to describe a static game. The matrix includes the basic elements of a game, i.e., players, strategies for each player, and corresponding utilities for each combination of the strategies. For example, the matrix in Fig. \ref{fig_extensive}(a) describes a simple normal-form game played by Blue and Red, each of whom chooses to either ``cooperate" or ``defect". The utilities of the players are denoted by the number in the corresponding colors. It can be observed that if Blue cooperates, Red should defect because defecting results in a higher utility than cooperating ($-1<\underline{0}$). In this case, the best response of Red is to defect. 
By analogy, if Blue defects, Red should defect instead of cooperating ($-5<\underline{-2}$).  On the other hand,  if Red cooperates, the best response of Blue is to defect because defecting  yields a higher utility than cooperating ($-1<\underline{0}$). By analogy, if Red defects, the best response of Blue is to defect ($-5<\underline{-2}$). It can be seen from Fig. \ref{fig_extensive}(a) that mutual defection is an NE where both players choose the best-response strategies and obtain better utilities, i.e., $(\underline{-2},\underline{-2})$. However, players can obtain higher utilities $(-1,-1)$ by cooperating instead of behaving selfishly. Therefore, the equilibrium of the static game could be  inefficient due to the players' selfishness.

An extensive-form representation often uses a decision tree to describe the sequence of {\color{color1} the players'} actions in the dynamic game. The game tree denotes players as leaves, {\color{color1} denotes} actions as branches, and {\color{color1} denotes} utility functions as the end of the paths. {\color{color1} The two-player example in Fig. \ref{fig_extensive}(a) can also be represented as an extensive form under the assumption that \textcolor{color1}{Blue} moves firstly and \textcolor{color1}{Red} moves secondly {\color{color1} as} shown in Fig. \ref{fig_extensive}(b). The {\color{color1} NE} of a dynamic game can be analyzed using backward induction. The backward induction firstly determines the best response of the last-moving player. Then, the best response of the next-to-last moving player is determined by considering the last player's action.  In the example shown in Fig. \ref{fig_extensive}(b), The best responses of players are obtained from the following steps: 1) if Red {\color{color1} is located} on the left sub-tree, it will defect because defecting yields a higher utility  than cooperating ($-1<\underline{0}$); 2) if Red is on the right sub-tree, it will defect because ($-5<\underline{-2}$); 3) given the best responses of Red (defect, defect), the best response of Blue is to defect because defecting results in a higher utility than cooperating ($-5,\underline{-2}$). Consequently, the NE of the game shown in Fig. \ref{fig_extensive}(b) is mutual defection.}

{\color{color1} The differences and comparisons between the static and the  dynamic games are concluded in Table \ref{tab_gameStaCompare}.
 Generally, the static game is represented as the normal form, and the dynamic game is represented as the extensive form. It should be noted that each type of game can be represented as either the normal form or the extensive form, theoretically. For example, the dynamic game shown in Fig. \ref{fig_extensive}(b) can be also represented as the normal form shown in Fig. \ref{fig_extensive}(a).} The {\color{color1} unique} difference between the static game and {\color{color1} the} dynamic game is that there is a time sequence in a dynamic game. That is, a player in a static game does not know {\color{color1} the} other players' actions before moving while a player in a dynamic game knows all of the preceding actions before making decisions. Besides, {\color{color1} players in a static game tend} to behave {\color{color1} selfishly} due to the stakes of {\color{color1} the} one-shot  {\color{color1}decision making}. Therefore, static games may lead to inefficiency (such as {\color{color1} in} the example {\color{color1} of} Fig. \ref{fig_extensive}(a). {\color{color1} The} players in a dynamic game could cooperate to obtain {\color{color1} an efficient NE} by playing repeatedly and learning from the history.

\begin{table}
	\scriptsize
	\caption{Main differences and comparisons between static and dynamic games}
	\label{tab_gameStaCompare}
	\renewcommand*{\arraystretch}{1}
	\begin{center}
		\begin{tabular}{|p{.12\textwidth}|p{.15\textwidth}|p{.15\textwidth}|}
			\hline
			\textbf{Game type}&\textcolor{color1}{\textbf{Static games}}&\textcolor{color1}{\textbf{Dynamic games}}\\
			\hline
			\textbf{Action sequence}&Players move simultaneously&Players move sequentially\\
			\hline
			\textbf{NE solution}&NE is self-enforcing&NE is obtained by backward induction\\
			\hline
			\textbf{Historical information}&\textcolor{color1}{There is no knowledge of} historical decisions before moving &Know all the historical decisions before moving \\
			\hline
			\textbf{Repeatability}&One-shot& One-shot or repeated\\
			\hline
			\textbf{Selfishness}&Players are selfish&Players may cooperate \\
			\hline	
			\textbf{General representation form}&Normal-form with matrix &Extensive form with a tree\\
			\hline	
			\textbf{Disadvantages}&The NE may be {\color{color1} inefficient}& The {\color{color1}NEs} may be not unique \newline  Require extra memories for storage of history actions\\
			\hline	
		\end{tabular}
	\end{center}
\end{table}


\vspace{6pt}
\subsubsection{Symmetric game {\color{color1} vs.} asymmetric game}
\label {sec_symmetric}
In a symmetric game, the utilities for adopting a particular strategy depend only on the strategies of the other players, not on who is playing. A game is symmetric if the utilities {\color{color1} cannot be} changed by only changing {\color{color1}the identities of the players}. As the example shown in {\color{color1} Fig. \ref{fig_extensive}(a)}, the game is symmetric because playing the strategy ``cooperate" against ``defect" receives identical utilities as playing strategy ``defect" against ``cooperate". On the other hand, a game is an asymmetric game if the strategies employed by players are different. {\color{color1} Decision making} in an asymmetric game {\color{color1} depends} not only on {\color{color1} the} other players' strategies but also on their types. {\color{color1} It is more appropriate to represent the symmetric game as the normal from ``$2\times 2$" matrix example of Fig. \ref{fig_extensive}(a) where each player is given the same changes without assuming the moving sequence \cite{prisner2014game}.}



\vspace{6pt}

\subsubsection{{\color{color1} Complete information} game {\color{color1} vs.}  incomplete information game}
\label {sec_complete}
The complete information game is under an essential assumption that all players have common knowledge of the game's elements. Specifically, the players know  who are playing,  what strategies are available for each player, and  how strategies of each player transform into the utility function. Furthermore, the common knowledge itself is known to each player, i.e., each player knows that {\color{color1} the} other players know the structure of the game. {\color{color1} On the other hand, the incomplete information game, also named Bayesian game, describes a game where
  players do not have common knowledge on the structure of the game. But each player has the belief on the types of the other players; this belief is a common knowledge shared among  the players. The structure of the Bayesian game is introduced in Section \ref{sec_NC} in detail.}

\subsection{Game Theoretical Models}
\label{sec_gameModelsinVN}

Although there are many different types of game-theoretical models in economics, we mainly introduce the games that have been employed to solve {\color{color1} VN problems} from the perspectives of classical games and non-classical games. {\color{color1} The} classical games are further generally classified into two main categories: non-cooperative and cooperative games. The game approaches falling into these categories are shown in Table \ref{tab_game} and are introduced in the following subsections. {\color{color1} It should be noted that different mathematical variables represent different elements or terms of the game models. The elements or  terms that have been defined in  Section \ref{sec_BgGTBasic} or those that have the same meaning will not be introduced or defined repeatedly in the following subsections.}

\begin{table*} 
	\caption{Summary and comparison of key characteristics, solutions, and applications of GT}
	\label{tab_game}
	\scriptsize
	\renewcommand*{\arraystretch}{1}
	\begin{center}
		\begin{tabular}{|m{.002\textwidth}|m{.002\textwidth}|m{.002\textwidth}|m{.002\textwidth}|m{.065\textwidth}|m{.01\textwidth}|m{.18\textwidth}|m{.1\textwidth}|m{.05\textwidth}|m{.25\textwidth}|m{.0001\textwidth}|m{.0001\textwidth}|m{.0002\textwidth}|m{.00001\textwidth}|}
			\hline
			\multicolumn{4}{|c|}{\textbf{Games}}&\textbf{Players}&\textbf{R}&\textbf{Key characteristics}&\textbf{Cooperation} &\textbf{Solution}&\textbf{Application in VNs}&\textbf{C}&\textbf{I}&\textbf{S}&\textbf{D}\\
			\hline
			\multirow{11}{0cm}{\rotatebox[origin=c]{90}{\textbf{Classical}}}&
			\multirow{10}{0cm}{\rotatebox[origin=c]{90}{\textbf{Non-cooperative}}}&\multicolumn{2}{c|}{\rotatebox[origin=c]{90}{\textbf{General}}}&Individual player& FR&Optimal strategies are selected based on {\color{color1} the} other players strategies (compete information) or on the beliefs on {\color{color1} the} other players' types (incomplete information) &Self-enforcing or incentives (e.g., pricing, reputation and credit)&NE& Solve the conflict problems resulted from restricted resources among selfish players  \begin{itemize} [leftmargin=3pt] \setlength{\itemsep}{0pt} \item Security protection \cite{sedjelmaci2015accurate,subba2018game}\item Privacy preserving \cite{fan2018network,wang2019optimization} \item  Routing optimization \cite{tian2017self,das2017new,hu2016novel,assia2019game,goudarzi2018non,goudarzi2019fair,suleiman2017adaptive}, \item MAC access \cite{li2019tcgmac,wang2018application,lang2019vehicle,al2017cooperative}\item {\color{color1} Power} control \cite{goudarzi2018non,goudarzi2019fair,hua2017game,sun2017non}\item VEC-enabled VNs \cite{zhang2019task,zhang2017mobile,zhao2019computation,yu2016optimal,tao2017resource,aloqaily2017fairness,brik2018gss}\item HetVNs \cite{xiao2018spectrum,zhao2019optimal}\item UAV networks \cite{xiao2018uav,alioua2018efficient}\end{itemize}&$\surd$& $\surd$&$\surd$&$\surd$\\
			\cline{3-14}
			&&\multirow{3}{0cm}{\rotatebox[origin=c]{90}{\textbf{Bayesian}}}&\rotatebox[origin=c]{90}{\textbf{Others}}&A nature and players &FR&Players do not have common knowledge of the game but can form beliefs on {\color{color1} the} other players' types&Self-enforcing or incentives&BNE&\begin{itemize} [leftmargin=3pt] \setlength{\itemsep}{0pt} \item Security and privacy-related problems
				\cite{sedjelmaci2014detection,behfarnia2019misbehavior,ying2015motivation,sedjelmaci2016intrusion}\item MAC access competition \cite{kwon2016bayesian} \end{itemize}&&$\surd$&&$\surd$\\
			\cline{4-14}
			&&&\rotatebox[origin=c]{90}{\textbf{Signaling}}&A sender and a receiver&FR&The sender with private information (which cannot be known for the receiver) moves first, based on which the receiver (whose information is public to the sender) moves sequentially&Self-enforcing or incentives&BNE&Resource competition among heterogeneous players (e.g., in HetVNs) \cite{mabrouk2016meeting}&&$\surd$&&$\surd$\\
			\cline{4-14}
			&&&\rotatebox[origin=c]{90}{\textbf{Auction}}&Bidders and an auctioneer&FR&A set of bidders compete for the item (resource) by valuating the item. The valuation of the item is private information of a bidder. &Bidding strategies&NE& Resource  competition among homogeneous players \cite{kumar2016spectrum, hui2019edge} &&$\surd$&&$\surd$\\
			\cline{3-14}
			&&\multicolumn{2}{c|}{\rotatebox[origin=c]{90}{\textcolor{color1}{\textbf{Bargaining}}}}&\textcolor{color1}{Bargaining} parities& FR& Two players negotiate a proportion of resources or determine feasible utility pairs&Disagreement point as threats &NBS &\begin{itemize} [leftmargin=3pt] \setlength{\itemsep}{0pt}\item Optimal routing  selection \cite{kim2016timed}\item Subcarrier or transmission allocation \cite{eze2019design} \item Power rate control \cite{chen2016information} \item Computation resources allocation \cite{huang2017distribute} \end{itemize}& &$\surd$&&$\surd$\\
			\cline{3-14}
			&&\multicolumn{2}{c|}{\rotatebox[origin=c]{90}{\textbf{Stackelberg}}}&Leader and follower&FR& A two-level game where a player with high priority acting as a leader moves first, and the rest acting as followers move sequentially &Self-enforcing or incentives&Backward induction&Solve the heterogeneity in  {\color{color1}decision making} among various players in the scenario where a player with higher priority must move firstly, based on which the {\color{color1} the} other players can decide their actions \cite{brahmi2019cyber,sedjelmaci2018generic,li2016control,aujla2017data,chahal2019network,alioua2019incentive,zhang2017optimal,zhou2018begin,liwang2019game,lin2019vehicle,zhou2018bandwidth} &&$\surd$&&$\surd$\\
			\cline{3-14}
			&&\multirow{4}{0cm}{\rotatebox[origin=c]{90}{\textbf{Potential}}}&\rotatebox[origin=c]{90}{\textbf{Others}}&Individual players&FR &Defines a global potential function to express players' incentives changing their strategies&Potential function& NE&Non-cooperative resource allocation in a distributed algorithm \cite{liu2018computaion,klaimi2018theoretical}& $\surd$&&$\surd$&\\
			\cline{4-14}
			&&&\rotatebox[origin=c]{90}{\textbf{Congestion}}&Individual players&FR&Utility of a player depends on the selected resource and the number of its competing opponents& Preferences order&NE& Congestion control \cite{belghiti20185g,chen2017congestion} &&$\surd$&$\surd$&\\
			\cline{2-14}
			&\multicolumn{3}{c|}{\rotatebox[origin=c]{90}{\textbf{Cooperative}}}&Coalitions&FR &  Players cooperate to form a coalitions, groups, or clusters to coordinate joint strategies to optimize and redistribute the collective
			utility &Externally-enforced agreement or contract (through threat)&Shapley value, Core &Cooperative resource allocation or sharing to maximize the collective utility.\begin{itemize} [leftmargin=3pt] \setlength{\itemsep}{0pt}\item Form clusters to detect or prevent misbehavior nodes \cite{mabrouk2018signaling,boudagdigue2018distributed,halabi2019trust,kumar2015intelligent}
				\item  Cluster-based routing \cite{wu2018computational,sulistyo2019coalitional,assia2019game}\item Resource allocation or sharing \cite{shah2018shapely, yu2015cooperative,hui2017optimal,hui2019game}\end{itemize}&$\surd$&$\surd$&$\surd$&$\surd$ \\
			\cline{1-14}
			\multirow{2}{0cm}{\rotatebox[origin=c]{90}{\textbf{Non-classical}}}&\multicolumn{3}{c|}{\rotatebox[origin=c]{90}{\textbf{Evolutionary}}}&Generations&LR&Players with limited-rationality have initial strategies and
			their offspring will inherit the same
			strategies&Reward or punishment of environment&ESS&\begin{itemize} [leftmargin=3pt] \setlength{\itemsep}{0pt}\item Untrustworthy vehicles remove \cite{tian2019evaluating} \item Cluster formation and optimization \cite{khan2018evolutionary} \item Optimal access or communication mode selection 	\cite{mekki2017proactive,wang2018mode,shattal2018channel}
				\item Spectrum allocation \cite{tian2019channel} \end{itemize}& &$\surd$&&$\surd$\\
			\cline{2-14}
			&\multicolumn{3}{c|}{\rotatebox[origin=c]{90}{\textbf{Matching}}}&Two sets of players&FR&Constructing stable matches among two sets of elements with ordering preferences&Externally-enforced agreement or contract&Gale-Shapley algorithm&\begin{itemize}[leftmargin=3pt] \setlength{\itemsep}{0pt}\item Task (or computation) offloading, allocating or sharing in MEC-VNs \cite{gu2019task,xu2018low,zhou2019computation}
				\item Transmission and reception beamwidths paring in millimeter-wave (mmWave) V2V communication \cite{perfecto2017millimeter} \end {itemize}&&$\surd$&&$\surd$\\
				\hline
			\end{tabular}
			Legend: C=Complete, I=Incomplete, S=Static, D=Dynamic, R=Rationality, FR=Full Rational, LR=Limited Rational
		\end{center}
	\end{table*}

\vspace{6pt}
\subsubsection{Non-cooperative games}
\label{sec_NC}

 Non-cooperative games study the conflict situations where rational players compete for limited resources, and their strategies have conflicting consequences. That is to say, the utility obtained by a player depends not only on his strategy but also on {\color{color1} the} other players' strategies. Each player acts selfishly and independently to maximize its  utility through updating its strategy according to {\color{color1} the} other players' actions. It should be noted that ``independently" does not mean there is no cooperation, but that any desired cooperation among players should be self-enforcing (e.g., through credit or reputation). {\color{color1} Formally, NE is the most common way to define the solution of the non-cooperative game.} Because the non-cooperative game can be easily defined by the {\color{color1} formal} structure, it has been widely used for solving the conflict problems resulted by restricted resources among selfish players in VNs, such as security protection \cite{sedjelmaci2015accurate,subba2018game}, privacy preserving \cite{fan2018network,wang2019optimization}, trust management \cite{mehdi2017game,fan2019trust}, routing optimization \cite{tian2017self,das2017new,hu2016novel,assia2019game,goudarzi2018non,goudarzi2019fair,suleiman2017adaptive}, MAC access \cite{li2019tcgmac,wang2018application,lang2019vehicle,al2017cooperative}, power control \cite{goudarzi2018non,goudarzi2019fair,hua2017game,sun2017non}. Furthermore, it has shown the ability to solve the conflict problems in next-generation VNs, such as VEC-enabled VNs \cite{zhang2019task,zhang2017mobile,zhao2019computation,yu2016optimal,tao2017resource,aloqaily2017fairness,brik2018gss}, HetVNs \cite{xiao2018spectrum,zhao2019optimal},  {\color{color1}and} UAV-assisted VNs \cite{xiao2018uav,alioua2018efficient}. However, the non-cooperative game is characterized by a drawback of the absence of learning or interaction among players. This weakness could be overcome by games with interactions (e.g., Stackelberg game, signaling game, auction game, and bargaining game) or incentive mechanisms such as rewards and punishments.

A \textit{\textbf{Bayesian game}} {\color{color1} $  \mathbf{G}_I=(Nature, \Theta, \mathcal{N}, \mathcal{S}, \mathcal{U})$}, also known as the incomplete information game, defines a situation where players do not have common knowledge of the structure of the game such as the utilities or preferences of {\color{color1} the} other players. In the Bayesian game, a new player named $Nature$ is introduced to assign types {\color{color1} $\Theta=\Theta_1\times\ldots\times\Theta_n$ to players, where $\Theta_{i}=\{\theta_{i}^1,\ldots,\theta_{i}^{N_\theta^i}\}$ is player $i$'s type set of size $N_\theta^i$ . $\Theta$ is based on a probability density function over players' types $f(\Theta)$, which is used to model the uncertainty of the type of the player} Although a player does not know {\color{color1} the} other players' strategies or utilities, he knows the precise way that Nature chooses these preferences. That is to say, $ f(\Theta) $ is common knowledge among the players. In the Bayesian game, player $i$'s strategy $s_{i}^j\in S_i$ is a map $s_{i}^j: \Theta_{i} \rightarrow S_{i}$ that defines an action for each possible type. Each can form his belief on {\color{color1} the} other players' types based on $ f(\Theta) $ to select the best strategies by maximizing {\color{color1} his} utility function {\color{color1} $ u_i\in \mathcal{U}:\mathcal{S}\times \Theta_i \rightarrow \mathbb{R}$}. Similar to the NE for the complete information game (Definition \ref{def_NE}), the NE for Bayesian game is defined as follows.
\begin{myDef}
	\label{def_BNE}
	A strategy profile $ s^{b*}=(s_{1}^{b*},\ldots, s_n^{b*}) $  is a Bayesian Nash equilibrium (BNE) if $s_{i}\left(\theta_{i}\right)= \arg \max _{s_{i}^{\prime}\in S_{i}} \sum_{\theta_{-i}\in \Theta_{-i}} f\left(\theta_{-i} \mid \theta_{i}\right) u_{i}\left(s_{i}^{\prime}, s_{-i}^{*}\left(\theta_{-i}\right), \theta_{i}, \theta_{-i}\right)$, {\color{color1} for each player $ i \in \mathcal{N}$ and for each $ \theta_i \in \Theta_i$,} where $s_{-i}^{*}\left(\theta_{-i}\right)$ denotes the best responses of {\color{color1} the} other players.
\end{myDef}

 Compared with complete information games, Bayesian games are more suitable for solving {\color{color1} the security or privacy-related problems}
 \cite{sedjelmaci2014detection,behfarnia2019misbehavior,ying2015motivation,sedjelmaci2016intrusion} {\color{color1} because a vehicle's information (such as characteristics, types, or preferences) is not completely exposed to the other vehicles.} Specifically, the private information of a vehicle (e.g., the identity and {\color{color1} the} location information) is protected and the attackers or malicious vehicles are concealed. The Bayesian game can help a vehicle form a rational conjecture on {\color{color1} its neighbor} vehicles' types so that it can take the best-response actions according to the prior probability. One possible drawback of Bayesian games is that it may be difficult for Nature to assign a correct prior probability to the {\color{color1} players'} types. Consequently, players in dynamic Bayesian games could obtain more reasonable prior conjectures by updating their beliefs. 
 
\begin{figure}[!hbt] 
 	\centering
 	\includegraphics[width=3.5in]{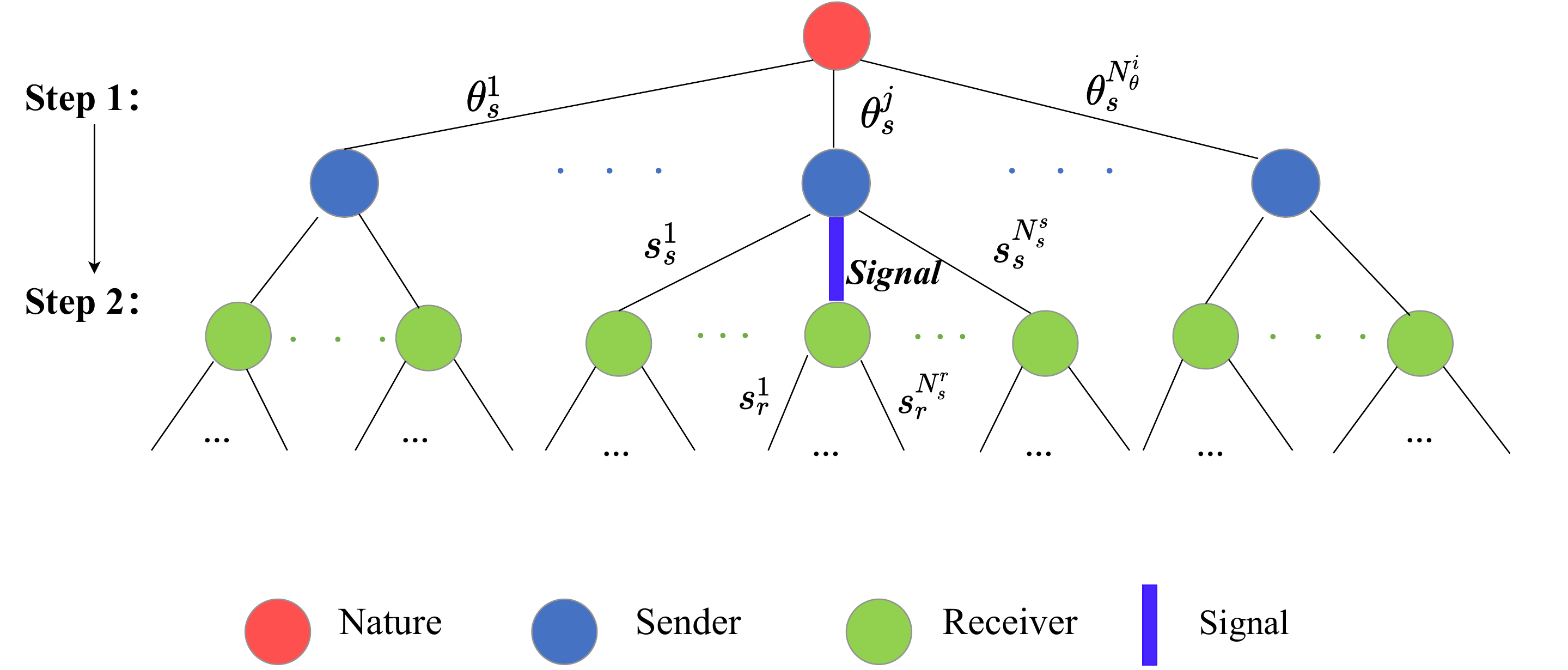}
 	\caption{\textcolor{color1}{Two-player signaling game.}}
 	\label{fig_signaling}
 \end{figure}

 A \textit{\textbf{signaling game }{\color{color1} $ \mathbf{G}_S=(Nature,Signal,\Theta, \mathcal{N}, \mathcal{S}, \mathcal{U})$}} is an example of the dynamic Bayesian game that deals with the lack of information among players. In the signaling game, the player that moves in the first step  $s\in \mathcal{N}$ is informed while {\color{color1} the} other players are not. {\color{color1} The key characteristic of this game is the ``signal" (or ``message") $Signal\in S_s=\{s_s^1,\ldots, Signal,\ldots,s_s^{N_s^s}\}\in \mathcal{S}$}, which is the strategy that is selected by player $s$ to inform {\color{color1} the} other players of the hidden information. The term signal often represents the cost that can be observed by all players {\color{color1} in the game. It} is used as an incentive to encourage or discourage the players with hidden information. Sometimes, it is also viewed as a criterion to differentiate players. For example, an employee with a high scholar degree conveys information about his ability and skill to {\color{color1} an employer}. As {\color{color1} shown} in Fig. \ref{fig_signaling}, a simple signaling game is played between two players: a sender with a set of possible types {\color{color1} $\Theta_s=\{\theta_s^1,\ldots,\theta_s^{N_\theta^i}\}$} and {\color{color1} a} receiver with one type \textcolor{color1}{$\theta_r$}. The sender's type is its private information and {\color{color1} cannot be} known to the receiver. Initially, Nature assigns a type $\theta_s^j\in \Theta_s$ to the sender according to a probability function $f(\Theta_s)$ that is the common knowledge, {\color{color1} which means that the receiver knows the types' probability but does not know} {\color{color1} the specific type}. The sender firstly learns its type $\theta_s^j$ and chooses a {\color{color1} strategy} from {\color{color1} the strategy set  $signal\in S_s$}. After observing the sender's action, the receiver can then decide its strategy from its strategy set {\color{color1} $S_r=\{s_r^1,\ldots,s_r^{N_s^r}\}$}. The critical difference between the signaling game and the Bayesian game is {\color{color1} that the} players {\color{color1} of the signaling games have} asymmetric information.  The signal is suitable for designing {\color{color1} the} incentive schemes for {\color{color1} the} players with asymmetric information. For example, it has been applied to detect and exclude malicious nodes in VNs \cite{haddadou2014job}. The scheme requires the cost of sending, i.e., the more malicious a player is, the higher {\color{color1} the} cost for sending.


An \textit{\textbf{auction game }} {\color{color1} $ \mathbf{G}_A=(Nature,\mathcal{B},\mathcal{V},\Theta, B,\mathcal{U})$} is also an example of Bayesian game that deals with the competition or allocation of an item by enforcing a specific set of rules among a group of bidders. {\color{color1} The players of the auction game are a set of bidders $\mathcal{B}=\{B_1,\ldots,B_n\}$. To win the auction, the bidders first valuate the item's valuation $\mathcal{V}=V_1\times\ldots \times V_n$, where $V_i=\{V_i^1,\ldots,V_i^{N_v^i}\}$ denotes the valuation set of the bidder $B_i$. } Auction games are generally classified into \textit{private value auctions} and \textit{ common value auctions}. In the private value auction, each bidder's valuation of the item is his private information and is unknown to other bidders. In the common value auctions, the valuation of the item is public to every {\color{color1} player}. However, a bidder's personal valuation of the item is private to {\color{color1} the} other bidders. {\color{color1} Similarly to the Bayesian game, Nature assigns types $\Theta$ to the bidders' valuations according to the probability distribution $f(\Theta)$. Then bidders submit their bids $Bid=\{b_1,\ldots, b_i,\ldots,b_n\}$ according to their evaluations of the item and their beliefs on their opponents' evaluations. That is to say, the  bid of bidder $i$ is a map $b_i: \theta_i\rightarrow \mathbb{R}$. The utility function of the winning bidder $B_j$ is calculated by $u_j(b_j,b_{-j})=v_j-b_j$.} Furthermore, it should be noted that the difference between the auction game and the signaling game is that {\color{color1} the} players in the auction game are homogeneous where each of them has private information but {\color{color1} the} players in the signaling game are heterogeneous where only the sender has private information. Auction games can be applied to resource competition or allocation among homogeneous players such as  vehicles \cite{kumar2016spectrum}.



A \textit{\textbf{bargaining game}} {\color{color1} $ \mathbf{G}_B=(\{buyer, seller\}, F, D, \mathcal{U}_B)$} is an example of a non-cooperative game {\color{color1} that models the  negotiation between two players who bargain over dividing the gains of a trade. The formal bargaining game consists of a set of two players (i.e., a $buyer$ and a $seller$), a feasibility set of the two players' possible agreement utilities $F\subseteq\mathbb{R}^{2}$, a disagreement point of the two players' disagreement utilities when the negotiation breaks down $D=\{d_{buyer},d_{seller}\}$, and a utility set of the two players' utilities $\mathcal{U}_B=\{u_{buyer},u_{seller}\}$. } The disagreement point is the key element directly affecting the Nash bargaining solution (NBS) \cite{nash1950bargaining}. The NBS is the unique solution to the bargaining game which should satisfy the following axioms:
\begin{itemize}[itemsep=3pt,topsep=3pt]
	\item Invariant to affine transformations or equivalent utility representations
	\item Pareto optimal
	\item Independent of irrelevant alternatives
	\item Symmetric
\end{itemize}

	Considering the above-mentioned formal bargaining game, it is proved in\cite{nash1950bargaining} that the solutions satisfying the above axioms are $(x,y)\in F$ which maximizes the expression {\color{color1} $(u_{buyer}(x)-u_{buyer}(d_{buyer}))(u_{seller}(x)-u_{seller}(d_{seller}))$}. {\color{color1} Bargaining} games can stimulate cooperation and agreement between the two players  through repeated negotiation. However, long-time bargaining could {\color{color1} lead to inefficiency} due to the increasing loss {\color{color1} of} delay. Both bargainers aim to reach an agreement {\color{color1} earlier} because their utilities will be discounted over time, especially in the time-sensitive VNs. The preferences over time are the driving force of the game. Therefore, the payers' preferences can be given as $\left(\delta_{i}\right)^{t}\cdot u_{i}\left(d_{i}\right)$, where $u_i$ is an increasing and concave function for any $0<\delta_{i}<1$. However, the ``early agreement" depends on the degree of uncertainty, that is, the quantity of information on the adversary's preferences. The incomplete information has a {\color{color1} negative} effect on the efficiency of the bargaining game \cite{cramton1984bargaining} {\color{color1} because the less information a vehicle has on its adversary's preferences, the longer time it will cost for negotiating an agreement on the bargaining.}  Regarding {\color{color1} the application of the bargaining game in VNs}, the bargaining game can be used to model the negotiation of resource allocation between an SP (that sets the resource budget) and a requester (that requests the resource). {\color{color1} The typical negotiations include the} subcarrier allocation \cite{eze2019design}, {\color{color1} the} power rate allocation \cite{chen2016information}, and {\color{color1} the} computation resources allocation \cite{huang2017distribute}.

A \textit{\textbf{Stackelberg game}} {\color{color1}$ \mathbf{G}_S=(\{l,\mathcal{F}\}, \mathcal{S}_{S}, \{u_l,\mathcal{U}_f\})$} is a {\color{color1}  two-level}  non-cooperative game with cooperation between two types of players, i.e.,  a leader $l$ with {\color{color1} a} higher priority and a set of followers $\mathcal{F}=\{f_1,\ldots,f_{Nf}\}$ with lower priorities. {\color{color1}The strategy set is denoted as $\mathcal{S}_S=S_l\times S_{f}$, where $S_l=\{s_l^1,\ldots, s_l^{N_s^l}\}$ denotes the strategy set of the leader and $S_f=S_{f}\times \ldots \times S_f^{N_f}$ denotes the strategy sets of the followers. $u_l$ and $\mathcal{U}_f=\{u_f^1,\ldots,u_f^{N_f}\}$ are utility functions of the leader and follower, respectively. In the Stackelberg game, the players act as follows: 1) the leader moves firstly; 2) the leader announces its strategy to the followers; 3) each follower $f_j\in \mathcal{F}$ chooses a strategy from its strategy set $s_{fj}^k\in S_{fj}=\{s_{fj}^1,\ldots, s_{fj}^{N_s^{fj}}\}$  after observing the leader's strategy; 4) each follower announces its strategy to the leader. Obviously, it is an extensive-form game incorporating with an evolution fashion. The NE can be solved by using backward induction where the calculation moves `backwards' from followers. The leader selects the best response strategy by anticipating the followers' best responses. Firstly, the best response function of each follower $j$ is calculated by $s_{fj}^*=\arg \max_{s_{fj}^k\in S_{fj}}  u_{f^j}(s_{fj}^k,s_l)$, where $s_l$ is the leader's strategy and is considered as a variable. Secondly, the best response of the leader can be calculated by considering each follower's best response as an input of the leader's utility function $s_l^*=\argmax_{s_l\in S_l,f_j\in\mathcal{F}} u_l (s_l,s_{fj}^*)$.} The one-leader multi-follower Stackelberg game can be further extended into multi-leader multi-follower Stackelberg game. The Stackelberg game is mostly used for solving the heterogeneity among various players \cite{brahmi2019cyber,sedjelmaci2018generic,li2016control,aujla2017data,chahal2019network,alioua2019incentive,zhang2017optimal,zhou2018begin,liwang2019game,lin2019vehicle,zhou2018bandwidth}. The advantage of the Stackelberg game is that it can be applied to the scenario where the leaders with higher priority must move firstly, and the followers react to the leaders' actions by deciding their optimal actions. However, a drawback of the Stackelberg game is that the Stackelberg  equilibrium could be a worse result than the NE due to the hierarchical decision process.

A \textit{\textbf{potential game}} {\color{color1} $\mathbf{G_P}=(\mathcal{N}, \mathcal{S},\mathcal{U},\Phi)$} is a subclass of the normal-form game that defines a global potential function $\Phi$ to express {\color{color1} the} players' incentives {\color{color1} of} changing their strategies. {\color{color1} The potential function is defined as $\Phi: \mathcal{S}\rightarrow \mathbb{R}$, which has the information on the utility of each player and quantifies the disagreement among players.} Therefore, it can be also interpreted  {\color{color1} as the} route leading to NE. The existence and convergence of NE can be easily proved based on the potential function \cite{liu2018computaion,klaimi2018theoretical,liu2018computaion,zhang2019task,zhao2019computation }. Every exact potential game with finite strategy sets always has a NE and provides the finite improvement property.

A \textit{\textbf{congestion game}} {\color{color1} $\mathbf{G_{Con}}=\left(\mathcal{N}, \mathcal{R},\mathcal{S},\left(c^{r_j}\right)_{r_j \in \mathcal{R}},\mathcal{U}\right)$} is a special case of the potential game where the utility of each player depends on the selected resource and the number of its competing opponents.  {\color{color1} $\mathcal{R}=\{r_1,\ldots,r_j,\ldots,r^{N_r}\}$} denotes a set of resource. The strategy of player $i$ is selecting a subset  $S_i\subseteq  R$ from the resource resource set. $\left(c^{r_j}(x)\right)_{r_j \in \mathcal{R}}$ is the payment for the resource $r_j$ if $x$ competing players also select $r_j$ as their strategies, and $u_{i}\left(s_{i}, s_{-i}\right)=\sum_{j \in s_{i}} c^{r_j}\left(x\right)$ denotes the utility of player $i$. It has been proved that any finite congestion game has a pure strategy equilibrium. The congestion game can solve the congestion problem in VNs by modeling the transmission flows where each selfish node aims to minimize the routing delay\cite{belghiti20185g,chen2017congestion}. In such a scenario, the possible routes can be viewed as resources, and transmission delay on the routs can be modeled as the node's costs (utility).  The costs depend on the selected routes and the number of {\color{color1} the} other players {\color{color1} choosing the same routes}.

\vspace{6pt}
\subsubsection {\textbf{Cooperative (coalitional) game} {\color{color1} $\mathbf{G_{Coo}}=\left(\mathcal{N},V\right)$}}
\label{sec_coopereative game}
A cooperative (or coalitional) game  focuses on how a set of players cooperate to form a \textit{coalition} to optimize and redistribute the collective utility $V$. The key distinguishing feature of a cooperative game is that {\color{color1} the} cooperative behaviors are external-enforced by an agreement or {\color{color1} a} contract, as opposed to the non-cooperative game where {\color{color1} the} cooperation among selfish players is self-enforced. Formally,  a cooperative game consists of an $n-$ player set $\mathcal{N}=\{1,\ldots,n\}$ and a characteristic function $V: 2^{n} \rightarrow \mathbb{R},\ V(\emptyset)=0$ that maps players to real values. The characteristic function is used to quantify the value or worth of coalitions in the cooperative game. Specifically, $V(\mathcal{N})=\sum_{i\in\mathcal{N}}u_i $ represents the value of the game $ \mathbf{G_{Coo}}$, which is the sum of each player's utility.  

\begin{myDef}
	\label{def_coalitionParition}
	A coalition partition is defined as the set $\Pi=\{C_{1}, \ldots, C_{l}\}$ where $\forall j, C_{j} \subseteq \mathcal{N}$ are disjoint coalitions, such that $\bigcup_{j=1}^{l} C_{j}= \mathcal{N}$
\end{myDef}
The coalitional game is non-cohesive if a set of disjoint coalitions are formed. Otherwise, it is a \textit{grand coalition} if a single coalition is formed by jointing all players together. The cooperative game focuses on the utility of the coalition rather than the individual player. 

Under the assumption that the grand coalition can be formed, different concepts are proposed for solutions of the coalitional game. The solution of the cooperative game is formally defined as an allocation vector $A\in \mathbb {R}^{n}$ where the gain of the coalition is allocated among its players fairly. Two popular solution concepts  solving the cooperative game (i.e., Shapley value and core) are introduced as follows:

 \vspace{6 pt}
\begin{itemize}[itemsep=3pt,topsep=3pt]
\item  The \textbf{Shapley value} presents a solution of distributing the total gain of the coalitional game to each player fairly.
\begin{myDef}
	\label{def_shapley}
	The Shapley value of the coalitional game $\left(\mathcal{N},V\right)$ is the amount of gain that allocated to each player $i\in \mathcal{N}$:
	\begin{equation}
	\label{eq_Shapley}
	A_i=\frac{1}{n!}\sum_{C\subseteq \mathcal{N} \backslash\{i\}}\frac{|C|!\left(n-|C|-1\right)}{n!}\left(V(C \cup\{i\})-V(C)\right),
	\end{equation}
	where $n$ and $|C|$ denote the size of the player set $\mathcal{N}$ and the size of the coalition $C$, respectively. 
\end{myDef}

\item The \textbf{core} of the cooperative game presents a situation where there is no coalition that has a value greater than the sum of its players' utilities.

\begin{myDef}
	\label{def_core}
	The core of the coalitional game $\left(\mathcal{N},V\right)$ is a set of allocation vector $A$:
	\begin{sequation}
	\label{eq_core}
	Core=\left\{A\in \mathbb{R}^{n}: \sum_{i \in \mathcal{N}} A_i=V(\mathcal{N}); \  \sum_{i \in C} A_{i} \geq V(C), \forall C \subseteq \mathcal{N}\right\}.
	\end{sequation}	
\end{myDef}
\end{itemize}

Cooperative games are often applied to situations where intelligent nodes aim to form alliances to maximize their group interests. {\color{color1} Coalitional games are used to help vehicles form clusters to avoid broadcast storm \cite{wu2018computational,sulistyo2019coalitional,assia2019game}.} Moreover, vehicles can form clusters to prevent selfish or malicious behaviors by using coalitional games \cite{mabrouk2018signaling,boudagdigue2018distributed,halabi2019trust,kumar2015intelligent}. Besides, coalitional games are used for resources allocation or sharing problems in VNs where players in the same coalition can share the resources \cite{yu2015cooperative,hui2017optimal,hui2019game}. A stable coalition with fair resource allocation for a coalitional game ensures that the coalition is immune to defections of players. One critical and difficult issue is designing an enforced agreement or commitment (through threat) to guarantee the stability of coalitions in VNs. An authority (e.g., base station or cloud) focusing on the collective utility maximization is required to enforce commitments.

 %


\vspace{6pt}
\subsubsection {Non-classical games}
\label{sec_NonClassicalgames}
Non-classical games generally include the evolutionary game and the matching game.

\begin{figure}[!hbt] 
	\centering
	\includegraphics[width=3.5in]{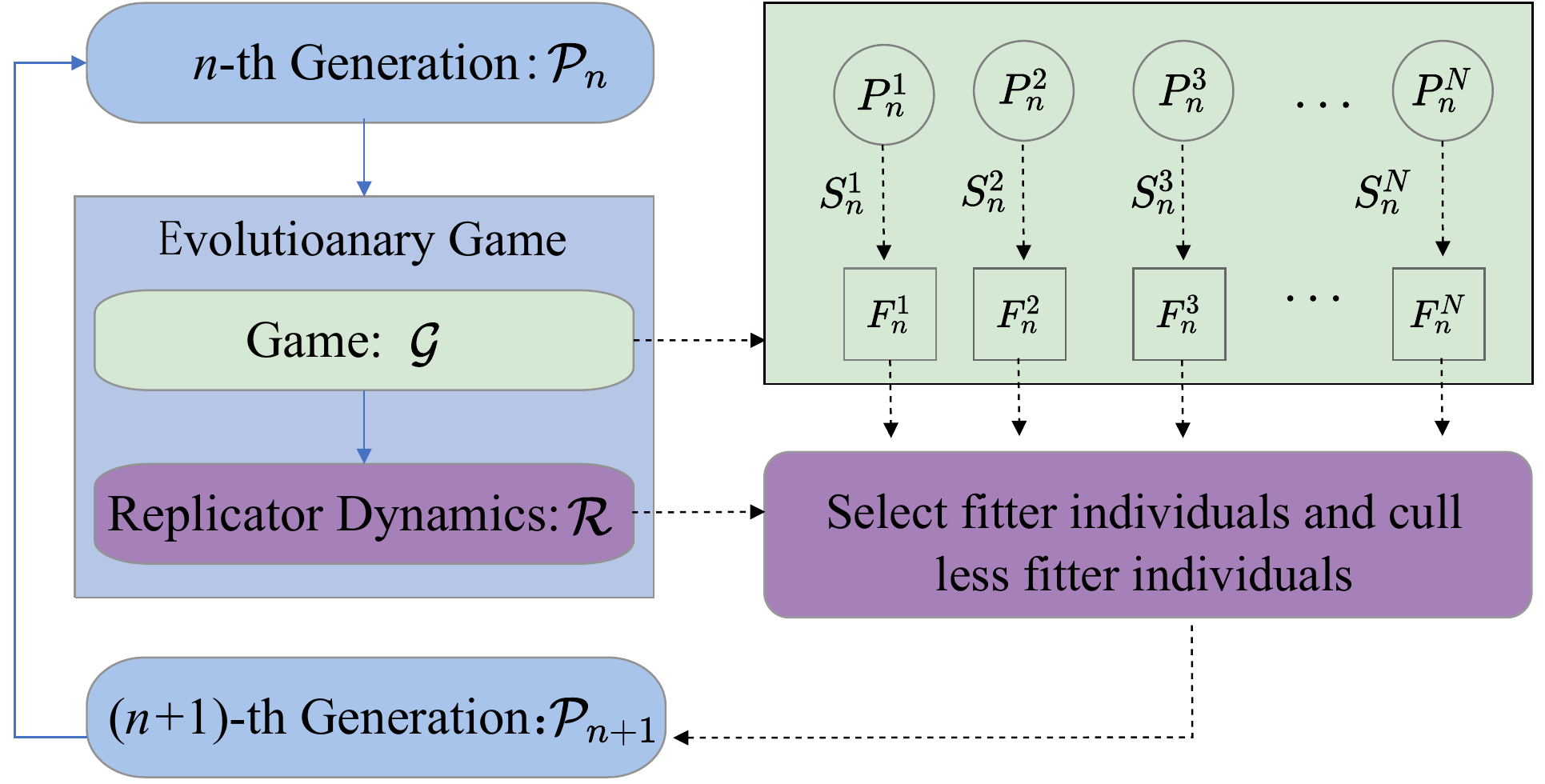}
	\caption{\textcolor{color1}{Evolutionary game.}}
	\label{fig_evolutionary}
\end{figure}

An \textit{\textbf{evolutionary game}} {\color{color1} $G_E=\left(\mathcal{P},\mathcal{G},\mathcal{R}\right)$} merges the classical game theory with the theory of population ecology, which includes the components of population {\color{color1} $\mathcal{P}_n=\{\mathcal{P}_1,\ldots,\mathcal{P}_n,\ldots\}$,} {\color{color1} the} game $\mathcal{G}$, and {\color{color1} the} replicator dynamics $\mathcal{R}$. The evolutionary game is played repeatedly among population in the following phases as shown in Fig. \ref{fig_evolutionary}. {\color{color1} Firstly, the $n$-th generation of population $\mathcal{P}_n=\{P_n^1,\ldots,P_n^N\}$ is input in the game system $\mathcal{G}=\{\mathcal{P}_n,\mathcal{S}_n,\mathcal{F}_n\}$. Secondly, the game rules are designed to model the competition among individuals according to their strategies $\mathcal{S}_n=\{S_n^1\times \ldots\times S_n^N\}$. Then, the utilities in unit of \textit{fitness} $\mathcal{F}_n=\{F_n^1,\ldots,F_n^N\}$ are calculated to quantify the production rate of individuals. Thirdly, the population heredity is fulfilled by replicator dynamics based on the resulting fitness values. Specifically, individuals with higher fitness values are selected to produce replica and those with lower fitness values are culled. Fourthly, superior behaviors are inherited and the new $(n+1)-$th population $\mathcal{P}(n+1)$  is produced. The game will be executed repetitively until an equilibrium is achieved.}

 The \textit{evolutionary stable strategy (ESS)} {\color{color1} is reached} after several generations if the youngest players' strategies are optimally adapted to the environment. An ESS is defined as a strategy that {\color{color1} it cannot be invaded by any alternative rational strategies once all players of a population adopt.} Therefore, an ESS can be viewed as a refinement to an NE. The difference between these two concepts is that players will stay in an NE for a short term but eventually deviate from it if there exist better strategies to make them better off. By contrast, the ESS assumes that natural selection excludes the strategies with lower utilities.

The main advantage of using the evolutionary game is that players with bounded rationality can choose their initial strategies (e.g., be honest or deceptive), and their offspring will inherit the same strategies. Then, the players will receive a reward or punishment from the environment. Players with high fitness value will be inherited by the next generation using the replicator dynamics process. 
However, one weakness of the evolutionary game is that it may lead to a high convergence time, which is unacceptable in the delay-sensitive {\color{color1} VNs}. Another issue should be aware {\color{color1} of} is the altruism. Although an altruistic behavior may be advantageous {\color{color1} from a collective perspective}, the selfish players who refuse to behave altruistically will obtain high fitness functions and be selected with {\color{color1} a} higher probability. In the worst case, selfishness may eventually overwhelm altruism. Therefore, it is critical to design incentive or punishment approaches for the environment to prevent selfish behaviors in the evolutionary game. In conclusion, by designing effective schemes of competition, selection, and heredity, the evolutionary game helps {\color{color1} non-rational players to dynamically learn from the environment and make individual optimal decisions.} For example, it has been used to remove the untrustworthy vehicles from networks through nature selection \cite{tian2019evaluating}, cluster formation \cite{khan2018evolutionary}, optimal access strategy or communication mode selection \cite{mekki2017proactive,jia2019bus,wang2018mode,shattal2018channel}, and spectrum allocation \cite{tian2019channel}.

A \textit{\textbf{matching game}} {\color{color1} $ \left(\{M_1, M_2\},O,\mathcal{M} \right)$} aims at constructing stable matches $\mathcal{M}$ among two sets of players,\textcolor{color1}{i.e.,} $M_1=\{m_1^1,\ldots, m_1^{n_1}\}$ \textcolor{color1}{and} $M_2=\{m_2^1,\ldots, m_2^{n_2}\}$. Each player in each set $m_i^j\in M_i,i\in\{1,2\}$ has an ordering preference to the players in the other set, i.e.,
{\color{color1} 
\begin{equation}
\label{eq_matchPreference}
\begin{aligned}
 &O(m_i^j)=\{m_{-i}^{k_1}, m_{-i}^{k_2},\ldots, m_{-i}^{k_n}\}, \\& \forall m_i^j \in M_i, \ m_{-i}^k \in M_{-i}, \ i\in\{1,2\},\ -i\neq i,
 \end{aligned}
\end{equation}
where $m_{-i}^{k_j}$ denotes the other player.} Matching can be classified into one-to-one, one-to-many, and many-to-many matching. A matching is stable if for each matching $(m_i^j,m_{-i}^k)\in \mathcal{M}$ there could not find another match $(m_i^{j'},m_-i^{k'})$ where both participants prefer each other over their current partners. The gale-Shapley algorithm gives the solution to finding stable matching. Solutions of matching games are not unique, and there could be more than one stable matching outcome. The matching game can be applied to task offloading, allocating or sharing in \textcolor{color1}{VEC-enabled} VNs \cite{gu2019task,xu2018low,zhou2019computation,gu2016exploiting}. The interactions between the tasks and edge nodes can be modeled as \textcolor{color1} {players} of the matching game. \textcolor{color1}{Each} task prefers to be offloaded with high QoS, resulting in a preference order of the edge nodes. Similarly, each edge has a preference list of the tasks because it prefers to provide resources at a low cost. 


\section{ Game theory meets vehicular networks}
\label{sec_GTMeetsVNs}
\textcolor{color1} {The intelligent nodes in VNs have stringent requirements on network performances}, such as 1) low latency, 2) high reliability, and 3) high security and privacy. With the explosive growth of vehicular devices, various services are emerging, including safety-related services, traffic management services, and infotainment services. \textcolor{color1}{It could be complex to design efficient mechanisms to satisfy various,  stringent, and possibly conflicting requirements for various intelligent nodes in VNs. GT has the ability to solve the difficult-to-model and conflicting problems of {\color{color1}decision making} through analyzing the complex interactions among multiple decision makers under the limited resources \cite{friedman1986game}.} These decision makers cooperatively or competitively take rational actions which may have conflicting consequences. The following key characteristics of GT have advantages for solving the corresponding {\color{color1} VN issues}.

\begin{itemize}[topsep=3pt, itemsep=3pt]
	\item  GT provides basic elements to model the characteristics of VNs and give solutions to the corresponding model. A mapping of GT elements to VN components is presented in Table   \ref{tab_mapVNGT}.
	\begin{itemize}[label=$-$,leftmargin=10pt,topsep=3pt, itemsep=3pt]
		\item ``Players" model intelligent nodes in VNs, such as vehicles, infrastructures, and VEC servers. The \textcolor{color1}{salient features of vehicles such as} the velocity, direction, power, and location can be modeled as the characteristics of the players.

		\item``Strategies" model the actions taken by intelligent nodes, such as security mechanism, network/server selection, resource price, and requested bandwidth.
		
		\item ``Utilities" evaluate the motivations or objectives of the intelligent nodes. The utility involves the quantification for 1) characteristics of vehicular environment such as the mobility pattern of vehicles, traffic density, and channel properties; 2)  requirements on network performances such as delay, throughput, reliability, and security; 3) capabilities of the intelligent nodes such as the computation, space, energy, and attack capabilities; and 4) the limited resources such as the bandwidth, battery life, and spectrum.
		
		\item Solutions of GT can deal with the possible conflicting {\color{color1}  {\color{color1}decision making}} in VNs, describing how the game is played, which strategy is  optimal, and the result. The solutions help the intelligent nodes dynamically decide the optimal strategies that satisfy their requirements or preferences {\color{color1} according to the situated vehicular context.}
	\end{itemize}
	
	\item  GT offers a powerful tool to model and analyze the complex interactions among intelligent nodes in VNs. The interaction is one of the efficient approaches to overcome the limited resources and restricted performance requirements. For example, the vehicles within the RSU server range compete for downloading the same resource in a low delay. The cooperative game enables the RSU lacking resources to coordinate and cooperate with another \textcolor{color1}{lacking-resource} RSU to obtain the shared resources.
	
	\item  GT gives an insight into the uncertainty in VNs. The incomplete information of GT precisely captures the uncertainty of the vehicular environment and a node's belief or conjecture on other nodes' strategies, especially the malicious nodes.

	\item  GT, such as the cooperative game or the evolutionary game, supports the scalability and flexibility of VNs because it allows the nodes to dynamically join or leave the ``coalition" or the ``population".
	
	\item GT solves problems induced by \textcolor{color1}{limited resources which is} one of the major concerns in VNs. GT offers efficient resource allocation and sharing methods to fully utilize the resources. Some models, such as the biding game, provide pricing-based mechanisms for efficient resource allocation to multiple competitive requesters.  \textcolor{color1}{Besides}, some models \textcolor{color1}{,} such as the cooperative or bargaining games \textcolor{color1}{,} enable intelligent nodes to negotiate satisfactory solutions for resource sharing.		
\end{itemize}

{\color{color1} In this survey, the GT solutions for VNs are classified as a layered taxonomy based on the layered structure of VNs. Specifically, the existing GT solutions for current VNs and next-generation VNs are categorized in layered structures corresponding to the protocol stack layered structure of the traditional VNs and the hierarchical structure of the next-generation VNs, respectively. The layered structure of the contents in this survey is shown in Fig. \ref{fig_fit}(b), where each section corresponds to a specific layer of the VNs, as shown in Fig. \ref{fig_fit}(a).}

\begin{table*}
	\caption{General mapping of elements in VNs to a game}
	\footnotesize
	\label{tab_mapVNGT}
	\renewcommand*{\arraystretch}{1.2}
	\begin{center}
		\begin{tabular}{|p{.15\textwidth}|p{.8\textwidth}|}
			\hline 
			\textbf{Game Components} &\textbf{Corresponding elements of VNs}\\ 
			\hline
			Players& The intelligent nodes in VNs, such as vehicles, RSUs, SPs, VEC servers, cloud servers\\
			\hline Strategies& Actions taken by the intelligent nodes, such as security protection mechanism, network selection, resource price, \textcolor{color1}{and requested bandwidth}\\
			\hline
			Utility/utility functions&Evaluations of the intelligent nodes' motivations or objectives, such as delay, throughput, \textcolor{color1}{and reliability}\\
			\hline
			Solutions& Describe how the game is played, which strategy is optimal, and give the optimal (usually equilibrium) strategies that satisfy the intelligent nodes' requirements or preferences\\
			\hline
		\end{tabular}
	\end{center}
\end{table*}

\section{Game Theory in Current VNs}
\label{sec_GTinVNs}

GT has been applied to many possible applications in VNs.  Related \textcolor{color1}{studies} on game-theoretic approaches in VNs are introduced from the perspectives of security protection and QoS guarantee in this section.

	\subsection{Game Theory for Security Protection in VNs}
	\label{sec_security}
		The features of VNs may result in security vulnerabilities that make the communication suffer from various types of internal and external attacks \textcolor{color1}{. These attacks will} lead to three main concerns in the design of secure VNs, i.e., security, privacy and trust. \textcolor{color1}{Accordingly,} this section introduces the use of GT in \textcolor{color1}{cyber security} \cite{mabrouk2018signaling,sedjelmaci2014detection,sedjelmaci2015accurate,subba2018game,brahmi2019cyber,sedjelmaci2018generic}, \textcolor{color1}{location privacy}  \cite{fan2018network,wang2019optimization,ying2015motivation,al2018teaming,zhang2017otibaagka,zhang2019pa,mukherjee2019efficient}, and trust management \cite{halabi2019trust,boudagdigue2018distributed,kumar2015intelligent,mehdi2017game,fan2019trust,haddadou2014job,tian2019evaluating} in VNs.

	\begin{table*}
	\scriptsize
	\caption{Game models for cyber security in VNs}
	\label{tab_intrusion}
	\renewcommand*{\arraystretch}{1.08}
	\begin{center}
		\begin{tabular}{|p{.02\textwidth}|p{.055\textwidth}|p{.19\textwidth}|p{.08\textwidth}|p{.015\textwidth}|p{.01\textwidth}|p{.015\textwidth}|p{.05\textwidth}|p{.018\textwidth}|p{.015\textwidth}|p{.12\textwidth}|p{.03\textwidth}|p{.03\textwidth}|p{.01\textwidth}|}
			\hline 
			\multirow{2}{1cm}{\textbf{Ref}}&\multirow{2}{2cm}{\textbf{Game}}&\multirow{2}{3cm}{\textbf{Player and Strategy}}&\multirow{2}{2cm}{\textbf{Solution}}&\multicolumn{3}{c|}{\textbf{Role}}&\multirow{2}{1cm}{\textbf{Incentive schemes}}&\multirow{2}{2cm}{\textbf{L}}&\multirow{2}{2cm}{\textbf{CL}}&\multirow{2}{2cm}{\textbf{Attacks defended}}&\multicolumn{3}{c|}{\textbf{Performance evaluation}}\\
			\cline{5-7}\cline{12-14}
			&&&&\textbf{IDS}&\textbf{IPS}&\textbf{IRS}&&&&&\textbf{DR}&\textbf{FPR}&\textbf{COV}\\
			\hline
			\cite{sedjelmaci2014detection} &Bayesian game&\textbf{ RSU}: $\left\{Detect, Wait \right\}$ \newline \textbf{Attacker}: $ \left\{Attack, Wait \right\}$&Mixed strategy BNE&$\surd$&$\surd$&$\times$&$\times$&$\times$&$\times$&False alert generation&98.2\%&1.39\%&8.79\\
			\hline
			\multirow{4}{1cm}{\cite{sedjelmaci2015accurate}} &\multirow{4}{1.5cm}{Non-cooperative game}&\multirow{4}{3.8cm}{\textbf{LIDS}  $\left\{Detect,Wait\right\} $ \newline \textbf{Vehicle members}:\newline $ \left\{Attack, Be normal\right\} $}&\multirow{4}{2cm}{Pure strategy NE}&\multirow{4}{0.5cm}{$\surd$}&\multirow{4}{0.5cm}{$\surd$}&\multirow{4}{0.5cm}{}&\multirow{4}{4cm}{Reputation}&\multirow{4}{1cm}{SVM}&\multirow{4}{1cm}{$\times$}&Selecting service\newline Blackhole&98.97\%& 1.05\%&\\
			\cline{11-13}
			 &&&&&&&&&&Packet duplication\newline Resource exhaustion&98.37\% &1.04\% &14.69\\
			\cline{11-13}
			 &&&&&&& &&&Wormhole&98.37\% &1.06\%&\\
			\cline{11-13}
			&&&&&&&&&&Sybil&96.31\% &0.81\%&\\
			\hline
			\multirow{4}{1cm}{\cite{subba2018game}}&\multirow{4}{1cm}{Non-cooperative game}&\multirow{4}{3cm}{\textbf{IDS}:$  \left\{Monitor, Not \ Monitor \right\}$ \newline \textbf{ Malicious vehicle}: \newline$ \left\{Attack, Wait\right\} $}&\multirow{4}{1.5cm}{Mixed strategy NE}&\multirow{4}{1cm}{$\surd$}&\multirow{4}{0.5cm}{}&\multirow{4}{0.5cm}{}&Reputation \newline Price&---&$\surd$&Selective forwarding&90.85\%&8.8\%&---\\
			\cline{11-13}
			&&&&&&&&&&Blackhole&94.11\%&4.6\%&\\
			\cline{11-13}
			&&&&&&&&&&Wormhole&86.45\%&7.1\%&\\
			\cline{11-13}
			&&&&&&&&&&DoS&94.94\%&8.6\%&\\
			\hline
				\cite{mabrouk2018signaling}& Coalition game \newline Signaling game &\textbf{CH}: $ \left\{Idle, Defend\right\} $  \newline\textbf{ Vehicle member}:\newline $ \left\{Attack, Cooperate\right\} $&Pure strategy and mixed-strategy BNEs &$\surd$&&&$\times$&$\times$&$\surd$&General attack&---&---&---\\
			\hline
			\cite{brahmi2019cyber,sedjelmaci2018generic} & Stackelberg game&\textbf{IDA}: Launch strategies of the secondary agents\newline \textbf{Secondary agents}: \newline$ \left\{IDS,IPS,IRS\right\} $&Mixed strategy NE&$\surd$&$\surd$&$\surd$&$\times$&$\times$&$\surd$&General attack&--- &2.9\%&40\\
			\hline
		\end{tabular}
	
	Legend: DR= Detection Rate, FPR=False Positive Rate, COV=Communication Overhead (KB), L=Learning, CL=Clustering
	\end{center}
\end{table*}

\vspace{6pt}
\subsubsection{ GT for cyber security}
\label{sec_intrusion}	

Buinevich et al. \cite{buinevich2019forecasting} list the \textcolor{color1}{top ten} severe cyber-attacks on VNs, which are categorized as threats to confidentiality, availability, and integrity. In confidentiality attacks, malicious nodes aim to access the \textcolor{color1}{cluster head (CH)} by eavesdropping on the communication within \textcolor{color1}{their radio ranges}. In availability attacks (e.g., denial of service (DoS)), the adversaries attempt to exhaust bandwidth resources by disturbing or jamming the communications. Attacks targeting {\color{color1} on} communication integrity try to inject false information, change the order of messages, or replay old messages. Consequently, it is essential to provide a reliable security framework for VNs to deter these cyber-attacks. The frameworks developed for VN’s cyber security can be classified into the following systems, i.e., intrusion detection system (IDS), intrusion prediction system (IPS), and intrusion reaction system (IRS). IDS \textcolor{color1}{and} IPS can reliably detect malicious vehicles, and are often used as a second line of defense after the cryptographic \textcolor{color1}{protection}. GT can be used to model the interaction between malicious nodes and \textcolor{color1}{intrusion decision agents}. Moreover, GT is proved to be useful in increasing detection or prediction accuracy and decreasing communication overhead.


 To predict the malicious behavior in VNs, Sedjelmaci et al. \cite{sedjelmaci2014detection} propose a Bayesian game-theoretic intrusion detection and prevention scheme to predict the future behaviors of monitored vehicles. The vehicles within the communication range of the monitored vehicle are called monitoring vehicles, who monitor the behavior of a monitored vehicle when it broadcasts an alert message. This paper considers three types of alerts which are post crash notification (PCN), road hazard condition notification (RHCN), and slow vehicle advisor (SVA). The monitoring vehicle can detect a malicious vehicle {\color {color1} by checking the information of position, speed, and lane in the beacon message} because the malicious node decelerates and changes lane after disseminating alert messages. The beacon message of a monitored vehicle $V$ includes its position $\left(x_{beacon_{v}},y_{beacon_{v}}\right)$, the message generation time $t_{beacon}$, and its $v_{speed\_beacon}$ speed. The alert message of $V$ includes the alert type (PCN, RHCN, or SVA), new position $\left(x_{alert_{v}},y_{alert_{v}}\right)$, alert generation time $t_{alert}$, and speed $v_{speed\_alert}$. The speed of $V$ between the alert generation and a subsequent beacon message can be obtained as:
 \begin{equation}
 \label{eq_monitoredSpeed}
Speed_{v}=\frac{\sqrt{\left(x_{\text {beacon}_{v }}-x_{\text {alert}_{v}}\right)^{2}+\left(y_{\text {beacon}_{v}}-y_{\text {alert}_{v }}\right)^{2}}}{t_{\text {beacon}}-t_{\text {alert}}}.
 \end{equation}

The monitored vehicle is considered as an attacker generating a false alert if the equation (\ref{eq_monitor}) holds. In this case, the monitored vehicle does not change its lane.
 \begin{equation}
 \label{eq_monitor}
 \left\{\begin{array}{c}
 Speed_{v} > v_{ {speed\_alert }} \\
 \text { or } \\
 v_{ speed\_beacon }\geq v_{ speed\_alert }
 \end{array}.\right.
 \end{equation}
 
 The RSU and the suspected vehicle start a Bayesian game if a malicious behavior of the monitored vehicle is detected. The RSU can perform \textit{Detect} or \textit{Wait} and the monitored vehicle can \textcolor{color1}{}\textit{Attack} or \textit{Wait}. As a result, it is proved that there is a mixed-strategy BNE $\left\{\text{RSU} \ \left(Detect ,p^{*}\right), \ \text{Attacker} \ (Attack, q^*)\right\}$. The monitored vehicle is a malicious vehicle when the probability $ q > q^*  $, and the RSU triggers its detection when $ p < p^* $. The suspected vehicle that is predicted as an attacker will be added to a blacklist by the RSU. When the proportion of malicious nodes is 40\%, this scheme is able to detect and predict malicious behaviors with high rate of detection $(>98\% )$ and low rate of false-positive $ (<2\%) $. However, the performance degrades significantly when the {\color{color1} the number of} malicious nodes exceed 40\%. It is benefited from the Bayesian game to model the RSU's incomplete information on the potential attackers. However, one weakness is that it only focuses on detecting the \textit{false alert's generation attack} without considering other types of {\color{color1} possible} attacks in VNs.

The work \textcolor{color1}{in \cite{sedjelmaci2014detection}} is further extended into a lightweight hybrid IDS \textcolor{color1}{in \cite{sedjelmaci2015accurate}} by considering both the mobility of vehicles and various types of attacks. Three intrusion detection agents are considered in the hybrid IDS framework, i.e., local intrusion detection system (LIDS) at the cluster member level, global intrusion detection system (GIDS)  at the CH level, and global decision system (GDS) at the RSU level. The vehicles are grouped into velocity-based clusters, where the node with the highest trust value is selected as the CH for each cluster. At the lowest layer, a game is designed for a CH and vehicles located at its radio range; the LIDS monitor can be activated if the cluster members reach an NE. At the medium layer, the GIDS in CH performs an SVM-based learning scheme to evaluate its members' trustworthiness values. At the highest layer, the GDS calculates the vehicles' reputations and classifies them into the trustworthy node, suspected node, and {\color {color1} attacker node}.  Compared with \cite{sedjelmaci2014detection}, this scheme has the ability to detect multiple types of possible attacks, including the {\color{color1} selective forwarding attack, black hole attack,  packet duplication attack, resource exhaustion attack, wormhole attack, and Sybil attack.} Besides security, communication quality is addressed using the clustering method. Furthermore, the {\color{color1}trustworthiness reputation} incentive scheme effectively prevents selfishness. Another advantage of this study is the learning scheme, which mitigates incomplete information and uncertainty in the environment.

 {\color{color1} The studies of \cite{sedjelmaci2014detection,sedjelmaci2015accurate} pay} little attention on communication and computation overheads caused by the IDS schemes. A multi-layered IDS is designed in \cite{subba2018game} to made a trade-off between collected information for intrusion detection and the high-volume IDS traffic in VNs. It constructs the similar hierarchical architecture as in \cite{sedjelmaci2015accurate} but designs the interaction between IDS and the malicious vehicle as a non-cooperative game. The strategy spaces of the CH and the malicious vehicle are $  \left\{Monitor, Not \ Monitor \right\}$ and $ \left\{Attack, Wait\right\} $, respectively. The mixed-strategy NE describes the uncertain strategies of the malicious nodes. By adopting the monitoring strategies based on NE, IDS traffic volume is reduced while maintaining a satisfactory detection rate and accuracy. This study also takes advantage of the reputation-scheme but further considers price schemes to encourage cooperation among players. However, this IDS scheme is only applied to the vehicles that stop at the traffic signal or move to the intersection at low speed. 

Both \cite{sedjelmaci2015accurate} and \cite{subba2018game} use the clustering method to increase the network scalability and decrease the communication overhead. However, the two major clustering concerns are forming a stable cluster and selecting the best CH. The latter problem has been discussed in these works where the CH is selected based on trust values. However, creating stable clusters is challenging since VNs are characterized by high mobility. The fixed boundary distance $R/2$ proposed in \cite{sedjelmaci2015accurate} lacks theoretic basis and adaptation. Mabrouk et al. \cite{mabrouk2018signaling} propose to use the coalitional game for stable cluster formation in the internal malicious nodes detection in VNs. As is discussed in Section \ref{sec_coopereative game}, the coalitional game provides the methodology for coalition formation and the solution (such as core and Sharply) for stability evaluation. After the forming clusters, each CH and its members play a Bayesian game with strategies $(Defend,Idle)$ and $(Attack, Cooperate)$.  Although this study gives a novel game structure with both cooperative and non-cooperative games, it lacks the experiment or numerical analysis for validation.

	 The studies in \cite{sedjelmaci2018generic,brahmi2019cyber}  extend the IDS or IPS to a more advanced cyber defense framework integrating the functions of IDS, IPS, and IRS. The cyber defense problem is formulated as a Stackelberg game played by an intrusion decision agent (IDA) and the secondary agents (i.e., IDS, IPS, and IRS). The IDA aims at balancing communication quality and security by deciding the optimal activation of IDS, IPS, and IRS agents. The NE is obtained when the secondary players execute the corresponding strategies activated by the IDS. This scheme provides new insight into the security game by integrating IDS, IPS, and IRS in an architecture. This study's significant advantage lies in that it can dynamically and adaptively activate the corresponding system (ID, IPS, or IRS) to detect, predict, or react against attacks. The Stackelberg game's hierarchy property enables a leader (IDA) to manage the followers' actions centrally. Besides, the trade-off between security and the overhead of IDS is considered in the utility function, enabling vehicles to {\color{color1} achieve secure and low-latency communications}. However, this scheme has the limitation incurred by the properties of Stackelberg games, {\color{color1} which means that} the hierarchical decision process could lead to worse Stackelberg NE. 
	  
	\begin{itemize}[itemsep=3 pt,topsep = 3 pt]
		\item \textbf{Summary and conclusion of GT for cyber security}
	\end{itemize}

The contributions on game-theoretical cyber security methods of VNs are classified in {\color{color1}Table \ref{tab_intrusion}}, covering the game models including non-cooperative, Bayesian, Stackelberg, and cooperative games as well as different security frameworks including IDS, IPS, IRS, and the hybrid of them. Besides, we compare the performance of above-mentioned IDS schemes in terms of the average detection rate, false positive rate, and communication overhead. We conclude the following aspects for designing optimal strategies to defend or predict attackers. 

\begin{itemize}[label=$-$,topsep=3pt, itemsep=3pt]
	\item  The Bayesian game is a powerful tool to describe incomplete information in an insecure environment, particularly the incomplete information on the attackers.
	\item The hybrid of IDS, IPS, and IRS in \cite{sedjelmaci2018generic,brahmi2019cyber} provides versatile security schemes for VNs than using one of the single architecture. The problem of the hybrid approach {\color{color1}is the difficulty} to adaptively activate the right system. Using the Stackelberg game, IDA acts as a leader to {\color{color1}activate} the corresponding actions according to the situated environment.
	\item The communication quality should be considered in the delay sensitive VNs because security schemes could {\color{color1}cause} extra computation overheads. The cooperative game is a promising approach to {\color{color1} decrease the communication overheads by constructing stable clusters where} malicious detection can be performed in each cluster. 
	\item The incentive mechanisms such as reputation and price are critical to stimulate cooperation and prevent selfishness. A possible direction of improving detection accuracy is the incorporation of learning scheme into GT because it can overcome the incomplete information and uncertainty in VNs. It can be observed from Table \ref{tab_intrusion} that using incentive and learning mechanisms could significantly increase the detection rate and decrease the false forwarding rate.
\end{itemize}

\vspace{6pt}
\subsubsection{GT for location privacy}
\label{sec_GTprivacy}
 With the increasing number of location-based applications, location privacy is one of the primary privacy-related {\color{color1} VN issues} because the leakage of the location means the disclosure of  {\color{color1} the} critical information on social privacy of drivers. Unlike the data privacy that can be protected by {\color{color1} cryptography}, the highly dynamic of VNs makes it difficult to protect the location privacy anywhere and anytime in real-time. {\color{color1} The} pseudonym change is a promising approach and has been widely applied by academia and industry for location privacy protection in VNs \cite{boualouache2017survey}. The pseudonym change occurs in the area of \textit{mix-zone} where {\color {color1} the vehicles} change their {\color{color1} pseudonyms} periodically to protect their location privacy.  {\color{color1} However, one weakness of the mix-zone is that vehicles in the mix-zone are unwilling to change their pseudonyms if they have high level of location privacy.} This is obvious because changing pseudonym leads to extra {\color{color1} overheads} on communication and energy. If there are not sufficient vehicles changing pseudonyms, the mix-zone becomes vulnerable to the adversaries. Therefore, how to encourage vehicles to change pseudonym is essential to guarantee the privacy of the mix-zone. GT provides the ability to encourage vehicles to cooperatively make the optimal pseudonym-change decisions according to the situated vehicular environment

 A user-centric location privacy scheme is firstly explored in \cite{freudiger2009non} to model the non-cooperative behaviors of the mobile nodes, i.e.,  $Cooperate$ (change its pseudonym) or $Defect$ (not change its pseudonym). This study uses an $n$-player non-cooperative game where mobile nodes collectively change their pseudonyms in mix-zone to maximize their utilities at the minimum cost. Both complete information and incomplete information games are analyzed. With NE, vehicles in a mix-zone can dynamically coordinate to change their pseudonyms according to neighbors' privacy levels or threshold strategies.  Although this study is mobile-node oriented, it presents a promising pseudonym change game for selfish local players to dynamically coordinate their pseudonym change decisions. This methodology has been extended to vehicular scenarios by considering the characteristics of VNs. Besides, it is worth noting that this study captures the mobile nodes' location privacy loss over time.

\textcolor{color1}{Ying et al. \cite{ying2015motivation} extends the pseudonym-change scheme proposed in \cite{freudiger2009non} to the vehicular context} to encourage vehicles in the mix-zone to change pseudonyms. The main advantage of \cite{ying2015motivation} is that vehicles form mix-zones dynamically when their pseudonyms are close to expiry. The dynamic-formed mix-zone \textcolor{color1}{is able to} adapt to the highly dynamic vehicular environment, \textcolor{color1}{which overcomes the fixation of the traditional mix-zones.} Besides, the reputation incentive mechanism is designed to encourage selfish vehicles to cooperate. \textcolor{color1}{Taking the behaviors of both passive and active attackers into consideration, the study of \cite{plewtong2018game} overcomes the limitation in \cite{freudiger2009non,ying2015motivation} that mainly focus on passive attackers.} The game is played by Byzantine attackers and defenders. Two Byzantine attackers that target location privacy in VNs are considered in \cite{plewtong2018game}, i.e., a \textit{naive} attacker who never cooperates and a \textit{stealthy} attacker who attempts to obtain the location information of a vehicle while not being detected. The defender {\color{color1}decides to  cooperatively change the pseudonym for location privacy or to greedily keep the old pseudonym for cost saving.} One common weakness of the studies in \cite{ying2015motivation,plewtong2018game} is that the complete information game lacks the conjectures on the other vehicles. In the real-world context, vehicles in the privacy-sensitive VNs do not know either strategies or utilities of their opponents.

The above-mentioned {\color{color1}studies} do not model the overhead caused by privacy protection schemes on communication quality. Wang et al. \cite{wang2019optimization} jointly consider the location privacy and the communication delay {\color{color1}to extend the basic non-cooperative game model in \cite{ying2015motivation} in VNs,} where vehicles in a mix-zone can decide to change or {\color{color1} not change} the pseudonyms.  The main advantage is that the vehicle's utility function is formulated by considering multiple aspects, including: 1) anonymity entropy, 2) Nakagami fading channel, 3) vehicle's mobility pattern, 4) path loss, and 5) channel contention. The NE strategies are obtained, which are the optimal transmit power of each vehicle and the optimal number of cooperating vehicles in a mix-zone. The vehicles can dynamically adjust their transmit power according to the environment and the strategies of the neighboring vehicles. However, the study in \cite{wang2019optimization} has the same disadvantage of lacking complete information as in \cite{plewtong2018game}.

Few studies {\color{color1} focus on integrating GT with} the group signature for location privacy preserving. Zhang \cite{zhang2017otibaagka} propose a security tool for cryptographic mix-zone (CMIX) \cite{julien2007mixzones} establishment in VNs by designing a game between challengers and adversaries. This scheme offers a high level of privacy to avoid outsider attackers because the messages broadcast in the mix-zone are encrypted. However, the secret key management such as group key allocation and update could cause extra computation and energy cost. Besides, the   maliciousness and selfishness of vehicles are not considered. In the worst case, all vehicles {\color{color1} in the CMIX} could be exposed if the group key distributor is a malicious vehicle.

\begin{table*}
	\scriptsize
	\caption{Summary of Location Privacy Games in VNs}
	\label{tab_privacy}
	\renewcommand*{\arraystretch}{1}
	\begin{center}
		\begin{tabular}{|p{.02\textwidth}|p{.05\textwidth}|p{.06\textwidth}|p{.1\textwidth}|p{.05\textwidth}|p{.02\textwidth}|p{.08\textwidth}|p{.2\textwidth}|p{.18\textwidth}|}
			\hline
			\textbf{ref}&\textbf{Game}&\textbf{Scheme}&\textbf{Privacy Metric}&\textbf{Adversary model}&\textbf{Info}& \textbf{Time-sensitivity}&\textbf{Advantages}&\textbf{Weaknesses}\\
			\hline
			\cite{freudiger2009non}&Non-cooperative game&Pseudonym change&Anonymity entropy&Passive&I\&C &Linear privacy loss over time&\begin{itemize} [leftmargin=3pt] \setlength{\itemsep}{0pt}\item Prior knowledge on opponents is considered \item Comprehensive solutions and proofs for NE and BNE \end{itemize}&Focuses on MANETs but does not consider VNs' characteristics \\
			\hline
 			\cite{ying2015motivation}&Non-cooperative game&Pseudonym change&Anonymity entropy&Global, passive, external&C&Not consider&\begin{itemize} [leftmargin=3pt] \setlength{\itemsep}{0pt}\item Dynamic mix-zone formation \item Using reputation scheme to avoid selfishness\end{itemize}&Lacks prior knowledge on opponents or adversaries\\
 			\hline	
 			\cite{plewtong2018game}&Non-cooperative game&Pseudonym change&Anonymity entropy\newline Anonymity entropy degree&Passive and \& active&C&Linear privacy loss over time&Considers both passive and active attackers&\begin{itemize} [leftmargin=3pt] \setlength{\itemsep}{0pt}\item Lacks prior knowledge on opponents or adversaries \item Fixed mix-zones\end{itemize}\\
 			\cline{1-8}
 			\cite{wang2019optimization}&Non-cooperative game&Pseudonym change&Anonymity entropy&Passive&C&Not consider&Formulates VN's channel characteristics and vehicles' dynamic patterns&\\
 			\hline
 			\cite{zhang2017otibaagka}&Non-cooperative game&Group signature&&Active&C&Not consider&\begin{itemize} [leftmargin=3pt]\item Enhanced privacy level\item Overcomes the weaknesses of conventional CMIX in terms of efficiency and security \end{itemize}&\begin{itemize}[leftmargin=3pt]\setlength{\itemsep}{0pt}\item Extra overhead caused by key distribution, encryption and decryption\item Not consider the maliciousness of insider vehicles\end{itemize}\\
 			\hline
		\end{tabular}
	\end{center}
Legend: Info=Information, I=Incomplete, C=Complete
\end{table*}

\begin{itemize}[itemsep=3pt, topsep=3pt]
	\item \textbf{Summary and conclusion of GT for location privacy protection}
\end{itemize}

The game theoretic methods on location privacy are compared and summarized in Table \ref{tab_privacy}. We give the conclusion from the following aspects.
\begin{itemize}[label=$-$,itemsep=3pt, topsep=3pt]
\item \textbf{Cooperation awareness:} The critical problem in designing the game-theoretic pseudonym change schemes is encouraging vehicles to cooperate due to the computation and energy cost resulted from pseudonym change. The pioneering work in \cite{freudiger2009non} designs a good game approach for modeling the ``cooperative" and ``defect" behaviors of vehicles in the mix-zone. Almost all the game-theoretic location privacy schemes inherit from this work. Furthermore, incentive schemes such as reputation \cite{ying2015motivation} can further offer stimulation to mitigate selfishness.  

\item \textbf{Dynamic mix-zone:} The mix-zone should be dynamically formed to adapt to the highly dynamic VNs. Therefore, the game {\color{color1} should} be dynamic.
	
\item \textbf{Incomplete information:}  We believe the Bayesian game  is more appropriate to be used \textcolor{color1}{for location privacy preserving} because it \textcolor{color1}{captures} the fact that vehicles lack complete information on privacy problems.

\item \textbf{Location privacy metric:} Location privacy metric is the key to design effective and secure utility functions. The most used privacy metrics are concluded as follows:

\begin{itemize}[label=$\ast$,itemsep=0pt, topsep=0pt]
\item \textit{Anonymity entropy} {\color{color1} estimates the prior knowledge of an attacker on the target vehicle. It quantifies the uncertainty of identifying the target vehicle among the vehicles in the mix-zone by using information theory. It is formulated as:}
\begin{equation}
\label{eq_AnonymityEntropy}
P=\sum_{d=1}^{|S_A|} p_{d} \log _{2} p_{d}-\left(-\log _{2}p_d\right),
\end{equation}
where $p_{d}$ is the probability of vehicle $d$ {\color{color1} that is} being tracked. This is the most commonly used metric {\color{color1} to evaluate} location privacy.

\item \textit{Degree of anonymity} {\color{color1} is normalized in the range of $[0,1]$ by dividing the maximum anonymity entropy of the mix-zone.} It is formulated as:
\begin{equation}
\label{eq_DegreeAnonymity}
D=\frac{P}{P_{\max}}
\end{equation}
More efforts should be put into the VN-fitted location privacy evaluation. {\color{color1} Furthermore,} the privacy loss \cite{freudiger2009non,plewtong2018game} is important to be considered in the {\color{color1} location privacy model due to} the time-sensitive VNs. Besides, although the vehicular context such as the fading channel and vehicle {\color{color1}dynamics} are discussed in \cite{wang2019optimization}, the privacy quantification is essentially independent with the vehicular context model. We think the difficulty of this problem mainly lies in defining precise location privacy in the vehicular environment.	
\end{itemize}	
\end{itemize}

\vspace{6pt}
	\subsubsection{GT for trust management}
	\label{sec_GTtrust}
	Trust management {\color{color1} aims to guarantee the secure communications in VNs by  preventing the false information of dishonest vehicles} \cite{tian2019evaluating,halabi2019trust}. Current trust evaluation mechanisms can be categorized as {\color{color1} entity-oriented approach} that identifies the trustworthiness of the transmitter, {\color{color1}data-oriented approach} that identifies the trustworthiness of transmitted messages, and {\color{color1}hybrid approach} that incorporates the entity-oriented and data-oriented approaches \cite{zhang2011survey}.
	 
\vspace{3pt}
{\color{color1} \textbf{ \textit{a) Entity-oriented approaches}}}
\vspace{3pt}

 Halabi and Zulkernine \cite{halabi2019trust} design a distributed trust cooperation game to protect the data integrity of vehicles. The players in the game are the vehicles aiming to form trustworthy vehicular coalitions to exchange information securely. The strategy of each player is whether to join in a new coalition. The structure of the vehicular coalition is defined as:
 \begin{equation}
 \label{eq_halabi2019trust}
\Pi=\left\{V C_{m}, m \in \mathbb{N}^{*}, m \leq M\right\},
 \end{equation}
 where {\color{color1} $\mathcal{V}$ is a set of vehicles, $\mathcal{M}$ is the maximum number of coalitions, and} $V C_{m} \subseteq \mathcal{V}$ denotes a disjoint vehicular coalition that satisfies $\bigcup_{m=1}^{M} V C_{m}=\mathcal{V}$ and $ V C_{m} \cap V C_{m^{\prime}}=\emptyset, \forall m^{\prime} \neq m$. Furthermore, the utility function of each vehicle $V_i$ in the coalition $\phi$ is given as:

 \begin{equation}
 \label{eq_halabi2019trust1}
U_{i}(\phi)=\left\{\begin{array}{ll}
Trust_{i}(\phi) & \text { if } \phi \notin h_{i} \\
0 & \text { otherwise },
\end{array}\right.
 \end{equation}
 where $h_i$ {\color{color1} is} a history visiting set of $V_i$ in which the identities of previously visited coalitions are stored. The trustworthiness level ${Trust}_{i}$ is an important property \textcolor{color1}{for vehicles to} construct their preferences over vehicular coalitions. It is defined as a real value function ${ Trust }_{i}: V C_{m} \rightarrow[0,1]$ of the coalition $V C_{m} \in \Pi$: ${Trust}_{i}\left(V C_{m}\right)=\prod_{j \in V C_{m}} T_{i j}$, where $T_{i j}$ is the trust value of vehicle $V_i$ with respect to $V_j$.  $T_{i j}$  is updated according to a Beta function. In this way, the coalitions will only be visited once in the process of coalition formation. Based on the utility function, each vehicle can move to the coalition with {\color{color1} a higher trust level.} For example, a vehicle shows high preference for  $\phi$ if $U_{i}(\phi)>U_{i}(\phi^{'})$. The punishment in this work is assigning low trust value to the new-coming vehicles to prevent attackers joining new coalitions regularly. This is an important advantage that overcomes the limitation of the general punishment schemes where new-coming players are neglected. In this work, the newcomers can decide to join a new preferable coalition each time the cooperative game is initiated. As a result, the compromised benign vehicles can be reduced.  Another advantage of this approach is that vehicles in the same coalition are not required to re-compute the trustworthiness values each time a new message is received, resulting in the {\color{color1} reduced} computation overhead. One limitation of this research is that the update of the trustworthiness value is mainly based on the Beta distribution, which lacks flexibility in the uncertain VNs. The reputation mechanism can be taken into consideration to make the trustworthiness model more adaptive to the VNs.

\vspace{3pt}
{\color{color1} \textbf{\textit{ b) Node-oriented approach}}}
\vspace{3pt}	

 Trust is studied in \cite{kumar2015intelligent} by integrating with security schemes, including symmetric key encryption, hash-based message authentication, privacy preservation, and certificate revocation. In this research, a secure public key infrastructure (PKI) is designed based on the Bayesian coalition game and learning automata (LA). \textcolor{color1}{The structure of the coalitional game is defined as $\{C,LA,E\}$, which consists of a set of coalitions \textcolor{color1} {$C=\{C_1,\ldots,C_n\}$}, a set of LA players $LA=\{LA_1,\ldots,LA_n\}$, and the environment $E$.} The role of the environment is giving incentive feedback, i.e., penalty or reward, concerning the actions of players. After observing the  {\color{color1} feedback of the environment}, each {\color{color1} player} \textcolor{color1}{LA} updates the strategy vector and decides whether to join a new coalition to improve communication security. The advantage of learning is incorporated \textcolor{color1}{into} the game by modeling the interaction between LAs and environment. On the one hand, LAs are incorporated into the game to improve the solution's accuracy by learning from the environment. On the other hand, the LAs' actions can be punished or rewarded by the environment. The solution's accuracy can be improved by learning. However, one limitation is that the learning scheme is simple with a constant learning rate in a stationary environment, leading to low adaptation to the highly-dynamic vehicular contexts.


	Mehdi et al. \cite{mehdi2017game} propose a game-based trust model to identify the defender nodes and malicious nodes in VNs.  The following three metrics are designed to measure the identities of {\color{color1} vehicles}.
	\begin{itemize}[itemsep=3pt,topsep=3pt]
		\item \textit{Betweeness centrality} $C_m$ is quantified by the probability of 
			\item Trust level $E_p$ is quantified using the information provided by vehicles and RSUs. It is set high for vehicles with high priority (e.g., police cars) and vehicles that successfully report the trustworthiness of events.
			\item Node density $D_i$ is calculated based on the number of neighbors.
	\end{itemize}
	These metrics are calculated for each vehicle to measure its identity. A vehicle is tagged as a defender if it meets the threshold of each metric. The strategies of the defender and malicious node are defined as $  Detect /  not \ Detect  $ and $ Attack / not \ Attack $, respectively. An NE is obtained as the set of optimal values of betweenness centrality $C_M$, experience-based trust $E_T$, and node density $N_D$.  This algorithm can intelligently identify the attackers and defenders according to the three proposed metrics. By adopting the calculated NE strategy, the defenders in this game perform better than in the random security game \cite{alpcan2010security} in terms of throughput, re-transmission rate, and data drop rate.  However, the criterion of the threshold values for the metrics are not presented preciously in this work. Taking an example of the betweenness centrality, it is set as $0-1$ for successful communication and $-1-0$ for failed communication. This may cause failed trustworthiness identification of vehicles because threshold values are the key to successful identification between defenders and attackers. Another limitation is that the trust model is based on a highway scenario, and high-dense vehicle scenarios are not considered in the simulation. The node density is set as {\color {color1} 10, 40, 80, 100} on 3000 $\times$ 3000 m, i.e., $10^{-6} - 10^{-5} $ node/m, which is much less than the values of $0.001$ (4 nodes/km/lane) for \textcolor{color1}{fluent traffic} and $0.01$ (30 nodes/km/lane) for \textcolor{color1}{dense traffic} (given the \textcolor{color1}{lane width} is 10 feet) \cite{kuhlmorgen2015performance}. However, the density of vehicles has a significant impact on VNs performances. A vehicle located in an area with dense vehicles could be exposed to attackers with high possibility. Moreover, the vehicle may suffer from poor communication performances in {\color{color1} the environment}. In such a scenario, it is challenging for \textcolor{color1}{vehicles} to identify the trustworthiness of the high volume of \textcolor{color1}{received messages.}

\vspace{3pt}
{\color{color1} \textbf{\textit{c) Hybrid approaches}}}
\vspace{3pt}	

Compared with the node-oriented and data-oriented approaches, hybrid approaches \textcolor{color1}{provide} superior security levels by combining the advantages of the former approaches. In \cite{fan2019trust}, a game-theoretical trust model is developed based on a vehicle's reputation obtained from its communication history and the feedback of the RSU or its neighborhood. In the game, vehicle nodes fall into three categories, i.e., normal node cluster, selfish node cluster, and malicious node cluster. Vehicles in different clusters take different actions from the strategy set: ``receive", ``forward", and ``release". {\color{color1} As shown} in Table \ref{tab_fan2019trust}, each type of vehicle can take cooperative strategies or non-cooperative strategies.
In terms of the incentive approach, the RSU punishes the vehicles that refuse to forward in the next period of time and reward those who cooperate. Regarding the utility, {\color{color1} vehicles pay for the receiving, forwarding or releasing energy consumption.} The game is played repeatedly, and the total utility of node $i$ after $k$ rounds of the game is:
\begin{equation}
\label{eq_fan2019trust}
P F_{i}=\sum_{n=1}^{k} p f_{i} \cdot \omega_{n}
\end{equation}
where $p f_{i}$ is the difference between vehicle $i$'s reward and consumption in the $n-$th round, and the weight $0<\omega<0$ indicates the impact of {\color{color1}future behaviors of vehicle $i$} on the utility. The histories of vehicles' behaviors and the corresponding incentive policies are stored by the RSU. Compared with the general incentive mechanisms, the reputation mechanism in this work mitigates both selfishness and maliciousness. However, one disadvantage of this approach is that the trust construction and storage are based on RSUs, which poses extra requirements on the storage space and deployment of the RSU. For example, the RSU only stores the history tables of the vehicles that have been located in its coverage; the new comers' trust values may be difficult to calculate by checking the table. This limitation may be solved by constructing cooperation among RSUs for information sharing. Another disadvantage is that using the repeated game for the utility model may lead to a long delay.

\begin{table}
	\caption{Strategies of normal node cluster, selfish node cluster, and malicious node cluster in \cite{fan2019trust}}
	\label{tab_fan2019trust}
	\scriptsize
	\renewcommand*{\arraystretch}{1}
	\begin{center}
		\begin{tabular}{|m{.13\textwidth}|m{.13\textwidth}|m{.16\textwidth}|}
			\hline
		\textbf{Players}&\textbf{Cooperative strategies}&\textbf{Non-cooperative strategies}\\
			\hline
			Normal node cluster& Receive, forward, and release (when observing accident or traffic jam) &Receive or forward, but are not willing to release \\
			\hline 
			Selfish node cluster&Receive and forward & Receive but are not willing to forward \\
			\hline
			Malicious node cluster&Receive & Refuse to receive\\
			\hline
			\end{tabular}
		\end{center}
	\end{table}

Given the dynamics of vehicles and the openness of VN channels, trust relationships among vehicles are constructed distributively without relying on a central entity in \cite{haddadou2014job}. An infrastructure-independent trust model {\color{color1} $ DTM^2 $} is proposed based on the signaling game.
Three types of players, i.e., normal, selfish, and malicious vehicles, are considered in this work. {\color{color1} Similarly to the incentive scheme in \cite{fan2019trust}, this mechanism  mitigates malicious behaviors by encouraging selfish vehicles to cooperate.} However, as mentioned before, the reputation may take a long time to construct, making it difficult to eradicate the malicious nodes. To address this problem, this paper designs a credit-based incentive scheme adapting to the signaling. The signal with a cost corresponding to vehicles' roles can reward or punish the vehicles based on their behaviors. The {\color{color1} $ DTM^2 $} maintains credits and allocates the same amount of credit to vehicles that construct connections with the network. Each vehicle is assumed {\color{color1} to be} equipped with an inviolable trusted platform module (TPM). The vehicle uses its credits to pay for the signaling cost of sending. On the other hand, the vehicle will be rewarded with credits if its messages are identified as trustworthy. As a result, the selfish nodes are motivated to cooperate for earning credits by establishing the price for receiving messages. Besides, to detect the malicious nodes, $  DTM^2  $ calculates the sending cost by formulating signal values corresponding to the vehicles' behaviors. {\color{color1}Besides detecting malicious behaviors, this scheme is able to evict malicious nodes from the VN, which means that} 100\% of the malicious nodes can be excluded from the network without causing false-positive detection. One limitation is that this mechanism needs message acknowledgments from receivers. However, {\color{color1}broadcast messages, especially IEEE 802.11p-based messages, do not require acknowledged feedback in VNs given the possible high delay and network congestion caused by message acknowledgment.}

  Rather than design a new trust management model, Tian et al. \cite{tian2019evaluating} propose an evaluation scheme for the reputation-based trust management models. This evaluation approach applies an evolutionary game to model the malicious behaviors in the VNs. The game is {\color{color1} defined} as $\left(P, S, A_{t}, U\right)$, where the players $P$ are the dishonest vehicles, each player selects a deception intensity $x$ from its strategy set $S=\{x \mid 1 \leq x \leq 100, x \in Z\}$,  $A_{t} \in\left\{\left(a_{1}, \ldots, a_{i}, \ldots, a_{100}\right) \mid a_{i} \in[0,1], \Sigma a_{i}=1\right\}$ denotes the decision distribution of the population at time $t$, and $U$ is the utility function of an individual player, which is the sum of the false event messages of the vehicle. The natural selection is used to remove the dishonest vehicles from the network. Furthermore, the two key elements in the evolutionary game is modeled as:
  \begin{itemize}[itemsep=3pt,topsep=3pt]
 	\item Natural selection: the dishonest vehicles are removed from the population
 	\item Reproduction: new vehicles will join the network according to {\color{color1} the} replicator equation: $\dot{a}=a_{i}\left[U_{i}-\sum_{j=1}^{100} a_{j} U_{j}\right]$.
 \end{itemize}
 In the simulation, the evaluation is finally converged into the state where all the attackers with different malicious behaviors obtain optimal strategies. This indicates that the evaluation approach can depict the evolutionary process of the dynamic malicious attacks in VNs, which can quantify the effectiveness of the trust management scheme. The main advantage of this work is the novel thinking mode of dynamic simulation and prediction of the malicious behaviors, rather than new schemes for trust management. It overcomes the limitation of the schemes where certain types of attacks with predefined attacking behaviors are {\color{color1} assumed}. Therefore, the evaluation scenario of this study is closer to the real world by simulating the selection of dynamic and diverse attacking strategies. An interesting finding in this work shows that the convergence time adapts to VNs with varied error tolerance. That is to say, the system with higher error tolerance leads to a long convergence time, which is reasonable because a dishonest vehicle can survive long in a high error tolerance VN. However, this could not guarantee that the game can be converged in any case since there is no theoretic proof. A possible solution {\color{color1} may be setting a predefined threshold for convergence time to avoid long convergence duration}.
 

\begin{table*}
	\scriptsize
	\caption{Comparison of Trust Games in VNs}
		\label{tab_trust}
	\renewcommand*{\arraystretch}{1}
	\begin{center}
		\begin{tabular}{|m{.04\textwidth}|m{.02\textwidth}|m{.06\textwidth}|m{.02\textwidth}|m{.28\textwidth}|m{.2\textwidth}|m{.1\textwidth}|m{.08\textwidth}|}
			\hline
		\textbf{Schemes}&\textbf{Ref.}&\textbf{Game}&	\color{color1} {\textbf{RSU}}&	\textbf{Advantages}&\textbf{Disadvantages}&\textbf{Attack resistance}&\textbf{Incentives}\\
			\hline
		\textbf{ Node-oriented}&\cite{halabi2019trust}&Coalition game&$\times$&\begin{itemize} [leftmargin=3pt,itemsep=0pt,topsep=0pt] \setlength{\itemsep}{0pt}\item  Vehicles in the same coalition are not required to re-compute the trustworthiness \item  The newcomers can decide to join a new preferable coalition each time the game is initiated\end{itemize}& The quantification of the trust value $T_{i j}$ maybe insufficient&Malicious vehicles&Punishment of assigning low trust values to new comers\\
			\hline
			\textbf{Data-oriented}&\cite{kumar2015intelligent} &Bayesian {\color{color1} game}&$\times$&
			\begin{itemize} [leftmargin=3pt] \setlength{\itemsep}{0pt}\item The trust model provides more versatile protections such as security, authentication, confidentiality, and privacy \item The advantage of learning is incorporated in to the game by modeling the interaction between LAs and environment
			 \end{itemize}& The constant learning rate in a stationary environment may lead to low adaptation to the highly-dynamic vehicular contexts&Selfish {\color{color1} vehicles} and malicious vehicles&Punishment or rewards of environment\\
			\cline{2-8}
			& \cite{mehdi2017game}&{\color{color1} Non-cooperative game}&$\times$&\begin{itemize} [leftmargin=3pt] \setlength{\itemsep}{0pt}\item Identify the attackers and defenders intelligently \item Defenders perform better in terms of throughput, re-transmission rate, and data drop rate \end{itemize} &\begin{itemize} [leftmargin=3pt] \setlength{\itemsep}{0pt}\item  The criterion of the threshold values for the metrics are not presented preciously \item The high-dense scenario is not considered in the trust model\end{itemize} &  General attackers& Not considered\\
			\hline
			\textbf{\textcolor{color1}{ Hybrid}}&\cite{fan2019trust} &{\color{color1} Non-cooperative game}&$\surd$&The reputation scheme mitigates both selfish and malicious behaviors&\begin{itemize} [leftmargin=3pt] \setlength{\itemsep}{0pt}\item The reputation is based on the RSU which lacks the history tables of the new comers\item Repeated game- based reputation construction may cause long delay \end{itemize}& Selfish {\color{color1} vehicles} and malicious vehicles&Reputation\\
			\cline{2-8}
			&	\cite{haddadou2014job} &Signaling game&$\times$&Detect and evict malicious nodes from the network without causing false-positive detection&Acknowledgment messages in broadcast may cause high delay or network congestion&Selfish {\color{color1} vehicles} and malicious vehicles&Credit and reputation\\
			\cline{2-8}
			&\cite{tian2019evaluating}&Evolutionary game&$\surd$&\begin{itemize} [leftmargin=3pt] \setlength{\itemsep}{0pt} \item  The evaluation scenario is closer to the real world \item Simulate selection of dynamic and diverse attacking strategies\item The convergence adapts to systems with varied error tolerance \end{itemize}& Possible long convergence time  & Malicious vehicles&Reputation\\
			\hline
		\end{tabular}
	\end{center}
\end{table*}

	\begin{itemize}[itemsep=3pt,topsep=3pt]
		\item \textbf{Summary and conclusion of GT for trust management}
	\end{itemize}

In this subsection, the trust management games in VNs are presented from the perspectives of node-oriented, data-oriented, and hybrid approaches. Table \ref{tab_trust} presents a comparative summary of these studies. {\color{color1} Incentive approaches are crucial for trust management games to stimulate benign behaviors and prevent maliciousness or selfishness. Existing GT approaches and incentive schemes used in games are concluded as follows.}


\begin{itemize}[label=$-$,itemsep=3pt,topsep=3pt]
	\item The characteristic of cost-signal in the signaling game can be used as incentives among vehicles with different states \cite{haddadou2014job}. 
	\item The coalitional game provides an effective way for trust management in high dynamic VNs where a newcomer can dynamically choose its preferable trust coalition without recomputing the trustworthiness value. The malicious nodes can be excluded from the VNs through natural selection by using the evolutionary game \cite{tian2019evaluating}.
	\item The reputation and credit are the most popular incentive schemes used in games because they can offer long-term trust measurement for vehicles. These values can be stored in the TPM units of vehicles\cite{haddadou2014job} or in RSUs \cite{fan2019trust,tian2019evaluating} for historical value check in the future.
	\item Punishment or reward are alternative incentive schemes used in games. Low trust value can be assigned to the newcomers to prevent maliciousness \cite{halabi2019trust}. Vehicles can also interact with environment to obtain the punishment or reward \cite{kumar2015intelligent}.  

\end{itemize}

\subsection{Game Theory for QoS Guarantee in VNs}
\label{sec_GTQoS}
Exploiting available resources to maximize the overall QoS is another challenge in VNs. GT is able analyze the behaviors of nodes to help them decide their optimal strategies to optimize the network performance. {\color{color1} GT approaches have been extensively studied for QoS guarantee at different vehicular protocol stack layers, including transmit power and spectrum resource allocation at the PHY layer, access control at the MAC layer, routing at the network layer, and charging schedule at the higher layer.}

\vspace{6pt}

\subsubsection{GT for PHY-layer transmit power and spectrum resource allocation}
\label{sec_GTphy}
{\color{color1} In this section, we survey the application of GT in solving the PHY-layer problems of transmit power control and spectrum resource allocation.}

\vspace{3pt}
	\textbf{\textit{a) GT for transmit power control}}
\vspace{3pt}

Power or rate control is essential to avoid congestion caused by the communication uncertainties in VNs, such as unstable channels, dynamic vehicles, and varied environments. In VNs, vehicles periodically broadcast their status as basic safety messages (BSMs), also known as beacons, to make the neighbors aware of their {\color{color1} presences}. A higher power or rate is desired by a vehicle to disseminate the messages over a large distance. However, a high level of power could lead to collisions or congested communication in dense traffic environments, leading to performance and security degradation. Therefore, a good power or rate control scheme should adapt to varying environments, providing sufficient awareness on surrounding vehicles’ status while maintaining low congestion and highly secure communication. GT can help {\color{color1} vehicles to decide the optimal transmit power or rate according to the situated environment to obtain efficient communication performance. }


An earlier work of game theoretic method on power control includes \cite{chen2016information}, where the bargaining game is used for joint power and packet rate adaptation to avoid information congestion. The clustering scheme is employed in this study to decrease the strategy space. Each CH acts as a player {\color{color1} to determine} the cluster members' optimal transmission power and packet generation rate. The CHs select their strategies using a sequential bargaining tree where the leaf nodes decide the utilities. This scheme outperforms IEEE 802.11p in terms of queuing delay and throughput. However, the work only uses a bounded value to guarantee the end of the repetition. {\color{color1} It is obvious that this value should vary with different vehicular scenarios, which however is not discussed in this work.} Furthermore, the use of bargain tree leads to the limitations of repetition and sequence. On the one hand, the bargain tree specifies the move sequence of players, which {is not applied} to the distributed actions of CHs. Besides, the bargain is more suitable for the two-player game. The bargain with more players could cause long convergence time and high computational complexity. In conclusion, the idea of clustering power control of this paper provides novelty, but the coalitional game may be further exploited for the cluster formation.

In \cite{goudarzi2018non}, a non-cooperative game-based adaptive BSM power adjustment approach is designed for congestion control in VNs. In the game, a set of vehicles are defined as players who can select its strategy $p_{i}$ from a set of possible beacon powers $p_{i} \in \mathcal{P}_{i}=[1,100]$. The utility function is formulated based on the channel busy ratio (CBR) developed in \cite{chen2011mathematical}. The CBR parameter gives a well-grounded evaluation for successful information dissemination by jointly considering the BSM transmit power, vehicles' dynamic patterns (e.g., distance, speed, and acceleration), and  Nakagami channel conditions (e.g., Nakagami fading parameter and path loss exponent). The NE, i.e., the optimal BSM power, is theoretically proven to be existing and unique. Given the insufficiency of only controlling the transmit power, the authors further extend the beacon power control to a joint beacon power and frequency control game \cite{goudarzi2019fair}.  In conclusion, the NE of the adaptive power control guarantees satisfactory communication connection while controlling the congestion at the desired level. Joint transmit power and frequency control allows vehicles to adapt to the varied VNs more flexibly, especially in dense traffic. Furthermore, a player can also control its bandwidth share by controlling {\color{color1} the parameters of its utility function}. However, although the simulation results reveal that the transmit beacon power can converge to the NE, they only validate the convergence in certain cases. Further theoretical demonstrations or solutions are needed. Similarly, the two studies draw a common conclusion that the fairness bandwidth usage can be obtained by adjusting the parameters of the utility functions. This conclusion should be supported by further convictive theoretical proofs or solutions.

To address the fairness of the results, an adaptive transmit power control that uses the Shapely value is explored in \cite{shah2018shapely}. The study designs a cooperative congestion game {\color{color1} where vehicles within the carrier sense (CS) range act as players in the game}. The players cooperate to join and maximize the coalition's utility during a congestion event. Based on the Shapely model, the marginal contribution of the vehicle $i$ on the congestion is designed as:
\begin{equation}
\label{eq_shah2018shapely}
m c_{i}^{\sigma}(u)=u\left(t^{\sigma}(i) \cup i\right)-u\left(t^{\sigma}(i)\right)
\end{equation}
where $\sigma$ denotes the permutations, $u$ is the player's utility, and $t^{\sigma}$ is the set of all permutation. The fair utility allocation is given as the average marginal contribution: $\theta_{i}=\frac{1}{n !} \sum m c_{i}^{\sigma}(u)$. Then each player $i$ decreases its transmit power according to the fairness of power decrease:
\begin{equation}
\label{eq_shah2018shapely2}
fair_{t x p}=t x p_{c u r r e n t}-\theta_{i},
\end{equation}
where $t x p_{c u r r e n t}$ is the current transmit power of vehicle $i$. This power control scheme outperforms the schemes \cite{goudarzi2018non,goudarzi2019fair}  in terms of fairness and efficiency using the Shapely value model of the cooperative game. Furthermore, different from the traditional cooperative game, this study designs a novel approach by transferring the vehicles' marginal contributions on congestion into fair power decrease. This idea is inspired by the fact that decreasing the transmit power can narrow the CS range and further alleviate the congestion. However, this paper only model the vehicles in the CS range as players. The vehicles outside the CS range should also be included in the set of players because accumulative interference can lead to congestion.

Other studies \cite{hua2017game,sun2017non} employ GT to deal with trade-off problems in VNs through power control. Hua et al. \cite{hua2017game}
propose a repeated game-theoretic power control approach to make a trade-off between communication quality and power consumption. However, as introduced in Section \ref{sec_oneshot}, the main concern of the repeated game is the possible long convergence time, which is not discussed in this study. Sun et al. \cite{sun2017non} propose an adaptive transmit power and encryption block length control mechanism to balance QoS and security by allocating the limited computing resources in VNs. This study adopts {\color{color1} the non-cooperative game to  model} the two conflicting objectives as two abstract players, i.e., a ``communication player" who controls the transmit power and a ``security player" who decides the encryption block length. The optimal transmit power and encryption block length are obtained by calculating the NE value. However, this game model is simple because it only focuses on a single-vehicle without considering the competition or cooperative behaviors among multiple vehicles. Both the approaches in \cite{hua2017game,sun2017non} ignore the fairness of the result.

\vspace{3pt}
\textbf{\textit{b) GT for spectrum resource allocation}}
\vspace{3pt}

Dynamic and efficient GT approaches for spectrum resource allocation is crucial in VNs to overcome the selfishness of vehicles and the scarcity of spectrum resources.

Kumar et al. \cite{KUMAR201719} propose a spectrum handoff scheme for optimal network selection in VNs. An auction game is employed where vehicles act as bidders competing for accessing the network. The utility of each bidder is formulated based on multiple attributes and bidding cost. Numerical results demonstrated that this approach
enables the optimal network selection for spectrum handoff. However, how to extend this scheme to the dynamic and incomplete-information environment has not been considered.

{\color{color1} To overcome the limitation in \cite{KUMAR201719},}
Shattala et al. \cite{shattal2018channel} designed a DSRC-based dynamic spectrum access strategy to optimize the spectrum utilization in VNs. {\color{color1} Due to the advantage of the evolutionary game in terms of overcoming the bounded-rationality and selfishness among players, this approach uses the evolutionary game to construct the vehicles' competition on the limited spectrum resources. This work considers three strategy models that can be adopted by vehicles:} 1) always-consume user who consumes the spectrum resource blindly, 2) forage-consume user who sometimes consumes, and otherwise forages and uses the gathered information to choose where next to consume, and 3) social-forage-consume user who sometimes waits and allows other users exclusive access. The RSU first collects information of vehicles within its range and relays it to the cloud. Then, the cloud computes an optimal and stable strategy using the evolutionary game and sends the results to {\color{color1} the} RSU. Finally, the RSU broadcasts the recommended strategies to the vehicles within its radio range. The results show a significant improvement in the utility from 3\% to 136\% when most vehicles employ the evolutionary strategies recommended by the RSU. One limitation of this work is that it only relies on the DSRC protocol; vehicles cannot access various types of VNs flexibility.

{\color{color1}  Tian et al. \cite{tian2019channel} take advantage of the evolutionary game to overcome irrationality and selfishness of players.} The authors consider a  V2I communication scenario where exclusive-used and shared-used spectrum resources coexist. An evolutionary game is employed to model the evolution of selfish vehicles contending for the shared-use channels. {\color{color1} Compared with the study in \cite{shattal2018channel}, this approach integrates the incentive scheme in the evolutionary game to overcome the inefficiency of NE.} A dynamic channel pricing mechanism using marginal social cost is designed, which enables both vehicles and channels learn and update their strategies asynchronously. The NE of the spectrum allocation scheme is proved to be evolutionary stable and coincided with the social optimum. This study also relies on the single-protocol VNs and cannot be applicable to the VNs supported by multiple types of protocols.



\begin{itemize}[itemsep=3pt,topsep=3pt]
	\item \textbf{Summary and conclusion of GT for PHY-layer transmit power and spectrum resource allocation}
\end{itemize}

{\color{color1}
The game-theoretical schemes for power control \cite{chen2016information,goudarzi2018non,goudarzi2019fair,shah2018shapely,hua2017game,sun2017non} and spectrum resource allocation  \cite{KUMAR201719,shattal2018channel,tian2019channel} are summarized in {\color{color1} Table  \ref{tab_powercontrol}}. These studies are concluded as follows.

\begin{itemize}[label=$-$,itemsep=3pt,topsep=3pt]
	\item Adaptivity and dynamic are the main focuses in designing the strategies for power control and spectrum resource allocation in the dynamic VNs.
	\item Regarding the power control, existing studies focus on congestion control \cite{chen2016information,goudarzi2018non,goudarzi2019fair,shah2018shapely} and trade-off between the conflicting objectives \cite{KUMAR201719,shattal2018channel,tian2019channel}. One common weakness of most of these studies is the lack of fairness solutions.
	\item Regarding the spectrum resource allocation, the evolutionary game is able to overcome the bounded rationality of competitive vehicles \cite{shattal2018channel,tian2019channel}. The work in \cite{tian2019channel} achieves a more efficient NE solution by integrating the pricing incentive into the evolutionary game to enable the dynamic adaptation of VN channels. 
\end{itemize}
}


\begin{table*}
	\scriptsize
	\label{tab_powercontrol}
	\caption{\textcolor{color1}{Summary of GT for PHY-layer transmit power control and spectrum management in VNs}}
	\renewcommand*{\arraystretch}{1}
	\begin{center}
		\begin{tabular}{|m{.01\textwidth}|m{.02\textwidth}|m{.06\textwidth}|m{.18\textwidth}|m{.13\textwidth}|m{.13\textwidth}|m{.1\textwidth}|m{.2\textwidth}|}
			\hline
			&\textbf{Ref.}&\textbf{Game}&\textbf{Highlights}&\textbf{Strategy}&\textbf{Utility Metric}&\textbf{Solutions}&\textbf{Disadvantages}\\
			\hline
			&\multicolumn{7}{c|}{\textbf{Congestion control}}\\
			\cline{2-8}
			&\cite{chen2016information}&{\color{color1} Bargaining} game&Clustering-based information congestion control&Joint power and packet generation rate control&Channel load&Clustering and CH selection&\begin{itemize}[leftmargin=5pt]\setlength{\itemsep}{0pt}\item No theoretical bargaining solution \item Require moving sequence\item Possible long convergence time for more players situations $(>4)$ \end{itemize}\\
			\cline{2-8}
			\multirow{8}{3cm}{\rotatebox[origin=c]{90}{\textbf{Transmit power control}}}&\cite{goudarzi2018non}& Non-cooperative game&Adaptive beacon power adaption for congestion control&Beacon {\color{color1} transmit} power control&Channel load, channel busy ratio (CBR) and price function&NE&	\multirow{2}{3.6cm}{\begin{itemize}[leftmargin=5pt]\setlength{\itemsep}{0pt}\item Lack theoretical solutions or proofs for NE and fairness \item Lack cooperation or interaction among players\end{itemize}}\\
			\cline{2-7}
			&\cite{goudarzi2019fair}  &Non-cooperative game&Adaptive beacon power and rate adaption for congestion control &Joint beacon frequency and power control&Channel load, CBR, and price function&NE&\\
			\cline{2-8}
			&\cite{shah2018shapely}& Cooperative game& Fair and adaptive beacon power decreasing for congestion control &Adaptive transmit power control&Marginal contributions of vehicles towards congestion&Shapley value & Vehicles outside of CS range can cause accumulative interference\\
			\cline{2-8}
			&\multicolumn{7}{c|}{\textbf{Trade-off}}\\
			\cline{2-8}
			&\cite{hua2017game}&Non-cooperative game& Trade-off between {\color{color1} QoS and energy} &Adaptive transmit power control&Throughput, delay, and channel model&NE&\begin{itemize}[leftmargin=5pt]\setlength{\itemsep}{0pt}\item Possible long convergence time \item Lack fairness solutions \end{itemize}\\
			\cline{2-8}
			&\cite{sun2017non}&Non-cooperative game&{\color{color1} Trade-off} between QoS and security&Transmit power and encryption block length control&Channel capacity&NE&\begin{itemize}[leftmargin=5pt]\setlength{\itemsep}{0pt}\item Only single-vehicular model without competition or cooperation behaviors\item Lack fairness solutions \end{itemize}\\
			\hline\hline
			\multirow{8}{3cm}{\rotatebox[origin=c]{90}{\textbf{Spectrum resource allocation}}}&\cite{KUMAR201719}&Auction game&Spectrum handoff&Bidding&Multiple attributes and bidding cost&Winning bidder&\begin{itemize}[leftmargin=5pt]\setlength{\itemsep}{0pt}\item Not consider the dynamic situation\item Not consider the incomplete information\end{itemize}\\
			\cline{2-8}
			&\cite{shattal2018channel}&Evolutionary game& \begin{itemize}[leftmargin=5pt]\setlength{\itemsep}{0pt}\item Dynamic spectrum access\item Mitigated irrationality and selfishness\end{itemize}& \begin{itemize}[leftmargin=5pt]\setlength{\itemsep}{0pt}\item Always consume \item Forage consume \item Social-forage consume\end{itemize}&Throughput and packet delivery ratio&Stable evolutionary NE& \begin{itemize}[leftmargin=5pt]\setlength{\itemsep}{0pt}\item Possible inefficient NE 	\item Cannot be applicable to the multiple-protocol supported VNs \end{itemize}\\
			\cline{2-8}
			&\cite{tian2019channel}&Evolutionary game& \begin{itemize}[leftmargin=5pt]\setlength{\itemsep}{0pt}\item Dynamic strategy learning of vehicles and channels \item Mitigated  irrationality and selfishness\item Dynamic incentive of channel pricing  \end{itemize} & \begin{itemize}[leftmargin=5pt]\setlength{\itemsep}{0pt}\item Vehicles \begin{itemize}[leftmargin=5pt]\setlength{\itemsep}{0pt} \item Exclusively use \item Share \end{itemize}\item Channels\begin{itemize}[leftmargin=5pt]\setlength{\itemsep}{0pt} \item Channel pricing\end{itemize}\end{itemize}&Congestion
			degree&\begin{itemize}[leftmargin=5pt]\setlength{\itemsep}{0pt}\item Stable evolutionary NE \item Social maximum\end{itemize}& 
			Cannot be applicable to the multiple-protocol supported VNs \\
			\hline
		\end{tabular}
	\end{center}
\end{table*}

	\vspace{6pt}
\subsubsection{GT for MAC-layer access}
\label{sec_GTforMAC}

Similar to traditional wireless networks, the main concern for MAC scheme design is how to avoid the collision caused by vehicles' competitive access to the limited MAC-layer resource. Having the ability of modeling the competitive behaviors, GT has been applied by several studies for collision avoidance \cite{kwon2016bayesian,li2019tcgmac}, access priority \cite{wang2018application,lang2019vehicle}, and greedy behavior prevention \cite{al2017cooperative}. 


Kwon et al. \cite{kwon2016bayesian} use the Bayesian game for beacon collision alleviation at the MAC layer in urban VNs. Each selfish or cooperative vehicle makes individual decisions on power level or transmission probability based on its utility. The utility of each vehicle is constructed \textcolor{color1}{by incorporating} throughput rewards and transmission energy costs. The BNE of this game is proved to be existent. The experimental results further demonstrates that this approach achieves higher utilities and fairness than random access schemes. 

Collision alleviation is also studied in \cite{li2019tcgmac} by designing a MAC protocol that synthesizes time division multiple access (TDMA) and CSMA mechanisms. This scheme divides a time frame into two segments: a TDMA period for application data transmission and a CSMA period for slot declaration. GT is employed to maximize the usage of slots by deferring the nodes' choices when two or more nodes reserve the same slot during the CSMA period.

In \cite{wang2018application}, a concept named \textit{application value awareness} is proposed by assigning different values to packets according to the waiting time during the competition so that the nodes with higher packet values have a higher access probability. A cross-layer game model is then constructed for the vehicles to adjust their channel access probabilities based on their packet values. An equilibrium enables the successful transmission of message within a latency limit when the channel is nearly saturated. The inter-distance between vehicles and their roles are further considered in a medium access game in \cite{lang2019vehicle} to determine a player's priority to access the MAC-layer medium.

Al-Terri et al. \cite{al2017cooperative} proposed a GT-based scheme for greedy behavior prevention in VNs based on the CSMA/CA protocol. An incentive mechanism is proposed by designing group reputation and cooperative detection to encourage the selfish nodes to behave normally. Then the authors designed a tit-for-tat strategy under the threat of retaliation to impose cooperation among vehicles. The results reveal that this scheme can protect the communication against the ambiguous monitoring by preventing the greedy MAC-layer access. 

\vspace{6pt}
\subsubsection{\textcolor{color1}{GT for network-layer routing}}
\label{sec_GTrouting}
\begin{figure*}[!hbt] 
	\centering
	\includegraphics[width=6.6in]{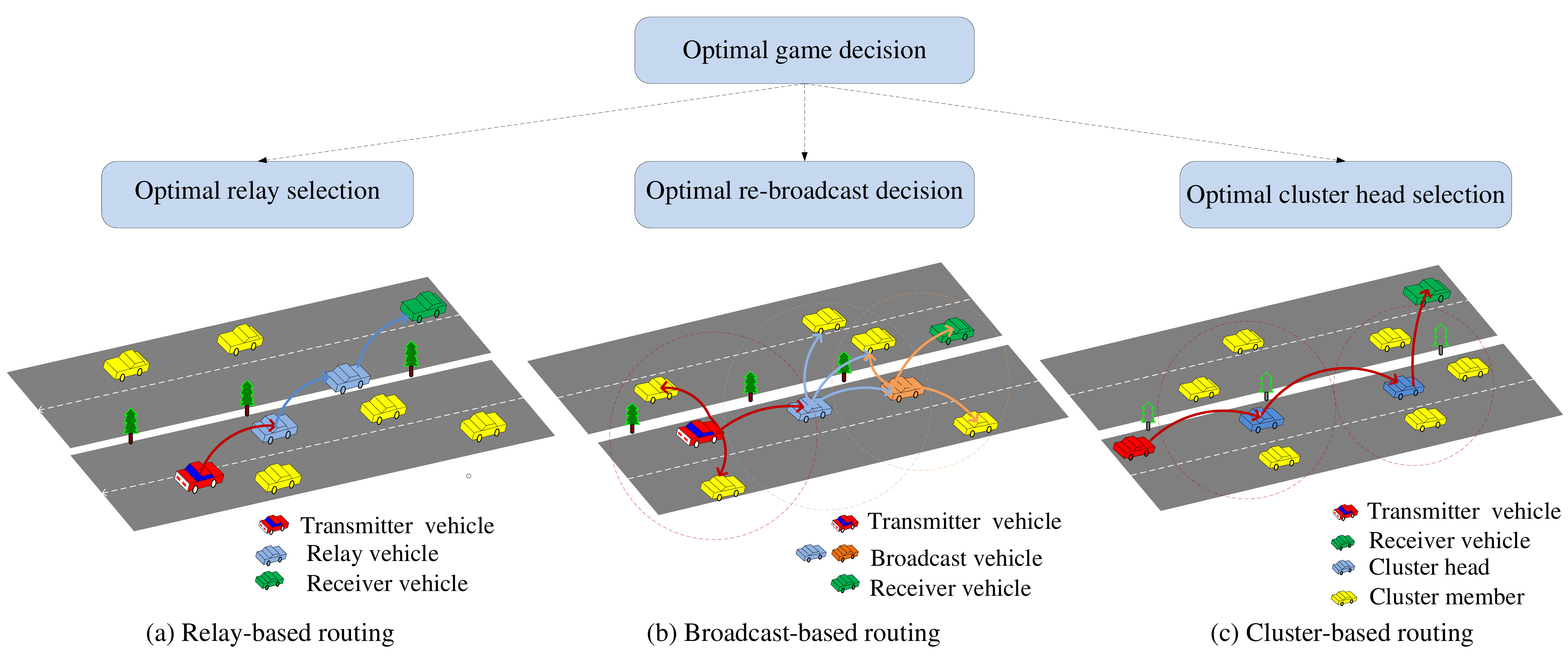}
	\caption{Routing schemes in VNs: (a) relay-based routing, (b) broadcast routing, and (c) cluster-based routing.}
	\label{fig_routing}
\end{figure*}

GT has been widely applied for mitigating the selfish behavior of packet forwarding in wireless networks. However, maintaining stable routes in VNs is challenging because of the dynamic nature of a mobile vehicle.  From the perspective of transmission mode, routing schemes for VNs can be categorized as relay-based, broadcast-based, and cluster-based routings, as illustrated in Fig. \ref{fig_routing}. From the view of network architecture, routing schemes can be categorized as centralized routing based on a central controller, decentralized routing based on V2V communication, or hybrid routing which incorporates the above two protocols. use the network is controlled by one center controller. The Centralized routing requires global network knowledge to control all processes which are housed in the central controller. On the one hand, the absence or impracticality of infrastructures could make the algorithms that require complete information (e.g., conventional global optimization algorithms) inapplicable \cite{tian2017self}. {\color{color1} On} the other hand, it may provide limited scalability and flexibility in the highly dynamic scenarios, especially in the environment with dense traffic where vehicles regularly move in and out of the infrastructure’s coverage. These characteristics makes it more suitable for the scenarios where the nodes in the center controller's coverage are stationary or move with low speed.  Regarding the decentralized routing protocols, they are self-adjusting and can adapt to dynamic and varied networks. Therefore, the decentralized approaches provide better scalability and flexibility. However, it has the weakness of {\color{color1}lacking complete information and knowledge on  environment}, which may limit the precise decision making {\color{color1}of} players. Most recent studies on routing in vehicular networks are based on the decentralized routing or the hybrid routing that combines the merits of the centralized and decentralized schemes.

\vspace{3pt}
\textbf{\textit{a) Relay-based routing}}
\vspace{3pt}


To overcome the limitation of a traditional bargaining game that lacks the complete information (in Section \ref{sec_BgGT}), Kim \cite{kim2016timed} proposed a new opportunistic routing scheme for dynamic VNs based on an iterative bargaining model. The game is modeled as $G=\{N, \mathbb{S},\textcolor{color1}{\left\{U_{1},\ldots, U_{n}\right\}}\}$, where $N$ is the number of players, $\mathbb{S}$ is the strategies of relay vehicle selection, $\mathbb{U}_i$ is the utility function of the player $i$, which is formulated by mapping the time and actions to the player-level satisfaction:
	
	\begin{sequation}
		\label{eq_kim2016timed}
		\begin{aligned}
		&	U_{i}\left(S_{i}(t), t\right)=\int_{0}^{t} U_{i}\left(S_{i}(t)\right) d t \cong \sum_{n=0}^{n=t} U_{i}\left(S_{i}(n)\right), \\
		 &\text { s.t., } \ S_{i}(t), S_{i}(n) \in \mathbb{S}_{i},
		 \end{aligned}
	\end{sequation}
	where $S_{i}(t)$ is the strategy of player $i$ at time $t$. The vehicle $i$  selects a most adaptable relay with the minimum link cost $PC_i$ that potentially incorporates more global information. The most adaptable route and relay nodes can be selected by making iterative decision makings using the sequential bargaining game. Specifically, the destination vehicle selects the most adaptable routing path $\Gamma$ according to (\ref{eq_Path_kim2016timed}).
	{ \color{color1}
	\begin{equation}
	\label{eq_Path_kim2016timed}
	\Gamma=\arg \underset{\{P_i, P_i\in \mathbb{S}\}}{\min} {\int_{t=\mathcal{T}_s}^{t=\mathcal {T}_{e}} \log \left(L- P_{k}\right) d t}
	\end{equation}
}
	where $ \mathbb{S} $ is the set of established routing paths, and $  P_i $ is the $ i $th routing path from the source vehicle to the destination vehicle. $\mathcal{T}_s$ and $\mathcal{T}_e$ are the packet forwarding start time and end time, respectively. The main novelty of this routing approach is that it overcomes the shortcoming of the traditional bargaining game that lacks full information by transforming it into a timed sequential bargaining approach. The estimated path cost is recursively forwarded until the most adaptable relays are selected, forming the final opportunistic multi-hop link.  One weakness is {\color{color1} that} this study does not prove the convergence of the iterative game during the time duration from $\mathcal{T}_{e}$ to $\mathcal{T}_{s}$, which may lead to an unlimited or long-delay result. The incentive approaches, such as rewards, can be further considered to stimulate the relay nodes to cooperate.

Tian et al. \cite{tian2017self} propose a self-organized relay selection method based on the non-cooperative game where the candidate relay nodes are treated as players aiming to optimize transmission quality with low energy consumption. The relay selection is mapped into a normal-form multi-player game:

\begin{equation}
\label {eq_tian2017self}
\mathcal{G}=\langle\mathcal{C}, \mathcal{J}, \mathcal{A}, \mathcal{U}\rangle,
\end{equation}
where  $\mathcal{C}$ is the space of the external states, $\mathcal{J}=\{j\mid j=1,2, \ldots, m\}$ is the set of the candidate relay nodes, $\mathcal{A}=\times_{j=1}^{m} \mathcal{A}_{j}=\mathcal{I}=\{i \mid i=1,2, \ldots, n\}$ is the action space for all players, which is the set of source nodes.
Furthermore, $\mathcal{U}=\left(u_{1}, u_{2}, \ldots, u_{m}\right): \mathcal{A} \rightarrow \mathbb{R}^{m}$ denotes a vector of players. The utility function $u_j$ for player $j$ is formulated by considering the trade-off between the received benefit and the energy or power cost, which is given as (\ref{eq_Utility_tian2017self}). 
		
\begin{equation}
\label{eq_Utility_tian2017self}
u_{j}\left(\boldsymbol{s}_{j}(t)\right)=f_{j}\left(\boldsymbol{s}_{j}(t)\right)-g_{j}\left(\boldsymbol{s}_{j}(t)\right), 
\end{equation}
where $f_j$ and $g_j$ denote the obtained benefit and the cost incurred by the cooperation of the relay node $j$ when adopting the strategy $s_j (t)$ over time $t$, respectively. Although the NE of this game can be obtained through the OPF,  it requires complete global network information. As is mentioned above, it could be inapplicable due to the absence or impracticality of the infrastructure in the dynamic VNs. Accordingly, a decentralized self-organized algorithm is designed for the NE solution. This algorithm enables the vehicles to adapt to the dynamic networks in a distributed and self-organized manner through rewarding the relay nodes. Specifically, the source vehicle uses the optimal transmission power allocation method to perform multi-relay cooperative communication. Moreover, the reward stimulation mechanism is considered in this study to encourage the \textcolor{color1}{relays} to cooperate. Another advantage of this study is the learning mechanism designed for the players to update their selection probabilities.  The authors proved that this algorithm converges to {\color{color1} an} NE when the learning rate $\delta$ is sufficiently low. Therefore, the decentralized self-organized algorithm enables payers to independently adapt  {\color{color1}decision making} behaviors {\color{color1} to} learn their NE strategies from the action-reward histories. This method outperforms the conventional schemes (i.e., stochastic relay selection and fixed relay selection) regarding energy consumption, energy benefit, and fairness. {\color{color1} The experiment of this study proves that the game scenario with 4 sources and 10 relays converges to the NE before the 400-th iteration. One weakness of this method is that this game is played repeatedly although it can converge to the NE. The work lacks further discussion that the method can guarantee a low delay without causing large computational overheads during the iteration.}

\textcolor{color1}{The authors in \cite{das2017new}} focus on a game-based relay selection from the perspective of MAC-level re-transmission. This game approach aims to mitigate transmission failure by selecting the cooperative node in the re-transmission.  This study  considers a network with $n$ vehicles $V=\left\{V_{1}, V_{2}, \ldots, V_{n}\right\}$ and a set of neighboring nodes $N_{i}=\left(V_{2}, \ldots, V_{n-1}\right)$. The neighboring nodes act as players and monitor the transmission from $V_1$ to $V_n$ and save a cope of the transmitted packet for re-transmitting in case of the failure. This study offers the novelty of incentive function, which is designed to allocate incentives to encourage relay nodes to participate. The amount of incentives that a neighbor node (relay) $V_k$ earns after re-transmitting the packet $\Delta_i$ to the destination node $V_j$ is:

\begin{equation}
\label{eq_das2017new}
v_{k}\left(G^{t}\right)=\left\{\frac{\Delta_{i} \cdot e_{k j}(G)}{\vartheta_{k j} \cdot \tau_{k i}(G)}, \quad\right. \text { if } \quad v_{i j}>0,
\end{equation}
where $\vartheta_{k j}$ denotes the relative speed between the relay $V_k$  and the destination $V_j$ and $\tau_{k i} $ denotes the relative distance between the relay $V_k$ and the transmitter $V_i$. The incentive function has the advantage of selecting the relay with a nearer distance to the transmitter and a closer relative velocity to the receiver. The VN is proved to be converged into a Nash network where no relay vehicle can improve its incentives by unilaterally change its strategy.  {\color{color1} One issue is that each player is required to copy and store the message transmitted from the source to the destination, leading to extra storage costs of the vehicles}. Moreover, this mechanism is based on the assumption that all the relay nodes in the neighboring set are trustworthy, which could cause potential attacks because all these nodes save a copy of the packet. Similarly to the issues mention above, this study neglects the demonstration of the convergence time of the Nash solution.

Suleiman et al. \cite{suleiman2017adaptive} employ the GT to demonstrate the incentive-compatibility of the routing protocol for highway applications in VNs. A fixed cellular network operator $ CO $ and a set of vehicles $ V $ constitute the players of the game in the highway scenario. On the one hand,  $ CO  $ chooses \textit{incentivizing cooperation} $  (IC) $ making credit-based bandwidth allocation or \textit{not incentivizing cooperation} $ (NIC) $ fixing bandwidth allocation. On the other hand, the strategy of any vehicle $  (AV_i \in V )  $ is Cooperation $ (C) $ relaying other vehicles’ messages or Not Cooperating $ (NC) $ dropping the messages. A one-shot game example with one operator  and two  vehicles is analyzed and expressed into an extensive form. The satisfied routing outcome $\{I C, C, C\}$ can be obtained. The study presents a novel routing protocol for highway applications. The main advantage is that it is capable of adapting to the mixed communication modes. More specifically, it can switch among {\color{color1} the} V2I routing, {\color{color1} the} V2V delay-tolerant routing, and {\color{color1} the} V2V direct routing.   Unlike the VN with a single communication mode, this study's incentive mechanism is also compatible with mixed communications. Specifically, the incentive is implemented through a credit-exchange system that is administered by the fixed network operator. Aiming at earning more credits paid by the transmitter, the relay vehicles are stimulated to share their memory and processing resources through V2V direct or delay-tolerant routing. Cooperative relay nodes owning high credit values are rewarded with extra bandwidth allocations, whereas selfish relays will be punished with less bandwidth. One limitation that should be aware of is that the credit management depends on the fixed network infrastructure, making the credit administration ineffective in an environment where there is no infrastructure, or it is impractical to install infrastructures.

		\begin{itemize}
			\item \textbf{Summary and conclusion of GT for  relay-based routing}
		\end{itemize}
	It can be found that {\color{color1} most of the} relay-based routing methods are modeled as {\color{color1} the} normal form non-cooperative games where {\color{color1} the} relay vehicles act as {\color{color1} the} players. Incentive approaches can be incorporated to stimulate cooperation and mitigate selfishness, such as rewards \cite{tian2017self}, incentive functions \cite{das2017new}, and credits  \cite{suleiman2017adaptive}. The learning scheme \cite{tian2017self} can play a positive role to overcome or mitigate incomplete information by learning from historical strategies. The {\color{color1} insecurity of the routing} incurred by the untrustworthy or malicious relays should be paid more attention, especially the relay-based routing where the message is transmitted through multi-hops. For example, the communication is under higher risk of attacks in \cite {das2017new} because each relay node stores a copy of the packet. Another problem that should be concerned is the convergence of Nash results. The multi-hop routing is constructed by iteratively updating the strategies to adapt to the dynamic VNs {\color{color1} \cite{kim2016timed, tian2017self,das2017new}}. Therefore, it is essential to prove the NE result is converged, and the convergence time should also be limited to an upper bound in the delay-sensitive VNs.

	\begin{table*}
	\footnotesize
	\label{tab_routing}
	\caption{Summary of routing games in VNs}
	\renewcommand*{\arraystretch}{1}
	\begin{center}
		\begin{tabular}{|m{.026\textwidth}|m{.08\textwidth}|m{.1\textwidth}|m{.15\textwidth}|m{.13\textwidth}|m{.09\textwidth}|m{.06\textwidth}|m{.015\textwidth}|m{.013\textwidth}|m{.013\textwidth}|m{.013\textwidth}|m{.014\textwidth}|}
			\hline
			\textbf{Ref.}&\textbf{Game}&\textbf{Players}&\textbf{Objective}&\textbf{Main Idea} &\textbf{Routing}&\textbf{\textbf{Scenario}}& \textbf{SP}&\textbf{PO}&\textbf{ID}&\textbf{AC}&\textbf{CM}\\
				\hline
				\multicolumn{12}{|c|}{\multirow{2}{4cm}{\textbf{ Relay-based routing}}}\\
				\multicolumn{12}{|c|}{}\\
				\hline
				\cite{tian2017self} &Non-cooperative game &Candidate relay nodes&Optimize transmission quality with low energy consumption&Incentive best relay selection& Decentralized&Urban&$\surd$&$\surd$&$\surd$&$\surd$&$\surd$\\	
				\hline			
				\cite{das2017new}&Non-cooperative game&Neighbor nodes&Retransmission in case of network failure&Best relay selection&Decentralized&$-$&&&&&\\
				\hline
				\cite{suleiman2017adaptive} &Static game&A fixed cellular network and vehicles&Credit-based bandwidth allocation&Best routing selection among V2I and V2V techniques&Hybrid&High-way&$\surd$&$\surd$&$\surd$&$\surd$&$\surd$\\
				\hline
				\cite{kim2016timed}&Bargaining game& Vehicles&Opportunistic routing selection&Best routing selection&Decentralized&$-$&$\surd$&&$\surd$&&\\
				\hline
				\multicolumn{12}{|c|}{\multirow{2}{4cm}{\textbf{ Broadcast-based routing}}}\\
				\multicolumn{12}{|c|}{}\\
				\hline
				\cite{hu2016novel}&Non-cooperative game&Vehicles&Delay minimization for emergency message broadcast & Rebroadcast&Decentralized&$-$ &$\surd$&&$\surd$&&$\surd$ \\
				\hline
				\cite{assia2019game}  &Non-cooperative game&  Vehicles& Reachability maximization and delay minimization &Rebroadcast&Decentralized&Grid-map&$\surd$&$\surd$&$\surd$&&\\
				\hline
				\multicolumn{12}{|c|}{\multirow{2}{4cm}{\textbf{ Cluster-based routing}}}\\
				\multicolumn{12}{|c|}{}\\
				\hline
				\cite{sulistyo2019coalitional}&Coalitional game&Vehicles &Cluster stability maintaining and link quality improvement&Optimum cluster formation and CH selection&Decentralized& City toll&$\surd$&&&&$\surd$\\
				\hline
				\cite{huo2016coalition} &Coalitional game & Vehicles&Trade-off between stability and efficiency of communication & Optimum cluster formation and CH selection &Decentralized&Crossroad&$\surd$&$\surd$&&&\\
				\hline 
				\cite{dua2017reidd}&Coalitional game &Vehicles&Broadcast storm mitigation&Optimal cluster formation and CH selection&Hybrid&Patiala city&$\surd$&&$\surd$&$\surd$&\\
				\hline
				\cite{khan2018evolutionary} &Evolutionary game&Clusters and vehicles& Throughput optimization for cluster\newline Routing stability maintaining &Optimum cluster formation and CH selection&Hybrid&Manhattan grid&$\surd$&&&&\\
				\hline
				\cite{wu2018computational} &Coalitional game&One-hop neighbors of RSU or CH &Multi-hop routing optimization&Dynamic cluster formation and CH selection&Hybrid&Freeway and street&$\surd$&&$\surd$&&$\surd$\\
				\hline				    
			\end{tabular}	
		\end{center}
		Legend: SP=Speed, PO=Position, ID=Inter-distance, AC=Acceleration, CM=Channel Model.
	\end{table*}

	
\vspace{3pt}
\textit{\textbf{b) Broadcast-based routing}}
\vspace{3pt}

In VNs, a broadcast storm may happen when the network is overwhelmed by {\color{color1} the} continuous broadcast or {\color{color1} the} rebroadcast traffic. The simplest broadcast scheme is {\color{color1} the} blind flooding \cite{tseng2002broadcast}, where each vehicle node rebroadcasts the received messages until all vehicles in the network receive the messages. However, in a high-density network, flooding will lead to a large amount of redundant information, which {\color{color1} could} result in collisions or congestion \cite{li2020influence,li2019drivers}. On the other hand, in an environment with sparse traffic, {\color{color1} a small number of forwarders may not} maintain a good connection or {\color{color1}} reachability of {\color{color1} the} communication. Therefore, cooperative forwarding is essential for vehicle nodes to cover the network with a low rebroadcast frequency.

Emergency message warning is highly time-critical and requires a more intelligent broadcast mechanism with low latency and high reliability. The proposed emergency message broadcast game in \cite{hu2016novel} considers vehicle density and link quality. In this scheme, each vehicle calculates the link quality {\color{color1} by} using the receiving power. Then, the probability of rebroadcasting the emergency message is determined based on the NE. The proposed scheme results in {\color{color1} lower} broadcast overhead and transmission delay than blind flooding.

   Assia et al. \cite{assia2019game} design an effective rebroadcasting protocol to maximize the reachability while minimizing delay and \textcolor{color1}{the} number of rebroadcasts in VNs. They present an adaptive algorithm by modeling cooperative forwarding as a volunteer dilemma game. In this game, the rebroadcasting decision of each vehicle is based on the received information such as speed and position. Simulation results show that this protocol outperforms prior work in terms of reachability, \textcolor{color1}{the} number of rebroadcasts, and delay, especially in congested areas.

 \vspace{3pt}
\textit{ \textbf{c) Cluster-based routing}}
 \vspace{3pt}

  Clustering can be considered as an efficient solution to enhance the connectivity of VNs. The coalitional game is considered as a potential approach for cluster formation. It has been used in \textcolor{color1}{MANETs} \cite{massin2017coalition} because the cluster concept is compatible with a coalition. However, limited communication resources and high mobility of vehicles make it difficult for them to maintain a balance between cluster stability and communication efficiency. The clustering protocols for VNs can be categorized as centralized clustering based on a central controller (e.g., roadside unit), decentralized clustering based on V2V communication, or hybrid clustering which incorporates the above two protocols.  However, centralized clustering has the weakness of inflexibility and low scalability.  Most of the recent GT-based clustering schemes for VNs aim to balance cluster stability and clustering efficiency using the decentralized clustering \cite{huo2016coalition, sulistyo2019coalitional} or hybrid clustering \cite{dua2017reidd,wu2018computational,khan2018evolutionary}.

	Focusing on crossroad scenarios, Huo et al. \cite{huo2016coalition} present a coalition game-based clustering strategy that considers both communication efficiency and cluster stability. The vehicle's strategy is to decide whether to move into a new coalition or stay in its current coalition. The coalition utility includes three metrics: velocity function $V_h$, position function $P_h$, and efficiency function \textcolor{color1}{$E_h$:} $U_{\text {hcluster}}=k_{1} \cdots V_{h}+k_{2} \cdots P_{h}+a \cdots E_{h}$. A vehicle decides to switch to a new coalition with higher social utility. Regarding the CH selection, the stability of the node $i$ is measured firstly to evaluate if it is eligible to play the role of the CH. If it is qualified, then the vehicle that can provide the largest available bandwidth is selected as the coalition's CH. This study designs a detailed switch rule for the coalition formation, which corresponds to the cooperative game's external-enforced agreement to prevent players from deviating from the coalition. Another advantage is using the stability metric for measuring the eligibility of the CH. This step guarantees cluster stability because the CH's stability affects the cluster's stability. Furthermore, the proof of the Nash-stable coalition convergence provides a theoretical basis for the coalitional game's stability. 

	To improve the communication link quality, a coalitional game-based distributed clustering method is proposed in \cite{sulistyo2019coalitional} for V2V communication. In this game, \textcolor{color1}{each vehicle decides to select} neighbor vehicles to establish V2V coalition with high link quality. The player's utility is formulated as a function of SNR, link lifetime, and velocity difference among vehicles to quantify the V2V link quality. The criteria for selecting the CH $i$ is based on the term electability, which is defined as the number of vehicles that intend to connect with $i$. Furthermore, the vehicle with a lower speed will be selected as the CH if two or more vehicles have the same electability. However, the cluster's stability, which is an important metric to measure the quality and efficiency of the cluster formation, is not proven or evaluated in this study.

GT is used by \cite{dua2017reidd} to design data dissemination protocol in VNs. \textcolor{color1}{A vehicle regularly updates the utility based on the weight of link between it and its neighbor}, which is computed by combing the parameters of  distance $D_{t}$, relative speed $RS_t$, relative acceleration $RA_t$, and angle $AG_t$ as  $W_{t}=D_{t} \times k_{1}+R S_{t} \times k_{2}+R A_{t} \times k_{3}+A G_{t} \times k_{4}$. The vehicles  dynamically form clusters based on the utilities by including the vehicles with stable links.  The vehicle node with the minimum value of traffic flow values (TFV) \cite{kumar2013alca} is selected as a CH because it has low mobility and can locate in the cluster for a longer time. The nearest RSU can join the game and plays as an assistant by executing the store-carry-forward technique when the route between the source and destination is not connected. This study also employs the advantage of the learning approach for TFV update.  The vehicles could obtain more global information by learning from the traffic and dynamic patterns of \textcolor{color1}{the} other vehicles.  Another novelty is that the learning rate $\lambda \in[0,1]$ is updated through reward or punishment mechanisms. However, the communication link quality is formulated using the traffic and dynamic patterns, which is insufficient because the link of VNs is also affected by channel parameters such as path fading exponent. The parameters in \cite{wu2018computational, sulistyo2019coalitional} can be further considered in the formulation.  Besides, the convergence or stability of \textcolor{color1}{NE} is needed to be proven or analyzed.

To improve channel contention efficiency at the MAC layer, Wu et al. \cite{wu2018computational} design a multi-hop clustering routing protocol for vehicle-to-roadside (V2R) communication based on the coalitional game. The game is represented as $(\mathcal{N}, v)$, where the set of players $\mathcal{N}=\{1, \ldots, N\}$ is consisted by one-hop neighbors, and $v$ is the coalition value. Players aim to form clusters $S_i\subseteq \mathcal{N}$, where the CH in each cluster is selected as a relay for multi-hop transmission. The value for the cluster $S_i$ is quantified based on the MAC-layer contending, {\color{color1} which is given by} $v(S)=\left(1-\frac{P(N-|S|)}{2}\right) \cdot N$, where $|S|$ denotes the cardinality of the coalition $S$, and $P$ denotes the collision probability. This study proposes a novel incentive scheme that an RSU or CH acts as a utility allocator and distributes utilities to its one-hop neighbors. Consequently, each vehicle would prefer to cooperate for data forwarding because only an RSU or CH can assign the utility. In terms of CH selection, a fuzzy logic algorithm brings benefits for dynamic CH selection by combining multiple metrics such as speed, signal quality, and vehicle movement pattern. Another advantage is using reinforcement learning (RL) for the reward allocation, which alleviates the limitations of incomplete information and limited rationality; this incentive mechanism helps vehicles select the route with maximum network performance by learning from the environment. The game is proved to be stable, and the core of the game is proved not empty.


The evolutionary game is employed in \cite{khan2018evolutionary} for cluster routing to improve the communication throughput globally because it can handle the limited rationality of players. The game framework is given as $G=\left\langle N, H, S, u_{C_{h}}\right\rangle$, where \textcolor{color1}{$N=\{1,2,\ldots,n\}$} is the population, $H=\{H_1,H_2,\ldots,H_j\}, \ H_j\subset N$ is the set of clusters acting as players of the game, and the utility function $u_{C_{h}}$ is defined as the total throughput of the entire cluster. The vehicle decides to be a CH or a cluster member. The replicator dynamics is used for CH selection. A replicator with higher \textcolor{color1}{throughput} will replicate faster. This study's main advantage is that it overcomes the limitation of frequent cluster reorganizing and CH nomination in the highly dynamic VNs. The evolutionary game can alleviate the bounded-rationality of players and provide global network knowledge to improve overall network routing performance. Furthermore, the existence of \textcolor{color1}{the ESS is proved analytically using the Lyapunov function} in this study. \textcolor{color1}{The vehicles can obtain higher throughput by adapting their strategies until the ESS is reached}. This study 
addresses the possible long convergence time of the evolutionary game by using the ESS to guarantee the stability and convergence of the outcome. The authors analyze that the overhead increases logarithmically with the complexity of $\mathcal{O}\left(N \log _{2} H\right)$. However, the intrinsic weakness of evolutionary games\textemdash selfishness could overwhelm altruism when selfish players refuse to behave altruistically\textemdash is not considered in this study.

\begin{itemize}[itemsep=3 pt,topsep = 3 pt]
	\item \textbf{Summary and conclusion of GT for Cluster-based Routing}
\end{itemize}
 Game theoretical clustering routing provides an approach to alleviate the broadcast storm by limiting the number of transmitted messages. However, the difficulty of the clustering approach is the stable cluster formation and CH selection in the highly dynamic VNs. The cooperation or coalitional game is suitable for the cluster-based routing because the cluster formation among vehicles can be easily modeled as: the players cooperate or coordinate to form coalitions (or alliance) with the aim of maximizing the collective utilities. Furthermore, the cooperation game provides the evaluation of stability of the cluster formation through the quantification such as the Shaplay value and the core (which are introduced in Section \ref{sec_coopereative game}). There are some key aspects that should be payed attention in the design of game-theoretical clustering routing mechanisms. 
	
	\begin{itemize}[label=$-$,itemsep=3pt,topsep=3pt]
	\item Regarding the quantification of the coalition utility value, besides the traffic and mobility patterns \cite{huo2016coalition,sulistyo2019coalitional,wu2018computational,dua2017reidd}, the physical-layer parameters (including SNR, path loss, and link lifetime) \cite{sulistyo2019coalitional} and the channel contention at MAC-layer \cite{wu2018computational} can be jointly considered for a more precise quantification. \item  CH selection is another major step in the cluster-based routing. Specific metrics mostly used for CH selection is the connectivity of links or stability of vehicles (with low mobility) \cite{huo2016coalition,sulistyo2019coalitional}. The study in \cite{huo2016coalition} gives a more effective selection approach by using the stability of the vehicle to evaluate the eligibility of the CH candidate firstly. Fuzzy logic algorithm \cite{wu2018computational} offers an advanced method to combine multiple metrics, which can be used for the difficult-to-formulate link quality quantification.
	\item Analyzing or proving the convergence or stability of the NE is an essential step to guarantee the stability of clusters in the highly dynamic VNs, which is also an element in the definition of the cooperative game.  Learning schemes \cite{dua2017reidd,wu2018computational} can be incorporated into the coalitional game to improve its effectiveness by providing more global information. 
	\item Evolutionary game can alleviate the frequent cluster reorganization and CH nomination, which is more suitable for the highly dynamic cases of cluster-routing. What should be noticed is the possible long convergence time of the ESS.
	\end{itemize}

\subsection{{\color{color1} Game theory} for energy management in EV-involved VNs}
\label{sec_GTenergy}
{\color{color1} Integrated with EVs and the intelligent energy entities such as smart communities (SCs), smart grids (SGs), the EV-involved VNs is emerging as a new paradigm for the future green ITSs. The energy requirement of EVs brings new challenges for energy management due to the uncertain location of the energy facilities such as charging stations and the highly dynamic of EVs. GT provides a useful tool to construct the interaction among these intelligent entities for efficient and dynamic energy management.}

Li et al. \cite{li2016noncooperative} design a price-driven distributed charging control method for EVs using {\color{color1} the} non-cooperative game. The EVs are considered as {\color{color1} the} selfish players aiming to minimize the charging cost under the  load constraints of feeders. It is proved that the game can converge superlinearly. The main advantage of the charging method is two-fold:  1) the delay of the energy management is low as the game is converged superlinearly; 2) the privacy of {\color{color1} the} EV is guaranteed by considering the EVs' information as private. However, only {\color{color1} considering} the interactions among vehicles may not enough because vehicles also communicate with other energy entities such as {\color{color1} the} SC {\color{color1} and the SGs}. To overcome this limitation, Zhang et al. \cite{zhang2016energy} propose a novel three-party architecture for the interaction among the power grid, EVs, and SCs. They suggest that  GT provides a powerful mathematical tool for the energy schedule or coordination among the three entities intelligently and cooperatively. However, the mechanism design and specific solutions are not provided. 

An auction game-based charging coordination scheme for EVs over a finite horizon is designed in \cite{zou2016efficient}. A set of EVs are players whose strategies are charging rate at a time interval. The bids of each player consists of a maximum amount of demand and the price, overcoming the limitation of the single divisible resource allocation where each player only has a two-dimensional bid. The utility function constructs the {\color{color1} trade-off} between the benefit for delivering the full charge and the penalty of not fully charging the EV over the horizon. One advantage of the scheme is it captures the effect of time and has an enforced incentive effect on preventing deviating. Another advantage is the that the charging coordination solution can achieve incentive compatibility, ensuring the truth-telling bids of all players and the efficiency of the resource allocation.

Wang et al.\cite{wang2019novel} propose an efficient and flexible energy management scheme for EVs in the VN with heterogeneous energy entities. The Stackelberg game models interactions among the three parties: main power grid, aggregators in the small grid, and EVs. Firstly, the main power grid acts as a leader who decides the energy price. Secondly, the aggregator plays as a leader of the local energy trade subgame, determining the amount of energy it desires to buy from the power grid. The aggregator then issues the selling price to EVs. Thirdly, EVs acting as followers decide to choose either the traditional energy, clean energy, or the hybrid of them. The NE solution gives optimal energy allocation strategies to EVs with different preferences on the energy sources. This EV charging scheme innovatively extends the traditional Stackelberg game to a three-level architecture, making it more adaptive to the complex energy VNs with heterogeneous and multiple energetic entities. Furthermore, the social relationship-based trust evaluation for the aggregator and credibility for EVs guarantees charging security. Another advantage is using a price-based incentive mechanism that promotes the interaction among the three parties. Finally, the weighted max-min fairness based approach has a positive effect on the fairness and efficiency of the energy allocation. However, the dynamic property of vehicles and the communication delay are not considered, which is vital in the real-time VNs.

Laha et al. \cite{laha2019game} employ a multi-leader and multi-follower Stackelberg game to model the interaction between EVs and smart-grid charging stations. Each charging station acts as a leader to maximize the revenue by setting the charging price; each EV acts as a follower deciding the preferred station with the lowest charging price given the charging price, time intervals, and location. A key advantage of this study is the consideration of location awareness in EVs' utility function, which captures the distant factor affecting the EVs' preference on choosing the charging stations. Furthermore, the follower-subgame is modeled as a matching problem, facilitating the obtaining of stable EV-station pairs according to their preferences.

To fully utilize the energy resources, Ye et al. \cite{ye2020motivational} focus {\color{color1} on} a V2V energy trading by using the Stackelberg game to facilitate the interaction between a seller EV and a buyer EV. This work overcomes the challenges of limited number and uncertain range of charging stations. The vehicle with idle energy coordinates satisfied energy sharing strategy with the vehicle desiring energy. However, this study only considers a one-to-one trading scenario, which is insufficient for the real vehicular scenarios. Besides, the seller (leader) and the buyer (follower) only coordinate once without feedback of follower. The bargaining or auction game could lead to a more precise outcome.

	\begin{itemize}[itemsep=3pt,topsep=3pt]
		\item \textbf{Summary and conclusion of GT for energy  management in EV-involved VNs}
	\end{itemize}
This subsection discusses the application of GT to the interaction among intelligent entities for EV-involved charging management. {\color{color1} The} Stackelberg game provides a well-suited framework to analyze the hierarchical energy trading interactions among the intelligent entities \textcolor{color1}{\cite{laha2019game,wang2019novel,ye2020motivational}}. With the emergence of the heterogeneous EV-involved VNs, only modeling the two-sided interaction between EVs and SG-enabled charging stations is insufficient \cite{laha2019game}. More complex interplay should be considered, such as the interactions among SGs (including the main power grid and micro-grid), SCs, and EVs \cite{zhang2016energy,wang2019novel}. The multi-leader and multi-follower Stackelberg game could be well-suited to model the complex interaction among multiple entities \textcolor{color1}{\cite{laha2019game}}. {\color{color1} Given the possible inefficient outcome of the Stackelberg game,} incentive mechanisms or other games with the incentives could be {\color{color1} applied}  to stimulate the interaction or {\color{color1} the} cooperation. For example, the matching approach can be incorporated into the follower-layer decision making to obtain the efficient and stable paring between SGs (or SCs) and EVs with different preferences. Besides, cooperation between EVs is critical to enable the resource sharing between EVs with idle resources and EVs with scarce energy, especially in the environment without charging stations. Another significant issue is the consideration of VN characteristics in the game modeling. The dynamic pattern of vehicles and the communication delay, as the salient features of VNs, have been overlooked by most of the existing work and could be modeled in EVs' utility functions. Furthermore, the privacy-preserving \cite{li2016noncooperative} and location awareness \cite{laha2019game}  are also essential in the interaction between EVs and SG-enabled charging stations. The incomplete information game could capture these uncertainties.

	\section{Game Theory in Next-generation VNs}
	\label{sec_GTin5G}

{\color{color1} The} next-generation 5G communication technology can potentially meet the demands of a number of connected IoT devices and various data-intensive applications. 5G provides low latency, high reliability, ubiquity, and energy efficiency in communications \cite{xiang20165g}. With the rigorous requirements of 5G, emerging technologies are integrated into VNs to promote their evolution toward 5G-based VNs. The recent contributions of GT to diverse emerging technologies that integrate with VNs, such as MEC, HetVNs, SDN, and UAV, are presented in the following paragraphs.

\subsection{Game Theory in HetVNs }
\label{sec_GTHetVN}

\begin{figure}[!hbt] 
	\centering
	\includegraphics[width =3.6in]{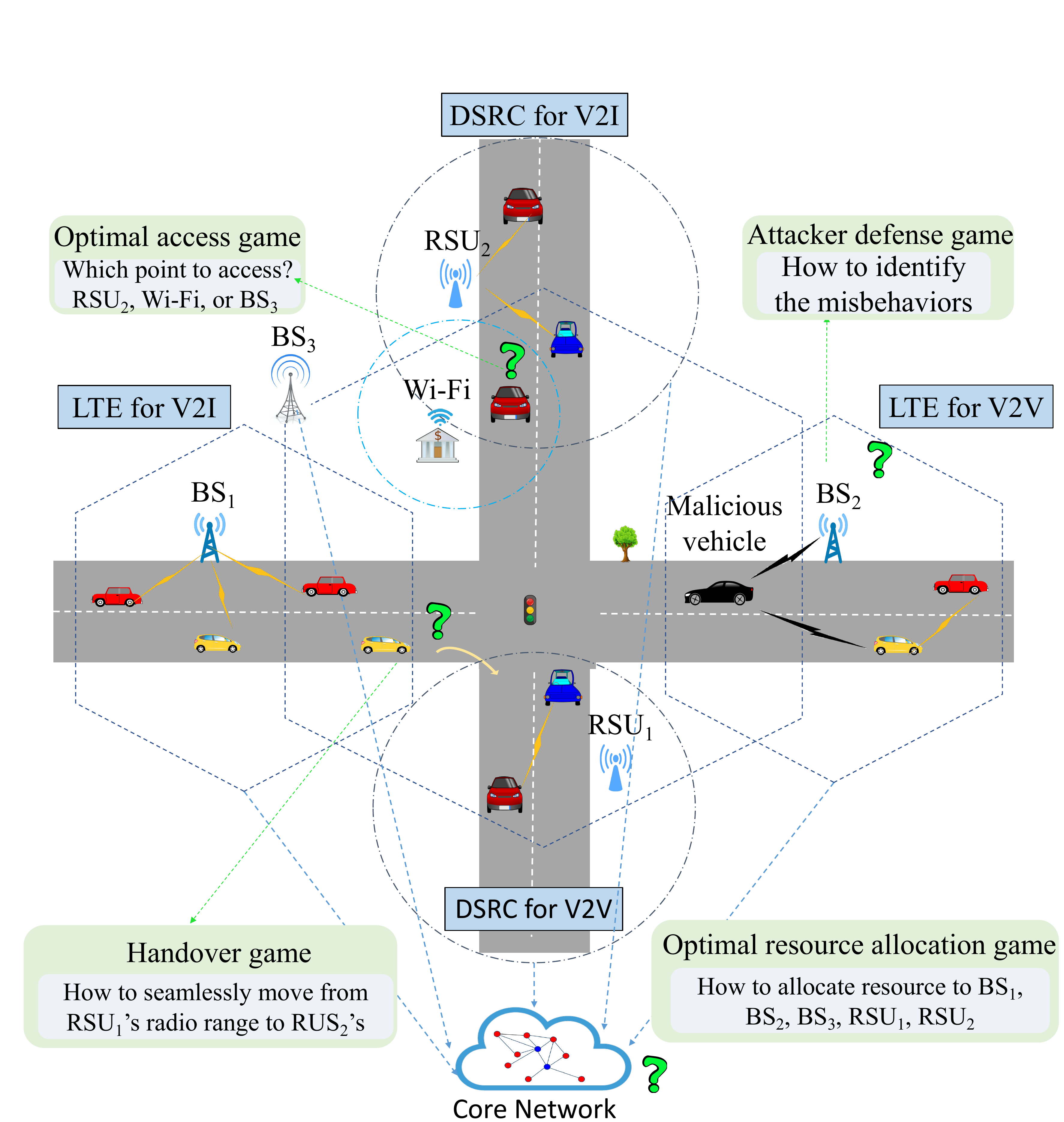}
	\caption{GT in HetVNs.}
	\label{fig_hetVN}
\end{figure}

Traditional VNs suffer from heavy communication and traffic load due to the \textcolor{color1}{ever-increasing} service demands. The  HetVNs thus is proposed to {\color{color1} use} heterogeneous access technologies to provide ubiquitous connections for vehicles. As shown in Fig. \ref{fig_hetVN}, {\color{color1} different types of communications that are supported by different communication technologies coexist in the HetVNs.} The dynamic vehicles and the integration of different technologies pose challenges to HetVNs. GT provides useful tools for vehicles or access points (RSU or BS) to decide the optimal strategies such as optimal resource management \cite{xiao2018spectrum}, efficient access decision \cite{mabrouk2016meeting,hui2017optimal,hui2019game,zhao2019optimal}, and security strategies \cite{sedjelmaci2017predict,chen2017congestion}.

\vspace{6pt}
\subsubsection{Resource sharing}

Xiao et al. \cite{xiao2018spectrum} investigate the spectrum resource reusing method for macrocell and RSUs in HetVNs. A non-cooperative game is used to allocate the macrocell's vacant spectrum resources to the RSUs to increase the system's performance and decrease the interference. The players are a set of RSUs in which RSU $r$'s strategy $A_{k}^{r} (k\in {1,2,\ldots,N_b})$ means $k$'s resource block is assigned to $r$; $N_b$ is the total quantity of the vacant spectrum bandwidth. The utility of RSU $r$ transmitting to the requested vehicle $i$ is formulated as the SINR {\color{color1} value}  $SINR_i^R(t)$ in the time interval $t$. The system's utility is {\color{color1} formulated as} the sum of RSU-VN utilities in the system. This work first invents a novel graph color-based algorithm for the strategy set formation based on RSUs' relative geographic positions.  Compared with the presupposed strategy set in the conventional studies, this work mitigates the competition between adjacent RSUs in advance before the game starts. Furthermore, the use of ``regret matching" provides a powerful iterative tool for obtaining the converged correlated equilibrium. One limitation is {\color{color1} that this work does not consider the selfishness of the macrocell} that may refuse to offer free vacant resources. The incentive for negotiation or cooperation between the macrocell and RSUs needs further research.


In \cite{hui2020collaborative}, the cooperation between the cellular base station (CBS) and RSU is studied for collaborative content delivery in the situation where RSUs could not deal with a large number of requests. A double auction game is used to motivate the cooperation between CBS and RSU which are viewed as the cooperative players in game. The RSU first selects the content that needs to be provided with the help of a CBS. Then the RSU sends a request to the CBS that covers it. The players aim to negotiate the optimal price for the content to maximize their utilities. The auction starts with the players' evaluations for the CBS-assisted content. As in the real-world, players estimate the evaluations of the content according to their situated context and their own capabilities. The agreement is achieved if players' biddings satisfy:
\begin{equation}
	\label{eq_het}
	b_{k, q^{*}}\left(v_{k, q^{*}}\right)<b_{j, q^{*}}\left(v_{j, q^{*}}\right),
\end{equation}
where the RSU $j$ selects the content $q^*$ to request the cooperation of the CBS $k$. Furthermore, $b_{j, q^{*}}$ and  $b_{k, q^{*}}$ are the biddings of RSU $j$ and CBS $k$, respectively. The biddings for the content depend on the evaluations estimated by players, i.e., $v_{j, q^{*}}$ for RSU $j$ and $v_{k, q^{*}}$ CBS $k$. The RSU decides to pay $\mathcal{J}_{j, q^{*}}$ to CBS $k$.

\begin{equation}
	\label{eq_het1}
	\mathcal{J}_{j, q^{*}}=\lambda b_{j, q^{*}}\left(v_{j, q^{*}}\right)+(1-\lambda) b_{k, q^{*}}\left(v_{k, q^{*}}\right),
\end{equation}
where $\lambda(0<\lambda<1)$ is the proportion of dividing the profits between the players. This work enables the RSU that cannot finish the delivery tasks to successfully ask for the cooperation of the CBS. The advantage of an auction game is to mitigate the selfishness of the CBS and stimulate the cooperation between the RSU and CBS. This work's main advantage is that the players can dynamically negotiate a satisfied price for the content so that the CBS agrees to help while both players can obtain the optimal profits. Furthermore, this auction game is a incomplete information game where the probability distribution is assigned to the service's evaluation. The conjecture on the evaluations precisely reflect the uncertainty of the \textcolor{color1}{RSU} (CBS) on the CBS's (RSU's) preferences of the service. The last but not least, this game presents the optimal theoretical bidding strategies, making the decision more precise and time-saving.



\vspace{6pt}
\subsubsection{Optimal network selection or access} Optimizing the connection for vehicles with the lowest cost is challenging in HetVNs, where different access technologies lead to different costs in terms of {\color{color1} downloading} latency and bandwidth utilization. An always-best-connected paradigm \cite{mabrouk2016meeting} based on the signal game is designed to help vehicles in HetVNs connect to an optimal network with low connection cost. However, this approach only considers {\color{color1} the} two-player competition in a limited geographical area in urban scenarios. In \cite{hui2017optimal,hui2019game}, the coalitional game is employed to design the optimal access decisions for vehicles. A novel evaluation on vehicles' download cost is constructed based on vehicles' interest degree on the latency or price of the requested content. The coalition formation algorithm is designed where each vehicle can select the optimal access network with the minimal cost based on the Nash stable partition. The advantages of this work is two-fold: 1) the interest level-based utility model captures vehicles' preferences of different contents, and 2) vehicles in the same coalition can download the requested contents cooperatively at low costs.


The studies mentioned above \textcolor{color1}{overlook} the unexpected massive handovers caused by \textcolor{color1}{highly dynamic of VNs. To address this problem,} Zhao et al. \cite{zhao2019optimal} propose an optimal non-cooperative game approach for network selection in HetVNs. Each vehicle acts as a player whose strategy is selecting DSRC, LTE, or WI-FI in the HetVNs to access. Vehicle's utilities are the evaluations of the three types of networks, which {\color{color1} are} formulated based on the parameters of delay, packet loss rate, and jitter.  Simulation results reveal that probabilistic strategies can adapt to the varying network with convergence. However, the system is tested in a relatively ideal environment where all vehicles are connected to DSRC networks in default and move at a constant speed. The highly dynamic character is one of the most significant difficulties for  {\color{color1}decision making} of network selection or access in HetVNs. {\color{color1} SPs need to accomplish the task continuously in a short delay when vehicles switching from one SP to another. }

\vspace{6pt}
\subsubsection{Security} 
 
Security in HetVNs is challenged by various potential attacks among heterogeneous intelligent nodes. {\color{color1}Sedjelmaci et al. \cite{sedjelmaci2017predict}} design an attack detection and prediction scheme based on GT to detect and predict the misbehavior of malicious vehicles in HetVNs. The attack–defense problem is formulated as a game between the attacker and the services center. Furthermore, the future behavior of monitored vehicles can be predicted based on the NE. {\color{color1}To guarantee the secure data transmission at intersections in HetVNs,} Chen et al. \cite{chen2017congestion} propose a congestion game-theoretical transmission control scheme with the aim of alleviating the channel congestion. A new concept of \textcolor{color1}{requirement of safety (RoS)} is proposed in this work for measuring the security level of services. One weakness of this work is that the game is based on complete information. In the real-world scenarios, the detector or the target vehicle has limited or no knowledge on the malicious nodes. The Bayesian game could be more appropriate for security problems in HetVNs.

\begin{itemize}[itemsep=3pt,topsep=3pt]
	\item \textbf{Summary and conclusion of GT for HetVNs}	
\end{itemize}

In this subsection, GT plays an advantageous role in HetVNs in the following aspects.

\begin{itemize}[label=$-$,itemsep=3pt,topsep=3pt]
	\item From the perspective of access points, GT stimulates them to coordinate for resource sharing. In the resource sharing game, existing studies \cite{xiao2018spectrum,hui2020collaborative} focus on using GT to construct cooperation or interaction among SPs. Interactive pricing strategies of auction game and bargaining game are promising to stimulate interaction and mitigate selfishness among SPs by negotiating a satisfying price for service. It can be concluded from the definitions in Section \ref{sec_NC} that the auction game is more appropriate to model the interaction among multiple SPs while the bargaining game is more suitable for the two-player situation.
	
	
	\item From the perspective of vehicles, GT helps them decide the optimal network access among the multiple communication technologies or avoid malicious attackers. GT is used to model the competitive \cite{mabrouk2016meeting,zhao2019optimal} or cooperative \cite{hui2017optimal,hui2019game} behaviors among vehicles. The utility function is often formulated as the evaluation of the HetVN performance so that vehicles can select the optimal network. However, little work considers vehicles' diverse preferences on different applications or services of heterogeneous networks in the utility models.

	
	\item  The GT application for handover management in HetVNs is still not mature. The game model for handover is expected to be complex due to the highly-dynamic and delay-sensitive VNs.
\end{itemize}


\subsection{Game Theory in VEC-enabled VNs}
\label{sec_GTVEC}
\textcolor{color1}{VEC} is an emerging architecture that offers cloud computing capabilities at the edge of the VNs to manage the computation-intensive and real-time tasks with short latency. However, the problems of task offloading, content delivery, and resource allocation or sharing are challenging in VEC due to the limited storage capacity of \textcolor{color1}{MEC}, especially in dense traffic environments where vehicles have quantity service demands. Furthermore, the lack of complete information between the tasks and edges makes the problem more complicated. GT, which is a powerful tool to deal with conflicts, can effectively address the  {\color{color1}decision making} concerns among multiple users who compete for limited resources. 

\vspace{6pt}
\subsubsection{Overview of VEC}

\begin{figure*}[!hbt] 
		\centering
	    \includegraphics[width =7.6in]{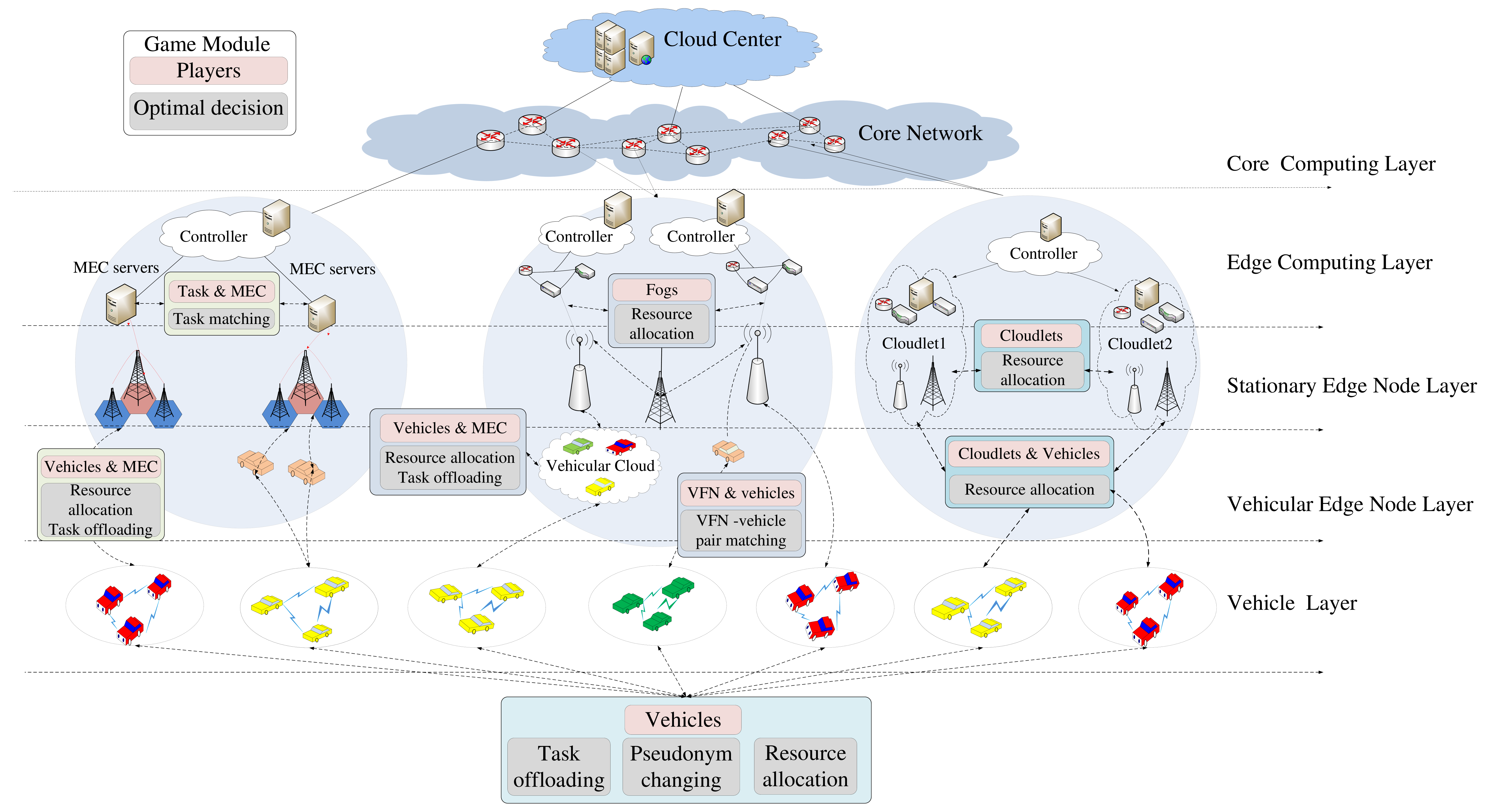} 
	    \caption{The architecture of VEC (primarily based on \cite{wang2018enabling}).} 
	     \label{fig_vec}  		
	\end{figure*}

As shown in Fig. \ref{fig_vec}, the three-layer VEC architecture consists of a cloud layer, an edge computing layer, and a vehicular layer. At the edge computing layer, the VEC implementations can be classified into mobile edge computing (MEC), fog computing (FC), and cloudlet computing \cite{dolui2017comparison}. At the vehicular layer, the vehicular cloud (VC), as an extension of edge layer architecture, plays an essential role in assisting the edge layer or cloud layer with effective resource management.

From the perspective of participating nodes, edge nodes are commonly characterized as stationary edge nodes and vehicular edge nodes (\textcolor{color1}{VENs}) , which are at the stationary edge node layer and vehicular edge node layer, respectively.

\begin{itemize}[itemsep=3pt,topsep=3pt]
	\item \textbf{Stationary edge nodes:} A roadside infrastructure such as RSU and BS connected to a MEC server or a fog server can serve as stationary nodes.  Besides, a cloudlet consisting of a set of roadside infrastructures also plays the role of an edge node that provides computation and storage capabilities to vehicles.
	
	\item \textbf{Vehicular edge nodes:} Vehicles with available computation resources and capabilities can form cloudlets or VCs for resource sharing.
	
	\item \textbf{Hybrid edge nodes:} Stationary and vehicular edge nodes coexist in the VEC.
	
\end{itemize} Table \ref{tab_edgenode} presents the types of edge nodes in VEC. The applications of GT in VEC are discussed from the aspects of MEC, FC, cloudlet, and VC receptively in the following subsections.

	\begin{table*}
		\footnotesize
		\caption{VEC edge nodes}
		\label{tab_edgenode}
		\renewcommand*{\arraystretch}{1}
		\begin{center}
			\begin{tabular}{|m{.16\textwidth}|m{.26\textwidth}|m{.28\textwidth}|}
				\hline
				\textbf{Categories} &\textbf{Types of VEC nodes}&\textbf{Related works}\\
				\hline
				Stationary Edge Nodes
				&Roadside infrastructure with MEC server & \cite{zhao2019computation},\cite{liu2018computaion}, \cite{zhang2017mobile},\cite{zhang2019task},\cite{zhang2017optimal}, \cite{hui2019edge},\cite{zhou2018begin},\cite{huang2017distribute}  \\
				\cline{2-3}
				&Roadside infrastructure with fog server& \cite{kang2017privacy},\cite{zhang2017resource}\\
				\cline{2-3}
				&Roadside infrastructure cloudlet&\cite{yu2016optimal}, \cite{tao2017resource} \\
				\hline
				Vehicular Edge Nodes&Vehicles& \cite{xu2018low},\cite{klaimi2018theoretical},\cite{aloqaily2017fairness},\cite{brik2018gss}, \cite{mekki2017proactive},\cite{liwang2019game}\\
				\cline{2-3}
				&SP cloudlet &		 \cite{yu2015cooperative}, \cite{lin2019vehicle}\\
				\hline
				Hybrid Edge nodes&MEC server and vehicles&\cite{gu2019task}\\
				\hline
			\end{tabular}
		\end{center}
	\end{table*}

	\vspace{6pt}
\subsubsection{MEC-implemented VEC}
	Several recent studies use GT to deal with the optimal  {\color{color1}decision making} for computation or task offloading \cite{zhang2017mobile,liu2018computaion,zhang2019task,gu2019task}  to reduce the computational delays and overheads. Furthermore, some works also consider resource allocation by jointly optimizing the task offloading for vehicles and the resource assignment for SPs \cite{zhang2017optimal,huang2017distribute,zhou2018begin,zhao2019computation}. Besides, content delivery with GT solution in MEC-implemented VEC is addressed in \cite{hui2019edge}. Table \ref{tab_MEC} summarizes the GT-based schemes in an MEC-implemented VEC.

	\vspace{3pt}
\textit{	\textbf{a)  \textbf{Computation or task offloading  game}}}
	\vspace{3pt}

	
	 We first conclude the basic system model commonly used by the recent studies on offloading  {\color{color1}decision making} in \textcolor{color1}{MEC-implemented} VNs. The network consists of a set of MEC servers  $\mathcal{M}=\{1,2,\ldots, M\}$, a set of vehicles $\mathcal{N}=\{1,2,\ldots, N\}$, and a set of tasks $\mathcal{T}=\{1,2,\ldots, T\}$. Each RSU is located along the {\color {color1} road} with the same coverage $L$. Concerning the computation task of vehicle $i$, it is usually characterized as $T_{i}=\left\{d_{i}, b_{i}, t_{i}^{m a x}\right\}, i \in \mathcal{N}=\{1,2, \ldots, N\}$, where $d_i$ is the  size of the tasks, $b_i$ is the amount of resources required to fulfill the task, and $t_i^{max}$ denotes the permissible delay of $T_i$.
	
	Zhang et al. \cite{zhang2017mobile} propose a cloud-based MEC offloading framework for VNs using adaptive computation transfer strategies with combined communication modes. The offloading for computation task is modeled as a general-form game where each player is a set of vehicles with the type-$i$ task. Each player $i$ uses mixed-strategy $P_{i,j}$, which is the probability of the vehicle set $i$ using the $j$-hop-away MEC servers. The players aim to minimize the offloading cost of task execution and data transmission with the constraint of delay. This study designs a game-theoretic framework for  MEC-implemented VEC. This framework provides an efficient and adaptive task offloading mechanism for vehicles to {\color{color1} execute the tasks locally, offload them to the direct MEC server via V2I communication, or offload them to the $j$-hop-away MEC server through predicted V2V relays.} Both the offloading cost and time consumption are reduced using this combination-mode offloading approach. {\color{color1} However, the NE is not proven is this study, which is essential to help the vehicles make the optimal offloading strategies in the real VNs.}

	
    Liu et al. \cite{liu2018computaion} model the  {\color{color1}decision making} for computation offloading among vehicles as a multi-user potential game where players are a set of vehicles moving with the constant speed. This study assumes a set of wireless channels as the strategies of each vehicle. The vehicle can execute the task locally or select a wireless path to offload it to a MEC server. The utility function is the \textcolor{color1}{computation overheads of local computing  and MEC offloading}. Each vehicle aims to minimize its computational overhead. This multi-user computation offloading method is proved to be a potential game, and a potential function is constructed. \textcolor{color1}{The existence and convergence of NE \textcolor{color1}{are} proved by using the potential function's advantages of disagreement quantification.}
    

The study \cite{zhang2019task} further extended the basic model in \cite{zhang2017mobile} to a more sophisticated model for more real scenarios of MEC-based VNs. The potential game is used to model the load balancing task offloading scheme. The game model of this study is structurally similar to that of \cite{zhang2017mobile} in terms of the following aspects: 1) \textcolor{color1}{The strategy of each player} is executing tasks locally or offloading them remotely; 2) \textcolor{color1}{the utility for each player is} formulated as the processing delay, and 3) the convergence of NE is proved with the help of the potential function. Compared with {\color{color1} the studies in \cite{zhang2017mobile,liu2018computaion}}, an advancement of this approach is that it considers more possible offloading strategies for vehicles, \textcolor{color1}{i.e.,} executing tasks locally, offloading them to the MEC server, and offloading them to the remote cloud. Furthermore, the vehicle's mobility patterns and the traffic are formulated in the utility using specific models instead of assuming as constants to accurately describe the actual VN scenarios. Concerning the system architecture, this study takes advantages of the SDN structure that can provide {\color{color1} the} enhanced centralized management and the fiber-wireless that combines low latency and high capacity of optical network. 


 	 Gu and Zhou \cite{gu2019task} focus on the task offloading in MEC-based VNs with incomplete information \textcolor{color1}{on channel states and mobility patterns.} This study designs a distributed task assignment \textcolor{color1}{scheme} by employing the matching game for task offloading. The matching game has the advantage of constructing a mutually beneficial relationship between two types of players with ordering preferences. In this game, both MEC servers and vehicles with excess computation capacities are viewed as edge nodes. The preference of each task is ordered by the computation delay, and that of each edge node is ordered by the transmission delay.  This offloading approach can provide better scalability because the matching decisions can be made distributedly using the local information.  Besides,  the utility function not only  aims to minimize the total delay of offloading (as in \cite{zhang2017mobile,liu2018computaion,zhang2019task}) but also takes into account the vehicle mobility and energy constraints.
  
 Wang et al. \cite{wang2020game} focus on the distributed task offloading for MEC-based VNs. A game-theoretical framework is designed to model the resource competition among vehicles. The competitive vehicles are acting as players who dynamically decide their offloading probability. The utility function is formulated based on the distance between vehicles and MEC servers and the application demands. The NE solution is theoretically proved to converge to a unique and stable outcome. Furthermore, the distributed offloading scheme brings flexibility and scalability. However, this study assumes a relatively simple scenario with only one MEC server and without considering vehicles' dynamic patterns.

			\begin{itemize}
			\item \textbf{Summary and conclusion of computation or task offloading game}
		\end{itemize}
	
		  In the studies of optimal  {\color{color1}decision making}, vehicles aim to offload tasks with minimum computational or processing overheads by choosing the optimal offloading decisions. Both non-cooperative and matching games can achieve this objective but with relative different ways. Non-cooperative games offer easily-expressed structure for the problem formulation: 1) modeling multiple vehicles as players, 2) modeling the offloading decisions as the strategies of vehicles \cite{zhang2017mobile,liu2018computaion}, 3) modeling \textcolor{color1}{the objectives as the utility functions of vehicles, and 4) giving NE solutions for the optimal decisions.} Matching game offers a distributed mechanism for paring tasks and \textcolor{color1}{edge nodes} according to their preferences. Regarding the strategies, the vehicle's strategy usually include executing the tasks locally \textcolor{color1}{and} offloading them to the $j$-th MEC server. The study of \textcolor{color1}{\cite{zhang2019task}} also considers an extra selection of offloading them  to the remote cloud. It can be indicated that the tasks with long delay constraint can be offloaded to the remote cloud.  
	\textcolor{color1}{The strategy that combines V2V and V2I communications is proven to be efficient in reducing delay and balancing load \cite{zhang2017mobile,zhang2019task}.}  The utilities of vehicles are usually formulated  as the minimization of the offloading cost (which is usually modeled as delay for task accomplishment) while satisfying the constraints such as latency  \cite{zhang2017mobile,liu2018computaion,zhang2019task,gu2019task} and energy consumption \cite{gu2019task}.  Based on the above studies, we conclude a relative comprehensive formulation for the offloading decision as (\ref{eq_offload}).
	
	\begin{sequation}
		\label{eq_offload}
		\begin{array}{l}
			\begin{aligned} 
			\textbf{	P:}&\min _{\left\{\lambda_{i}, g_{i}\right\}} \sum_{i=1}^{N} t_{i}\left(\lambda_{i}, g_{i}\right) \\&=\min _{\left\{\lambda_{i}, g_{i}\right\}} \sum_{i=1}^{N}t_{i}^{l o c} \cdot p_{\left\{\lambda_{i}=0\right\}}+t_{i, j}^{m e c} \cdot p_{\left\{\lambda_{i}=j\right\}}+t_{i}^{c} \cdot p_{\left\{\lambda_{i}=-1\right\}.}\\ \text { s.t. }  &C 1 :  t_{i}\left(\lambda_{i}, g_{i}\right) \leq t_{i}^{m a x}, \forall i \in \mathcal{N}, j \in \mathcal{M} \\ & C 2: E_i \left(\lambda_{i}, g_{i}\right)\leq E,  i \in \mathcal{N}, j \in \mathcal{M} \\ & C_n: Other \ constraints
			\end{aligned}
		\end{array}
	\end{sequation}
	Player $i$ aims to minimize the total delay for executing the task $t_{i}\left(\lambda_{i}, g_{i}\right) $, which jointly considers the overhead of local computing $t_{i}^{l o c}$, offloading to the $j$-th MEC $t_{i}^{l o c}$, and offloading to the cloud $t_{i}^{c}$.
	The vehicles $i$'s strategy profile $\{\lambda_{i} \in\{0,1,2, \ldots, M,-1\},g_{i} \in\{0,1\} \}$ includes executing the tasks locally $\lambda_{i}=0$, offloading them to the $i$-th MEC server $\lambda_{i}=j(1<j<M)$, offloading them  to the remote cloud  $\lambda_{i}=-1$, and using V2V ($g_{i} =0 $) or V2I ($g_{i} =1 $) communication modes. Besides, $C_1$ are $C_2$ are the constraints of maximum permissible delay $ t_{i}^{m a x}$ and  energy $E$. In terms of solutions, the potential function provides a good mechanism for NE. 


	\vspace{5pt}
\textit{	\textbf{b) \textbf{Joint task offloading and resource allocation game}}}
	\vspace{5pt}

			The studies in \cite{zhang2017mobile,liu2018computaion,zhang2019task,gu2019task,wang2020game} only focus on optimizing the offloading strategies without considering the resource management from the perspective of MEC servers. Joint task offloading and resource allocation games consider both \textcolor{color1}{offloading strategies for vehicles and resource allocation strategies MEC servers}.

The studies of \cite{zhang2017optimal,zhou2018begin} take advantage of the Stackelberg game to model the interaction between the two-hierarchical players, i.e., the vehicles and MEC servers. The price scheme provides an incentive to stimulate interactions between leaders and followers. Specifically, each MEC server's strategy is the resource price determination, and its utility is formulated as the total resource or service revenue. Concerning the vehicles,  they make optimal offloading decisions by dynamically reacting to the resource prices advertised by MEC servers. In \cite{zhang2017optimal} the resource competition among vehicles are modeled as a non-cooperative game as a sub-game at the follower-layer of the Stackelberg game. \textcolor{color1}{Zhou et al. \cite{zhou2018begin}} further integrate the concepts of energy efficiency, big data, QoS, and QoE into the joint task offloading and resource allocation by proposing a novel conceptual VEC architecture. However, the Stackelberg game is only used as a case study for this conceptual architecture without giving solutions. Further research is needed on the sophisticated models and the optimal solutions for this framework.

{\color{color1} The offloading approaches proposed in \cite{zhang2017optimal,zhou2018begin} mainly depend on the MEC servers.} Given the limited resource of MEC servers, Zhao et al. \cite{zhao2019computation} further put forward a complementary cloud-assisted MEC collaborative task offloading and resource allocation scheme.  Given the overall structure, this scheme is essentially a Stackelberg game where cloud-MEC servers and vehicles obtain the optimal decisions interactively and sequentially. At the \textcolor{color1}{low layer}, the potential game also facilitates the convergent NE solutions for offloading decisions. The potential game of this study is structurally similar \textcolor{color1}{to that of} \cite{zhang2019task} in terms of the players, strategies, and utility function. At the high-layer, the resource allocation is solved with a Lagrange function and a bisection search, leading to a decreased computational complexity. The Truncated Gaussian distribution used to model vehicles' dynamic patterns is more appropriate for the practical  VN environment. This cloud-assisted MEC  scheme mitigates the challenge of limited computation resources of MEC servers,  by integrating the higher hierarchical cloud server.

The security problem is also prominent in \textcolor{color1}{MEC-implemented} VNs, especially the trustworthiness of vehicles. There may exist dishonest or selfish vehicles using the MEC or cloud resources, leading to the overuse of the resources. Misbehaviors in a VEC environment \textcolor{color1}{are}  addressed in \cite{huang2017distribute}. A reputation management scheme provides a good  quantification to evaluate the trustworthiness of vehicles. The bargaining game offers an incentive mechanism to stimulate the collaboration among SPs and the vehicles. This scheme can prevent misbehaviors with low detection rates.

\begin{itemize}[itemsep=3pt,topsep=3pt]
		\item  \textbf{\textbf{Summary and conclusion of joint task offloading and resource allocation game}}
\end{itemize}

Joint task offloading and resource allocation games further supplement the offloading strategies of vehicles by including  MEC servers' resource allocation strategies. However, the problem becomes more complicated because two types players (i.e.,  vehicles and VEC servers) and larger strategy spaces are involved in the game. The Stackelberg game is one of the most commonly used model for constructing the interaction between the vehicles and MEC  servers due to its advantage of solving the heterogeneity among players of different hierarchies. Moreover, the price scheme offers an incentive to stimulate interaction between vehicles and MEC servers through resources trading. At the low level, vehicles' the optimal offloading decision making can be viewed as a sub-problem, which is similar with the studies of \cite{zhang2017mobile,liu2018computaion,zhang2019task,gu2019task}. Therefore, the model in (\ref{eq_offload}) can be used in the sub-problem formulation by considering the price resources. At the high level, the MEC servers aim to maximize its revenue given the constraint of the total computational resources. Therefore, we conclude the problems of vehicles' offloading  {\color{color1}decision making} and the MEC servers' resource allocation as the following formulations:
	\begin{sequation}
	\label{eq_jointoffload1}
	\begin{array}{l}
		\begin{aligned} 
			\bm{P_{v}:}&\min _{\left\{\lambda_{i}, g_{i}\right\}} \sum_{i=1}^{N} t_{i}\left(\lambda_{i}, g_{i}\right) -C_i\left(\lambda_{i}, g_{i}\right)=\min _{\left\{\lambda_{i}, g_{i}\right\}} \sum_{i=1}^{N}t_{i}^{l o c} \cdot p_{\left\{\lambda_{i}=0\right\}}\\&+t_{i, j}^{m e c} \cdot p_{\left\{\lambda_{i}=j\right\}}+t_{i}^{c} \cdot p_{\left\{\lambda_{i}=-1\right\}}-C_{i, j}^{m e c} \cdot p_{\left\{\lambda_{i}=j\right\}},\\
			\bm{P_{m}:}&\max _{\mathcal{F}} \sum_{i\in \mathcal{N}_{0}} C_{i, j}^{m e c} \cdot p_{\left\{\lambda_{i}=j\right\}},\\
	 		\text { s.t. }  \ & C 1 :  t_{i}\left(\lambda_{i}, g_{i}\right) \leq t_{i}^{m a x}, \forall i \in \mathcal{N}, j \in \mathcal{M}, \\ 
			& C 2: E_i \left(\lambda_{i}, g_{i}\right)\leq E,  i \in \mathcal{N}, j \in \mathcal{M},\\
			 & C3: \sum_{i=1}^{F} f_{i}^{m e c} \leq \mathcal{F}, \forall f_i^{mec} \in \mathcal{F},\\
			 & Cn: Other \ constraints,		
		\end{aligned}
	\end{array}
\end{sequation}
where $C_{i, j}^{m e c}$ is the cost of computation resource the vehicle $i$ paying for the MEC server $j$, $\mathcal{F}=\left\{f_{1}^{m e c}, f_{2}^{m e c},\ldots,f_{F}^{m e c}\right\}$ is a set of resources allocation, and $C3$ is the constraint of total computational resource of MEC servers.
Other metrics are defined in (\ref{eq_offload}).  In terms of the resource bottleneck of the MEC servers, the cloud-assisted MEC offloading and resource allocation is a promising solution to complement this limitation, especially in the computation-incentive scenarios.

\vspace{3pt}
\textit{\textbf{c) Edge content dissemination game}}
\vspace{3pt}

   The authors of \cite{hui2019edge} develop an auction game-based two-stage model for edge content dissemination in urban VNs. First, the provider uploads the content to the edge computing devices (ECDs), where the content is temporarily cached. Then, a two-stage game is designed for an ECD to select the optimal relay. In the first stage, an auction game is developed for ECDs to select relay vehicles with high transmission capability. Then the cached contents are delivered to these candidate vehicles through V2I communication. In the second stage, these candidates relay the contents to the destination vehicle. Simulation results show that this incentivized content dissemination scheme outperforms other conventional methods.
      
		\begin{table*}  
		\caption{Summary of GT for MEC-implemented VEC}
		\label{tab_MEC}
        \scriptsize
		\renewcommand*{\arraystretch}{.6}
		\begin{center}
			\begin{tabular}{|m{.018\textwidth}|m{.16\textwidth}|m{.055\textwidth}|m{.06\textwidth}|m{.13\textwidth}|m{.12\textwidth}|m{.07\textwidth}|m{.001\textwidth}|m{.001\textwidth}|m{.001\textwidth}|m{.001\textwidth}|m{.001\textwidth}|m{.001\textwidth}|m{.001\textwidth}|m{.001\textwidth}|}
				\hline
					\multirow{7}{2cm}{\textbf{Ref}}&\multirow{7}{1cm}{\textbf{Objective}}&\multicolumn{4}{c|}{\multirow{4}{1cm}{\textbf{Game}}}&\multirow{7}{1cm}{\textbf{Servers}}
					 &\multirow{7}{2pt}{\rotatebox[origin=c]{90}{\textbf{V2V}}}&\multirow{7}{2pt}{\textbf{\rotatebox[origin=c]{90}{\textbf{V2I}}}}&\multirow{7}{1cm}{\rotatebox[origin=c]{90}{\textbf{Mobility}}}&\multirow{7}{1cm}{\rotatebox[origin=c]{90}{\textbf{QoS}}}&\multirow{7}{1cm}{\rotatebox[origin=c]{90}{\textbf{Security}}}&\multirow{7}{1cm}{\rotatebox[origin=c]{90}{\textbf{Energy}}}&\multirow{7}{1cm}{\rotatebox[origin=c]{90}{\textbf{Cost}}} &\multirow{7}{1cm}{\rotatebox[origin=c]{90}{\textbf{Continuity}}} \\
					&&\multicolumn{4}{c|}{}&&&&&&&&&\\	 
					&&\multicolumn{4}{c|}{}&&&&&&&&&\\	 
					&&\multicolumn{4}{c|}{}&&&&&&&&&\\	 
			     \cline{3-6}
					&&\multirow{3}{1cm}{Type}&\multirow{3}{1cm}{Player}&\multirow{3}{1cm}{Strategy}&\multirow{3}{1cm}{Utility}&&&&&&&&&\\
					&&&&&&&&&&&&&&\\	 
					&&&&&&&&&&&&&&\\
				\hline
				 \multicolumn{15}{|c|}{\textbf{ Computation or Task Offloading  Game}}\\
				 \hline
	                \cite{zhang2017mobile}&\begin{itemize}[leftmargin=3pt]\setlength{\itemsep}{0pt}\item Reduce latency and offloading cost \item Make optimal offloading decisions \end{itemize} & Non-cooperative game&Vehicles&\begin{itemize}[leftmargin=5pt]\setlength{\itemsep}{0pt}\item Execute tasks locally  \item Offload them to MEC servers\end{itemize}& \begin{itemize}[leftmargin=5pt]\setlength{\itemsep}{0pt}\item Average offloading costs of vehicles \item Tasks processing delay \end{itemize}\item&\begin{itemize}[leftmargin=3pt]\setlength{\itemsep}{0pt}\item RSUs with MEC servers \item A cloud server\end{itemize}&$\surd$&$\surd$&$\surd$&$\surd$&&&&\\
	           \cline{2-15}
	           					\cite{liu2018computaion} &\begin{itemize}[leftmargin=3pt]\setlength{\itemsep}{0pt}\item Decrease computation overhead \item Select a proper wireless channel for each vehicle\end{itemize}&Potential Game&Vehicles&Wireless channel selection for offloading&Uplink data rate&MEC servers &&$\surd$&&$\surd$&&&&\\
	           \cline{2-15}    
                   	 \cite{zhang2019task}&\begin{itemize} [leftmargin=3pt]\setlength{\itemsep}{0pt} \item Load balancing \item Minimize the processing delay of all the computation tasks under delay constraints \item Jointly optimize the selection of MEC servers and vehicles’ task offloading decisions \end{itemize}&Non-cooperative game\newline \newline Potential game&Vehicles&\begin{itemize} [leftmargin=5pt] \setlength{\itemsep}{0pt}\item Computes locally\item Offloads to MEC server \item Offloads to cloud server\end{itemize} &Task’s execution time and data transmitting time&\begin{itemize}[leftmargin=3pt]\setlength{\itemsep}{0pt}\item RSUs with MEC servers \item A cloud server\end{itemize}&$\surd$&$\surd$&$\surd$&$\surd$&&&&\\     
     		    \cline{2-15}
             		  \cite{gu2019task} &\begin{itemize} [leftmargin=5pt] \setlength{\itemsep}{0pt}\item Minimize the task offloading delay  \item Satisfy energy consumption of vehicles and edge nodes \end{itemize}&Matching game&Task and edge nodes&Task matching &\begin{itemize}[leftmargin=3pt]\setlength{\itemsep}{0pt}\item Task: offloading delay\item Edge nodes: energy consumption \end{itemize}&\begin{itemize}[leftmargin=5pt]\item RSUs with MEC servers\item Vehicular edge nodes\end {itemize}&$\surd$&$\surd$&$\surd$&$\surd$&&$\surd$&$\surd$&\\
                 \hline  	
                 	 \multicolumn{15}{|c|}{\textbf{ Joint Task Offloading and Resource Allocation Game}}\\
                 \hline 
                        \cite{huang2017distribute} &\begin{itemize} [leftmargin=3pt] \setlength{\itemsep}{0pt}\item Provide security protection using reputation management \item Optimize resource allocation for SPs \item Optimize computation offloading for vehicles  \end{itemize}&Bargaining game&SP with MEC servers and vehicles &\begin{itemize}[leftmargin=3pt]\setlength{\itemsep}{0pt}\item SP: Resource budgets \item Vehicles: Amount of requesting computation resources \end{itemize}&\begin{itemize}[leftmargin=3pt]\setlength{\itemsep}{0pt}\item SP: Sum of resource budgets \item Vehicles: total processing delay of computation and cost for service\end{itemize}&\begin{itemize}[leftmargin=3pt]\setlength{\itemsep}{0pt}\item RSUs with MEC servers \item BSs with MEC servers \end{itemize}&$\surd$&$\surd$&&$\surd$&$\surd$&&$\surd$&\\
                \cline{2-15}
	          	 	 \cite{zhao2019computation}& \begin{itemize}[leftmargin=3pt]\setlength{\itemsep}{0pt}\item Optimize computation offloading for vehicles \item Optimize resource allocation of MECs and cloud\item Decrease the system complexity without loss of the performance\end{itemize}&Potential game \newline Stackelberg game &Vehicles&\begin{itemize}[leftmargin=3pt]\setlength{\itemsep}{0pt}\item Computes locally\item Offloads to MEC server \item Offloads to cloud server\end{itemize}&\begin{itemize}[leftmargin=3pt]\setlength{\itemsep}{0pt}\item Task processing delay \item Cost of computation resource \item Normalization factor \end{itemize}&\begin{itemize}[leftmargin=3pt]\setlength{\itemsep}{0pt}\item A MEC server \item A cloud server\end{itemize}&&$\surd$&$\surd$&$\surd$&&&&\\
	            \cline{2-15}
			         \cite{zhang2017optimal} &  \begin{itemize}[leftmargin=3pt]\setlength{\itemsep}{0pt}\item Optimal resource allocation \begin{itemize} [leftmargin=10pt]\setlength{\itemsep}{0pt}\item Backup resources sharing among MEC servers \item MEC servers dynamically  assign their resources to vehicles \end{itemize} \item Satisfy the delay constraints of tasks \end{itemize}&Stackelberg game&MEC servers and vehicles& \begin{itemize}[leftmargin=3pt] \item\setlength{\itemsep}{0pt} Vehicles: Choose the offloading target MEC servers \item MEC servers: \begin{itemize} [leftmargin=3pt] \item the amount of the resources bought from the BCS \item price of their resources sold to vehicles \end{itemize} \end{itemize}&\begin{itemize}[leftmargin=3pt] \setlength{\itemsep}{0pt}\item Vehicles: total time cost task offloading \item MEC servers: revenue of the resources\end{itemize}& \begin{itemize}[leftmargin=3pt]\setlength{\itemsep}{0pt}\item RSUs with MEC server \item A cloud server\end{itemize}&&$\surd$& &$\surd$&&&$\surd$&\\
			 \cline{2-15}
					 \cite{zhou2018begin}&\begin{itemize}[leftmargin=3pt]\setlength{\itemsep}{0pt}\item Dynamic resources allocation to guarantee: \begin{itemize}[leftmargin=8pt]\setlength{\itemsep}{0pt}\item Energy efficiency \item QoS guarantee \item QoE guarantee \end{itemize} \item Optimal computation offloading \end{itemize}&Stackelberg game&SPs and vehicular users&\begin{itemize}[leftmargin=3pt]\setlength{\itemsep}{0pt}\item SP: Service provision revenue and electricity consumption \item Vehicles: proportion  of resources to offload\end{itemize}&\begin{itemize}[leftmargin=3pt]\setlength{\itemsep}{0pt}\item SP: service provision revenue \item Vehicles: edge computing time and electricity cost \end{itemize}&\begin{itemize}[leftmargin=3pt]\setlength{\itemsep}{0pt}\item RSUs with edge controllers \item A cloud server \end{itemize}&$\surd$&$\surd$&$\surd$&$\surd$&&$\surd$&$\surd$&\\
			 \hline
			 \multicolumn{15}{|c|}{\textbf{Edge Content Dissemination Game}}\\
		 	\hline
					\cite{hui2019edge}&Select the relay vehicles to satisfy different transmission requirements&Auction game& ECD and vehicles&\begin{itemize} [leftmargin=3pt]\setlength{\itemsep}{0pt}\item ECD: Candidate relay vehicles selection \item Relay vehicles: Optimal bid selection \end{itemize}&\begin{itemize} [leftmargin=3pt]\setlength{\itemsep}{0pt}\item ECD: Transmission capabilities of the\item  Vehicles: Bid and relay cost
				\end{itemize}&RSUs with ECDs&$\surd$&$\surd$&$\surd$&$\surd$&&&$\surd$&\\
		  \hline		
			\end{tabular}
		\end{center}
	\end{table*}

	\vspace{6pt}
\subsubsection{FC-implemented VEC}
	Similar to MEC, FC enables edge computation by implementing a decentralized computing infrastructure based on heterogeneous and collaborative fog nodes (e.g., access points, routers, and IoT gateways) placed at any point in the architecture between the end-user devices and the near-user edge devices  \cite{dolui2017comparison}. As vehicular fog is also a resource-intensive component with limited resources, GT can be used for proper and fair computation and resource allocation for FC-implemented VEC in recent research. Table \ref{tab_FC} summarizes the  GT-based schemes of computation or task offloading and resources allocation in FC-implemented VEC.

Klaimi et al. \cite{klaimi2018theoretical} introduced a new concept of vehicular fog computing (VFC) for resource allocation by deploying the vehicle resources at the network edge. The system model consists of a set of parked electrical vehicles as fog resources and the demands of local mobile applications as the input data. The application demands are classified into high priority (HP) and low priority (LP) according to the application type and the maximum waiting time. The VFC resource allocation is formulated as a potential game where vehicles in the fog are players. The \textcolor{color1}{strategy set} is defined as a quadruple $\left(\pi_{0}, \pi_{1}, \pi_{2}, \pi_{3}\right)$, \textcolor{color1}{where $\pi_{0}$, $\pi_{1}$, $\pi_{2}$, and $\pi_{3}$ denote not satisfying any demands, satisfying only HP demands, sharing the computation (50\%) between HP and LP demands, and  satisfying only LP demands, respectively.} The utility function is formulated to minimize the CPU and energy consumption while maximizing QoS for application demands. This study proposed a novel VFC concept to satisfy the demands of different applications using the underutilized resources of parked vehicles. {\color{color1} The main advantage of deploying the VFC is that the resource-hungry mobile applications can be executed locally within low delay by decreasing the number of demands redirected to the remote cloud. } The advantage of potential function is also used for the NE solution. However, the fogs may lack the reward mechanism to attract new vehicles to join because there are vehicles leaving at any time. Besides, modeling the players' strategy as a discrete quadruple is insufficient. A better assumption could be sharing {\color{color1} $x\% \ (x\in[0-100])$} computation capacity to LP demands.

A more recent work \cite{sutagundar2019resource} uses the matching game for resources allocation in FC-based VNs from a perspective of assigning {\color{color1} tasks} to resources. The system model consists of three levels {\color{color1} which are} a vehicular cluster of cooperative vehicles, a set of fog nodes, and a central cloud. Each fog has a set of virtual machines (VMs) resources acting as players in the game. In each fog, a fog broker is assumed to make the optimal task assignments for VMs once it receives the \textcolor{color1}{vehicles' requests.} The employment of VM provides an intuition for service continuity guarantee because the VM can migrate to another fog if the vehicle is connected to another RSU. However, this study only provides the basic idea without designing the detailed game model (i.e., strategy, utilities, and NE solution) nor the VM migration approach.


One common weakness of the above studies on FC-implemented VEC is the lack of incentive mechanisms to encourage vehicles to serve as fogs and share resources  \cite {klaimi2018theoretical,sutagundar2019resource}. To mitigate this limitation, Xu et al. \cite{xu2018low} design a low-latency and massive-connectivity FC framework. Although the basic idea of this work is similar to  that of \cite{klaimi2018theoretical}, it designs a more detailed VFC framework and is more suitable for 5G communication. This study assumes a more comprehensive network with a set of vehicles, a set of UEs, and a set of tasks with different sizes, required resources, and delay constraints. The communication parameters such as channel fading and \textcolor{color1}{vehicular} mobility patterns are also considered in the delay formulation. A significant advantage of this study is using the matching game to offload UEs' tasks to the fog vehicles that have ideal resources. The matching game provides a self-organizing mechanism to deal with the heterogeneous preferences between tasks and resources. Another advantage is  that the price incentive mechanism is employed to construct stable matching between UEs and vehicles. This approach is further extended to a computation resource allocation and task assignment scheme \cite{zhou2019computation} by incorporating contract scheme with price and matching approaches. The contract mechanism provides a powerful incentive to stimulate vehicles in the fog to share resources for task offloading.


Preservation of location privacy in FC-implemented VEC is studied in \cite{kang2017privacy}. The authors present \textit{ pseudonym fogs}, each of which consists of a set of roadside infrastructures, and is deployed on nearby vehicles. These pseudonym fogs interactively play a real-time pseudonym change game to allocate pseudonym resources to nearby vehicles. This scheme improves the vehicles' location privacy with reduced pseudonym management overheads. One disadvantage of this scheme is that it is not applied to situations with sparse vehicles due to the difficulty of forming the pseudonym fogs.

	\begin{itemize}[itemsep=3 pt,topsep = 3 pt]
		\item  \textbf{\textbf{Summary and conclusion of GT for FC-implemented VEC}}
	\end{itemize}

	In VFC-based VNs, most studies focus on the problem of computation resource allocation and task assignment\cite{klaimi2018theoretical,sutagundar2019resource,xu2018low,zhou2019computation}.	The matching game shows its potential to provide self-organizing solutions for the heterogeneous preferences among players (e.g., UEs' tasks and fog's resources). Concerning the incentive mechanism, both contract and price schemes can stimulate interaction.  Price schemes can stimulate or enhance the interaction between servers and vehicles through resource trading. By comparison, the contract mechanism offers more powerful incentives. It not only can force interaction but also can prevent selfishness. Therefore, it can be indicated that the combination of price and contract \cite{zhou2019computation} could further optimize the incentive.

\begin{table*} 
	\scriptsize   
	\caption{Summary of GT for FC-implemented VEC}
	\label{tab_FC}
	\renewcommand*{\arraystretch}{.6}
	\begin{center}
		\begin{tabular}{|m{.018\textwidth}|m{.14\textwidth}|m{.05\textwidth}|m{.05\textwidth}|m{.14\textwidth}|m{.1\textwidth}|m{.09\textwidth}|m{.001\textwidth}|m{	.001\textwidth}|m{.001\textwidth}|m{.001\textwidth}|m{.001\textwidth}|m{.001\textwidth}|m{.001\textwidth}|m{.005\textwidth}|m{.001\textwidth}|}
			\hline
				\multirow{7}{2cm}{\textbf{Ref}}&\multirow{7}{1cm}{\textbf{Objective}}&\multicolumn{4}{c|}{\multirow{4}{1cm}{\textbf{Game}}}&\multirow{7}{1cm}{\textbf{Servers}}
			&\multirow{7}{2pt}{\rotatebox[origin=c]{90}{\textbf{V2V}}}&\multirow{7}{2pt}{\textbf{\rotatebox[origin=c]{90}{\textbf{V2I}}}}&\multirow{7}{2pt}{\textbf{\rotatebox[origin=c]{90}{\textbf{V2P}}}}&\multirow{7}{1cm}{\rotatebox[origin=c]{90}{\textbf{Mobility}}}&\multirow{7}{1cm}{\rotatebox[origin=c]{90}{\textbf{QoS}}}&\multirow{7}{1cm}{\rotatebox[origin=c]{90}{\textbf{Security}}}&\multirow{7}{1cm}{\rotatebox[origin=c]{90}{\textbf{Energy}}}&\multirow{7}{1cm}{\rotatebox[origin=c]{90}{\textbf{Cost}}} &\multirow{7}{1cm}{\rotatebox[origin=c]{90}{\textbf{Continuity}}} \\
			&&\multicolumn{4}{c|}{}&&&&&&&&&&\\	 
			&&\multicolumn{4}{c|}{}&&&&&&&&&&\\	 
			&&\multicolumn{4}{c|}{}&&&&&&&&&&\\	 
			\cline{3-6}
			&&\multirow{3}{1cm}{Type}&\multirow{3}{1cm}{Player}&\multirow{3}{1cm}{Strategy}&\multirow{3}{1cm}{Utility}&&&&&&&&&&\\
			&&&&&&&&&&&&&&&\\	 
			&&&&&&&&&&&&&&&\\		
			 \hline
			 		\cite{sutagundar2019resource}&\begin{itemize} [leftmargin=3pt]\setlength{\itemsep}{0pt}\item Prediction of resources required and the
					availability of resources \item Fair and efficient fog resources allocation for vehicles  \end{itemize}&Non-cooperative game&VMs&The amount of resources allocated&Estimation of amount of required resources&\begin{itemize} [leftmargin=3pt]\setlength{\itemsep}{0pt}\item RSUs with VM-based fog servers \item A cloud server\end{itemize}&&$\surd$&&$\surd$&$\surd$&&&$\surd$&\\
			\hline
					 \cite{xu2018low}&\begin{itemize} [leftmargin=3pt]\setlength{\itemsep}{0pt}\item Relive overload on BSs \item Reduce total network delay \end{itemize}&Matching game& Fog vehicles and general vehicles&Pair matching between fog vehicles and general vehicles&\begin{itemize} [leftmargin=3pt]\setlength{\itemsep}{0pt}\item Total delay \item Price for using the resources of fog vehicles \end{itemize}& Fog vehicles& &&$\surd$&$\surd$&$\surd$&&&&\\		
			\hline
					\cite{zhou2019computation}&\begin{itemize} [leftmargin=3pt]\setlength{\itemsep}{0pt}\item Minimize total delay \item Motivate vehicles to share resources using contract incentive \item Optimal task assignment \end{itemize}&Matching game& Fog vehicles and general vehicles&Pricing-based Pair matching between fog vehicles and general vehicles&Expected social welfare&Fog vehicles&&&$\surd$&$\surd$&$\surd$&&&$\surd$&\\
		   \hline
		   			 \cite{klaimi2018theoretical} &\begin{itemize} [leftmargin=3pt]\setlength{\itemsep}{0pt}\item Minimize the latency and optimize the  utilization of computation and energy \item Allocate fog resources dynamically  \end{itemize}&Potential game&Fog vehicles&\begin{itemize} [leftmargin=3pt]\setlength{\itemsep}{0pt}\item Do not satisfy any type of demands \item Satisfy only HP demands \item Share the computation \item Satisfy only LP demands  \end{itemize}&\begin{itemize} [leftmargin=3pt]\setlength{\itemsep}{0pt}\item CPU and energy utilization \item Task processing time \end{itemize}&Parked fog vehicles&&&$\surd$&&$\surd$&&$\surd$&$\surd$&\\
		   	\hline
		   			\cite{kang2017privacy} &Secure pseudonym changing management with low overhead&Non-cooperative game&Vehicles&Change pseudonym or maintain current pseudonym&\begin{itemize} [leftmargin=3pt]\setlength{\itemsep}{0pt}\item Vehicle-side entropy \item Pseudonym change cost\end{itemize}& \begin{itemize} [leftmargin=3pt]\setlength{\itemsep}{0pt}\item Pseudonym fogs with roadside infrastructures\item A cloud server\end{itemize}&$\surd$&$\surd$&&$\surd$&$\surd$&$\surd$&&$\surd$&\\
             \hline
		\end{tabular}
	\end{center}
\end{table*}

\vspace{6pt}
\subsubsection{ Cloudlet-implemented VEC} 
As shown in Fig. \ref{fig_vec}, a cloudlet in VNs is a trusted cluster of roadside infrastructures (e.g., RSU and BS) with available resources \cite{satyanarayanan2009case}. GT provides the ability to help the cloudlets share the available resources to the nearby vehicles by stimulating the cooperation among infrastructures in the same cloudlet, or stimulating the cooperation among different cloudlets. Table \ref{tab_Cloudlet} summarizes the recent studies employing GT in cloudlet-implemented VNs.

The authors of \cite{yu2015cooperative} propose a coalition game model to stimulate cloud SPs to cooperatively form coalitions for sharing their idle resources. First, each SP evaluates its revenue and decides whether to join a coalition in the cloud. Then, in each coalition, the two types of players act as 1) the $inviters$ who would like to rent resources $R_{from}$ and 2) the $invitees$ who would like to lease out their resources $R_{to}$. Using a two-sided matching game, the $inviters$ and $invitees$ match their demands effectively and form stable coalitions cooperatively. The coalition is proved to be stable, where the PO is the only stable solution. 

To demonstrate the resource allocation of cloudlets, Yu et al. \cite{yu2016optimal} define a new paradigm in 5G-enabled VNs, an enhanced cloud radio access network, which integrates geo-distributed cloudlets with SDN and D2D technologies. The authors exploited the matrix GT for cloudlet resource allocation and gave explicit solutions for global optimization. An RSU cloud resource allocation scheme based on a non-cooperative game is designed in \cite{tao2017resource}. To deal with the low efficiency of convergence at NE and to achieve a near Pareto optimal flow allocation, the previous one-shot game is extended to a repeated game where punishment for misbehaviors is considered to avoid selfish vehicles that do not cooperate.

Different from the studies mentioned above, vehicles are viewed as a dynamic extension of cloudlets, and share computation resources to cloudlets in \cite{lin2019vehicle}. The authors design a Stackelberg game to stimulate vehicles to trade in their computation resources with a cloudlet SP. After obtaining the vehicles' states from the cloud server, the cloudlet SP first sets the optimal discriminatory price for the trade based on the computation demands of the vehicles. Then, each vehicle decides the computation amount it will buy from the cloudlet SP according to the given price. The existence and uniqueness of NE are proved, and the solution has good scalability for more participating vehicles.

\begin{table*} 
	\scriptsize    
	\caption{Summary of GT for Cloudlet-implemented VEC}
	\label{tab_Cloudlet}
	\renewcommand*{\arraystretch}{.4}
	\begin{center}
		\begin{tabular}{|m{.02\textwidth}|m{.15\textwidth}|m{.05\textwidth}|m{.08\textwidth}|m{.12\textwidth}|m{.15\textwidth}|m{.04\textwidth}|m{.001\textwidth}|m{	.001\textwidth}|m{.001\textwidth}|m{.001\textwidth}|m{.001\textwidth}|m{.001\textwidth}|m{.001\textwidth}|}
			\hline
				\multirow{10}{2cm}{\textbf{Ref}}&\multirow{10}{1cm}{\textbf{Objective}}&\multicolumn{4}{c|}{\multirow{4}{1cm}{\textbf{Game}}}&\multirow{10}{1cm}{\textbf{Servers}}
			&\multirow{10}{2pt}{\rotatebox[origin=c]{90}{\textbf{V2V}}}&\multirow{10}{2pt}{\textbf{\rotatebox[origin=c]{90}{\textbf{V2I}}}}&\multirow{10}{1cm}{\rotatebox[origin=c]{90}{\textbf{Mobility}}}&\multirow{10}{1cm}{\rotatebox[origin=c]{90}{\textbf{QoS}}}&\multirow{10}{1cm}{\rotatebox[origin=c]{90}{\textbf{Energy}}}&\multirow{10}{1cm}{\rotatebox[origin=c]{90}{\textbf{Cost}}} &\multirow{10}{1cm}{\rotatebox[origin=c]{90}{\textbf{Continuity}}} \\
			&&\multicolumn{4}{c|}{}&&&&&&&&\\	 
			&&\multicolumn{4}{c|}{}&&&&&&&&\\	 
			&&\multicolumn{4}{c|}{}&&&&&&&&\\	 
			\cline{3-6}
			&&\multirow{6}{1cm}{Type}&\multirow{6}{1cm}{Player}&\multirow{6}{1cm}{Strategy}&\multirow{6}{1cm}{Utility}&&&&&&&&\\
			&&&&&&&&&&&&&\\	 
			&&&&&&&&&&&&&\\
				&&&&&&&&&&&&&\\
					&&&&&&&&&&&&&\\
						&&&&&&&&&&&&&\\
			\hline
				\cite{yu2016optimal}&\begin{itemize} [leftmargin=3pt]\setlength{\itemsep}{0pt}\item Resource sharing  among the geo-distributed cloudlet \item Improve resource utilization and reduce power consumption \end{itemize}&Non-cooperative game&RSU cloudlets &The amount of resources allocated&Resources  Utilization \begin{itemize} [leftmargin=3pt]\setlength{\itemsep}{0pt}\item CPU resource \item memory resource \item bandwidth resource \end{itemize}& RSU cloudlets &&$\surd$&&$\surd$&$\surd$&$\surd$&\\
			\hline
				 \cite{tao2017resource} &Effective flow rate assignment&Non-cooperative game&Vehicles&Flow rate allocated to each vehicle&Transmission efficiency&RSU cloudlets &&$\surd$&&$\surd$&&$\surd$&\\
		   \hline
		       \cite{yu2015cooperative}&Stimulate cloud SPs to form coalitions for resources sharing& Coalitional game & $inviters$ and $invitees$ &\begin{itemize} [leftmargin=3pt]\setlength{\itemsep}{0pt}\item $inviters$: rent resources \item $invitees$: lease out resources \end{itemize}& \begin{itemize} [leftmargin=3pt]\setlength{\itemsep}{0pt}\item $inviters$: \begin{itemize} [leftmargin=3pt]\setlength{\itemsep}{0pt}\item satisfaction function \item rent payment\end{itemize} \item $invitees$: \begin{itemize} [leftmargin=3pt,itemsep=0pt]\setlength{\itemsep}{0pt}\item satisfaction function\item rental income\end{itemize} \end{itemize}&SP cloudlets&$\surd$&$\surd$&&$\surd$&&$\surd$&\\
		     \hline
			     \cite{lin2019vehicle}&Stimulate vehicles to share computation resources to cloudlets&Stackelberg game&A cloudlet SP and individual vehicles&\begin{itemize} [leftmargin=3pt]\setlength{\itemsep}{0pt}\item Cloudlet SP: discriminatory price\item Vehicles: computation trading amount\end{itemize}&\begin{itemize} [leftmargin=3pt]\setlength{\itemsep}{0pt}\item Cloudlet SP: cost of renting \item Vehicles: benefit of the trading\end{itemize}&SP coudlets&$\surd$&&$\surd$&$\surd$&&$\surd$&$\surd$\\
			  \hline
	\end{tabular}
\end{center}
\end{table*}
\vspace{6pt}
\subsubsection{VC-implemented VEC}

Parked vehicles with idle resources can be viewed as an extension of RSUs to improve the efficiency of resource utilization. For example, the resources of parked vehicles are idle, while the vehicles in congested areas compete for insufficient resources. In this context, VC is emerging as a promising solution for the exploitation of underutilized resources by integrating VNs with CC. However, resource management in VC is more challenging than traditional CC because high mobility of vehicles may lead to ``cloud resource mobility". Moreover, communication in a VC is more vulnerable to attacks than in traditional VNs because a large number of users share the resources. GT has been employed by recent studies in the  decision making of resource management or security protection for VC-implemented VEC. Table \ref{tab_VC} summarizes the works that apply GT in VC-implemented VEC.

Mekki et al. \cite{mekki2017proactive} explore a hybrid wireless network access scheme in VC networks based on an evolutionary game. Vehicles either cooperate to form a temporary cloud to share resources or act alone to access the conventional cloud. Each vehicle can change its strategies from the conventional cloud to the VC or inversely according to its utility. To model the evolution of vehicles' strategies, the authors developed an evolutionary game that allows vehicles to determine their access to conventional or VCs.

The authors of \cite{liwang2019game} develop an opportunistic V2V computation offloading scheme based on a two-player Stackelberg game. In this game, each vehicle acts either as an SP with idle computation resources or a requester who has a computation-intensive task that can be carried out locally or offloaded to nearby providers. The utility functions involve the following metrics: vehicular mobility, V2V communication duration, computational capabilities, channel conditions, and service costs with both complete and incomplete information. This method can determine the appropriate offloading rate, optimal SP selection, and ideal pricing strategies of SPs.

To handle service management in VCs, Aloqaily et al. \cite{aloqaily2017fairness} propose a cooperative distributed game between vehicles and SPs. This game aims at maximizing the SP's social utility while ensuring the vehicle's QoE, considering service allocation efficiency and fairness among SPs. Considering the different preferences of SPs and consumers, Brik et al. \cite{brik2018gss} propose a distributed game-based approach to model the interaction between consumer vehicles and provider vehicles for effective service allocation. In this game, SPs rent out their resources, and each consumer vehicle selects the optimal SP vehicle who can provide him services with satisfactory QoS and cost.

\begin{itemize}[itemsep=3 pt,topsep = 3 pt]
	\item  \textbf{\textbf{Summary and conclusion of GT in VEC-enabled VNs}}
\end{itemize}

GT applications in MEC-implemented, FC-implemented, cloudlet-implemented, and VC-implemented VEC are discussed in this subsection. We conclude the major findings, pros and cons, and the possible indicating solutions.

\begin{itemize}[label=$-$,itemsep=3pt,topsep=3pt]
	\item   Stackelberg game and matching game are promising models in the VEC-assisted VNs, both of which have the advantages for dealing with the heterogeneity in VNs. However, they are suitable for different scenarios. The most salient feature of the Stackelberg game is the ``first move" advantage of the leader, which makes it appropriate for the decision making in the multi-hierarchical networks. It is suitable for jointly offloading and resource allocation situation where an MEC or cloud server announces its strategy first. The vehicles sequentially react to the leader's action with their offloading decisions. However, the resources or tasks are assumed to be homogeneous without preferences in the Stackelberg game regarding the resource or task allocation. In comparison, the matching game is applicable for dealing with the heterogeneous preferences among players in a self-organizing fashion. For example, it can be used for task assignment between tasks of UEs and idle resources of vehicles\cite {xu2018low,zhou2019computation}. Each UE has a preference list of all vehicles, and each vehicle has a preference list of all UEs. The mutual beneficial task-resource (or UE-vehicle) pairs can be found using the matching game.
	
	\item Regarding the incentive schemes, prices are most used for offloading or resource/task allocation to stimulate interaction between MEC servers and vehicles. The contract mechanism can powerfully force interaction and prevent selfishness. The combination of contract and price mechanisms could perform better than using one alone \cite{zhou2019computation}.
	
	\item A novel and promising way to relieve the limitation of resources is to cluster the edge nodes with idle resources into vehicle-fogs \cite{klaimi2018theoretical}, RSU-fogs  \cite{tao2017resource}, or VCs \cite{mekki2017proactive,aloqaily2017fairness,brik2018gss,liwang2019game}. However, very little research studies the cooperation among players in the fogs, cloudlets, or VCs except \cite{yu2015cooperative}. The stability of the edge clusters should be considered because vehicles may frequently join or leave fogs, cloudlets, or VCs. Given these concerns, the coalitional game could be a possible tool to construct cooperative clusters dynamically.
	
	\item Another problem that should be addressed is the security of VEC-assisted VNs, which has been concerned by little research. The security of communication could be more vulnerable in VEC networks due to resource sharing. For example, dishonest vehicles could offload the resources excessively and freely. The privacy of vehicles in the vehicular fog or VCs could be tracked, especially location privacy. The reputation or credit schemes could be possible solutions to the trustworthiness quantification and detection. The location privacy-preserving schemes in Section \ref{sec_GTprivacy} could be further adopted in the resource allocation approaches.
	
\end{itemize}

\begin{table*} 
	\scriptsize    
	\caption{Summary of games for VC-implemented VEC}
	\label{tab_VC}
	\renewcommand*{\arraystretch}{.5}
	\begin{center}
		\begin{tabular}{|m{.019\textwidth}|m{.15\textwidth}|m{.063\textwidth}|m{.08\textwidth}|m{.13\textwidth}|m{.15\textwidth}|m{.04\textwidth}|m{.0001\textwidth}|m{.0001\textwidth}|m{.0001\textwidth}|m{.0001\textwidth}|m{.0001\textwidth}|m{.0001\textwidth}|m{.0001\textwidth}|}
			\hline
				\multirow{10}{2cm}{\textbf{Ref}}&\multirow{10}{1cm}{\textbf{Objective}}&\multicolumn{4}{c|}{\multirow{4}{1cm}{\textbf{Game}}}&\multirow{10}{1cm}{\textbf{Servers}}
			&\multirow{10}{2pt}{\rotatebox[origin=t]{90}{\textbf{V2V}}}&\multirow{10}{2pt}{\textbf{\rotatebox[origin=t]{90}{\textbf{V2I}}}}&\multirow{10}{1cm}{\rotatebox[origin=t]{90}{\textbf{Mobility}}}&\multirow{10}{1cm}{\rotatebox[origin=t]{90}{\textbf{QoS}}}&\multirow{10}{1cm}{\rotatebox[origin=t]{90}{\textbf{Energy}}}&\multirow{10}{1cm}{\rotatebox[origin=t]{90}{\textbf{Cost}}} &\multirow{9}{3cm}{\rotatebox[origin=t]{90}{\textbf{Continuity}}} \\
			&&\multicolumn{4}{c|}{}&&&&&&&&\\	 
			&&\multicolumn{4}{c|}{}&&&&&&&&\\	 
			&&\multicolumn{4}{c|}{}&&&&&&&&\\	 
			\cline{3-6}
			&&\multirow{6}{1cm}{Type}&\multirow{6}{1cm}{Player}&\multirow{6}{1cm}{Strategy}&\multirow{6}{1cm}{Utility}&&&&&&&&\\
			&&&&&&&&&&&&&\\	 
			&&&&&&&&&&&&&\\
				&&&&&&&&&&&&&\\	
					&&&&&&&&&&&&&\\	 
			\hline
				 \cite{mekki2017proactive} &Optimal wireless network access &Evolutionary game&Conventional vehicle and VC member&\begin{itemize} [leftmargin=3pt]\setlength{\itemsep}{0pt}\item Access to conventional cloud \item Cooperate to be member in VC\end{itemize}&\begin{itemize} [leftmargin=3pt]\setlength{\itemsep}{0pt}\item Conventional Vehicle: \begin{itemize} [leftmargin=3pt]\setlength{\itemsep}{0pt}\item Amount of data received \item Payment for the service\end{itemize} \item VC member: \begin{itemize} [leftmargin=3pt]\setlength{\itemsep}{0pt}\item Data downloaded from other VC members \item Handover cost\end{itemize}\end{itemize}& \begin{itemize} [leftmargin=3pt]\setlength{\itemsep}{0pt}\item VC with eNB \item A cloud server \end{itemize}&$\surd$&$\surd$&&$\surd$&&$\surd$&$\surd$\\
			\hline
				 \cite{liwang2019game} &\begin{itemize} [leftmargin=3pt]\setlength{\itemsep}{0pt}\item Determine appropriate offloading rate of requestors \item Select the appropriate computation SP\item Identify the ideal pricing strategy for SP\end{itemize} &Stackelberg game&A task vehicle and one of the server vehicles&\begin{itemize} [leftmargin=3pt]\setlength{\itemsep}{0pt}\item Task vehicle: the amount of offloaded data\item Service vehicle: service price\end{itemize}& \begin{itemize} [leftmargin=3pt]\setlength{\itemsep}{0pt}\item Task vehicle:  \begin{itemize} [leftmargin=3pt]\setlength{\itemsep}{0pt}\item task processing delay \item payment for service \end{itemize} \item Service vehicle: service price\end{itemize}&VC with RSU&$\surd$&&$\surd$&$\surd$&&$\surd$&\\
			\hline
				\cite{aloqaily2017fairness} &\begin{itemize} [leftmargin=3pt]\setlength{\itemsep}{0pt}\item Fair resources allocation \item QoS and QoE guarantee \end{itemize}&Cooperative game&Vehicular driver and SP&Cooperate or reject to cooperate&Social welfare function&VC&$\surd$&&&$\surd$&&$\surd$&\\
			\hline
				\cite{brik2018gss}&Select the optimal provider vehicles to meet the QoS of consumer vehicles &Non-cooperative game&Consumer vehicle and provider vehicle&\begin{itemize} [leftmargin=3pt]\setlength{\itemsep}{0pt}\item Consumer vehicle: consume or not \item  Provider vehicle: offer or not  \end{itemize}&\begin{itemize} [leftmargin=3pt]\setlength{\itemsep}{0pt}\item Data throughput \item Successful execution ratio \item Execution duration \item Execution price\end{itemize}&VC&$\surd$&&$\surd$&$\surd$&$\surd$&&\\
			\hline
		\end{tabular}
	\end{center}
\end{table*}

\subsection{Game Theory in UAV-assisted VNs}
\label{sec_UAV}
\begin{figure}[!hbt] 
	\centering
	\includegraphics[width =3.5in]{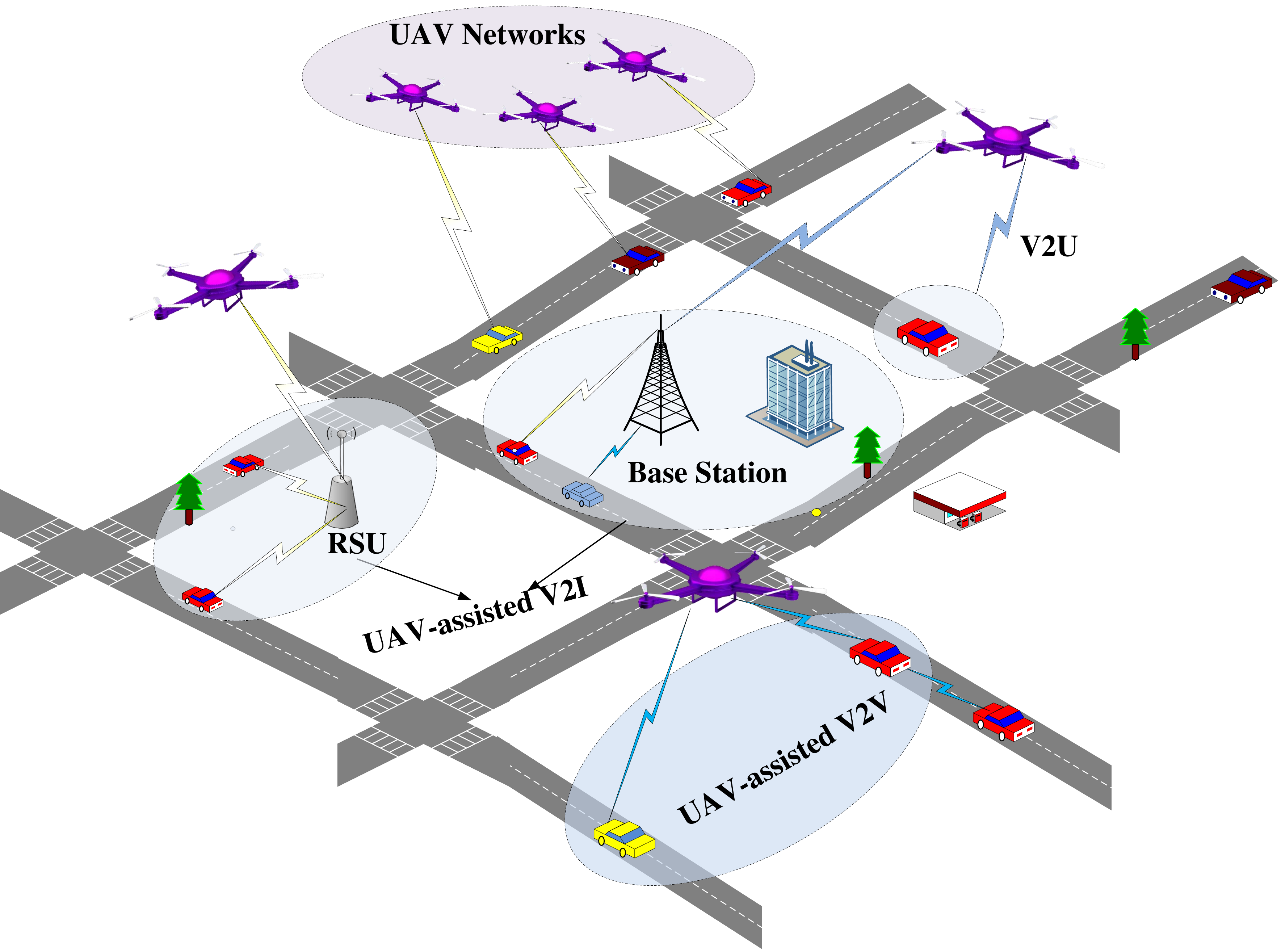}
	\caption{Architecture of UAV-assisted VNs.}
	\label{fig_uav}
\end{figure}

As an important and emerging element of IoT, UAV-assisted VNs can improve infrastructure coverage, enhance network connectivity in resource-limited scenarios, and reduce deployment cost. As shown in Fig. \ref{fig_uav}, UAV enables three types of communication: UAV-assisted V2V communication, UAV-assisted V2I communication, and UAV networks \cite{shi2018drone}. UAV-assisted VNs becomes more complex due to the dynamic of both UAVs and VNs, the limited energy of UAVs, larger scaled networks, and more heterogeneous network structure. GT provides mathematical framework to deal with the complicated interaction or competition among multiple nodes in UAV-assisted VNs.

The security problems such as the potential UAV attackers \cite{xiao2018uav} and the attacks on UAVs \cite{sedjelmaci2016intrusion} are studied in UAV-assisted VNs. Xiao et al. \cite{xiao2018uav} propose an anti-jamming relay game between UAVs and jammer for jamming resistance in UAV-assisted VNs. In this game, the UAV determines whether to relay the OBU's messages to an RSU beyond the jammer, and the smart jammer attempts to attack by adjusting its jamming power. Moreover, NE is derived for UAVs to select the best relay strategy to resist jamming according to the channel condition and transmission cost. The security problems of intrusion detection and attacker ejection in UAV-assisted VNs are discussed in \cite{sedjelmaci2016intrusion}. A Bayesian game is adopted to formulate the attack–defense interactions between IDS and attackers for detecting attacks with high accuracy and low overheads. 

Computation offloading problems are addressed in \cite{alioua2018efficient,zhao2020novel}. Alioua et al. \cite{alioua2018efficient} formulate the computation offloading as a two-player sequential game to balance the computation offloading delay and energy waste in emergency situations for UAV-assisted VNs with less or no infrastructure.  However, this scheme may not operate at full capacity in the infrastructure-existence scenario, where the infrastructures with computation capacity could also assist the task offloading.  Zhao et al. \cite{zhao2020novel} further design a novel multi-player offloading structure that incorporates the MEC capability into the UAV-assisted VNs. Each vehicle player intends to offload the tasks to the UAV or MEC server based on the quality of the communication link. This approach enables vehicles to execute complex communication tasks real-timely with the minimum overheads on execution time and energy.  

The mode selection scheme proposed in \cite{wang2018mode} is based on an evolutionary game. Each vehicle in the game selects the best communication mode from V2I, V2V, and V2U in UAV-assisted VNs to maximize transmission reliability while minimizing resource utilization cost.

	\begin{itemize}[itemsep=3 pt,topsep = 3 pt]
		\item \textbf{Summary and conclusion of GT for UAV-assisted VNs}
	\end{itemize}

The application of GT in UAV-assisted VNs mainly includes the aspects of security \cite{xiao2018uav,sedjelmaci2016intrusion}, computation offloading \cite{alioua2018efficient,zhao2020novel}, and network selection \cite{wang2018mode}. As discussed in Section \ref{sec_security}, the Bayesian game tends to be more appropriate for solving the security problems in UAV-assisted VNs due to is ability of capturing the incomplete information among players. Regarding the task offloading problems, existing methods employ sequential games to formulate hierarchical decision making \cite{alioua2018efficient,zhao2020novel}, which could be less efficient due to the lack of interaction. The Stackelberg game could be adopted in future studies due to its advantage of modeling interaction among the hierarchical players.

\subsection{Game Theory in SDN-based VNs}
\label{sec_GTSDN}
By decoupling network management from message transmission, SDN technology facilitates the efficient utilization of network resources. Existing SDN technologies include three types of architectures, i.e., the VN-based architecture, the cellular network-based architecture, and the hybrid architecture. The hybrid architecture can leverage the uncertain latency of ad-hoc networks and the cost of cellular networks. Li et al. \cite{li2016control} focus on leveraging the latency of vehicles and the costs of cellular networks. A two-period Stackelberg game is designed for the interaction between the controller and vehicles. The NE of the game gives the best strategies of the controller and the vehicles, i.e., the optimal bandwidth allocation and the number of packets to be sent through the cellular network. Besides, an intelligent network selection scheme for data offloading is designed in \cite{aujla2017data} using a single-leader multi-follower Stackelberg game. The authors of \cite{chahal2019network} employ a two-stage Stackelberg game the optimal network selection in SDN-based VNs with heterogeneous wireless interfaces. In \cite{li2016network}, the assignment of network virtualization solutions in SDN-based VNs is modeled as a non-cooperative game, from which the PE is obtained. Alioua et al. \cite{alioua2019incentive} address the edge-caching problem in SDN-IoV by modeling the interaction between content provider (CP) and multiple network operators (MNO) as a Stackelberg game. The CP is responsible for publishing the quantity of content to be cached. The MNO responds to the CP using the caching price.

\section{Lessons learned}	  
\label{sec_lesson}
\begin{figure}[!hbt] 
	\centering
	\includegraphics[width=3.5in]{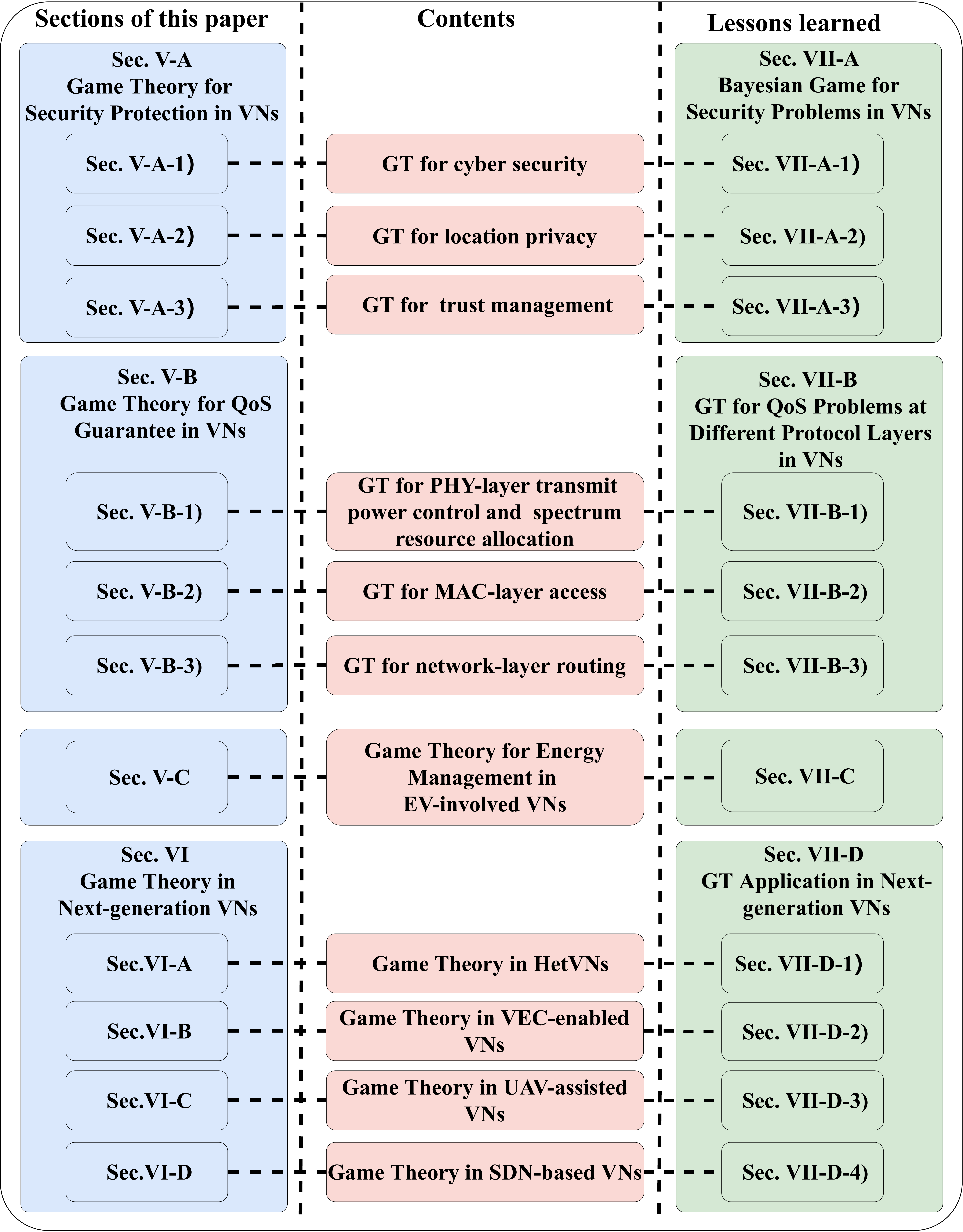}
	\caption{Relationship of the lessons learned and the sections in this survey.}
	\label{fig_lessons}
\end{figure}

 The key lessons learned and some recommendations could be concluded in the following subsections. The visual relationship of the lessons learned from each section in this survey is illustrated in Fig. \ref{fig_lessons}.

\subsection{Bayesian Game for Security Problems in VNs}

Generally, the intelligent nodes in VNs lack complete information due to the sensitivity to security or privacy. Therefore, it is recommended to use the Bayesian game to model the uncertain information on the attackers. The lessons learned from GT approaches for security problems in VNs as discussed in Section \ref{sec_security} can be summarized as follows.

\subsubsection{GT for cyber security}
The hybrid of IDS, IPS, and IRS is expected to provide versatile security schemes for VNs than using one of the single architecture. By combining the Stackelberg game with the Bayesian game, the IDA can act as a leader to activate the corresponding actions to provide the optimal cyber security in the uncertain VNs.

\subsubsection{GT for location privacy}

When designing the pseudonym change schemes for location privacy, the cooperation awareness is critical to be incorporated with the Bayesian game to mitigate selfishness in the uncertain VNs.

\subsubsection{GT for trust management}
Incorporated with the Bayesian game, the reputation and credit are promising incentive schemes for trust management in VNs. These mechanisms not only mitigate selfishness and maliciousness in VNs but also provide long-term trust detection. However, researchers should be mindful that constructing precise reputation or credit evaluations for vehicles may be a long-term task.

\subsection{GT for QoS Problems at Different Protocol Layers in VNs}
We highlight the following lessons learned from the perspective of GT for QoS guarantee at different protocol layers of VNs as analyzed in Section \ref{sec_GTQoS}.

\subsubsection{GT for PHY-layer transmit power control and  spectrum resource allocation}

\begin{itemize}[itemsep= 3 pt,topsep = 3 pt]
	\item The game strategies for power control or spectrum allocation should be adaptive and dynamic. The game solutions should provide fairness.
	
	\item The evolutionary game is able to overcome the bounded rationality and selfishness among competitive vehicles. Besides, integrating incentive schemes such as channel pricing in the evolutionary game could further overcome the generic inefficiency of  NE.
	
\end{itemize}

\subsubsection{GT for MAC layer access}
\begin{itemize}[itemsep= 3 pt,topsep = 3 pt]
	\item The Bayesian game is proven to be powerful for MAC-layer collision avoidance by modeling the competitive access of vehicles.
	
	\item Mathematical evaluations of the access priority are essential for vehicles to decide the optimal access strategy to avoid collisions.
	
	\item Incentive schemes can be further incorporated into the Bayesian game to prevent greedy vehicles that possess the limited MAC-layer resources.
\end{itemize}

\subsubsection{GT for network-layer routing}

\begin{itemize}[itemsep= 3 pt,topsep = 3 pt]
	\item Compared to the relay-based and broadcast-based routing schemes, the clustering routing provides flexibility to alleviate the broadcast storm in the highly dynamic VNs. The coalitional game is a powerful tool to help vehicles dynamically form stable groups to avoid collisions induced by broadcast storms.
\end{itemize}

\subsection{Game Theory for Energy Management in EV-involved VNs}

\begin{itemize}[itemsep= 3 pt,topsep = 3 pt]
	\item The multi-leader and multi-follower Stackelberg game has demonstrated its potential in modeling the interaction among hierarchical nodes in EV-involved VNs. 
	Exiting work only study the interactions between EVs or among several energy entities. The interaction among various heterogeneous entities in EV-involved VNs need further investigation but could be complex.

	\item The price-based incentive mechanism plays a critical role in the Stackelberg game 
	to stimulate the interaction among the energy entities and improve the efficiency of charging scheduling.
	
\end{itemize}

\subsection{GT Application in Next-generation VNs}
Having surveyed the literature on GT approaches for next-generation VNs in Section \ref{sec_GTin5G}, the lessons learned from each subsection are concluded as follows.

\subsubsection{GT in HetVNs}
\begin{itemize}[itemsep= 3 pt,topsep = 3 pt]
	\item The auction game and the bargaining game are promising to stimulate cooperation among  SPs for resource sharing in HetVNs. The auction game is more appropriate to model the interaction among multiple SPs while the bargaining game is more suitable for the two-player situation.
	
	\item The cooperative game tends to be more powerful in terms of decreasing cost since vehicles in the same coalition can collaboratively make optimal access decisions with low cost.
	
	\item Most of the studies only consider the VN performances in the evaluation of utility functions. Although the interest degree of a vehicle on the requested content is considered in \cite{hui2017optimal,hui2019game}, they only focus on the vehicle's interest on the latency of the content. Diverse preferences of vehicles on different applications or services of heterogeneous networks should be further considered in designing the utility models of vehicles in HetVNs.
\end{itemize}

\subsubsection{GT in VEC-enabled VNs}
\begin{itemize}[itemsep= 3 pt,topsep = 3 pt]
	\item The Stackelberg game is promising for joint task offloading and resource allocation in VEC-enable VNs by constructing sequential interaction among VEC servers and vehicles.
	
	\item The matching game provides powerful solutions for task offloading in VEC-enabled VNs. Compared with the Stackelberg game, the matching game is more applicable for dealing with the heterogeneous preferences among players. Specifically, vehicles and VEC servers who have heterogeneous preferences on each other can form mutually beneficial relationships in a self-organizing manner.
	
	\item The combination of contract and price mechanisms can powerfully force cooperation between VEC servers or between vehicles to facilitate the idle resource sharing.
\end{itemize}

\subsubsection{GT in UAV-assisted VNs}
\begin{itemize}[itemsep= 3 pt,topsep = 3 pt]
	\item More heterogeneous players emerge in the UAV-assisted VNs. Researchers should carefully construct GT models, where various players, strategies, and preferences should be managed efficiently with low delay.
	
	\item Regarding the problems involving the communication between UAVs and vehicles, existing work focuses on adopting the general sequential game. As a special sequential game, the Stackelberg game seems to be powerful in terms of building the hierarchical interaction between the heterogeneous players. 

	\item The limited battery lifetime is the significant characteristic of the UAV, which is considered by little  work for the game model construction.
	
\end{itemize}

\subsubsection{GT in SDN-based VNs}
\begin{itemize}[itemsep= 3 pt,topsep = 3 pt]
	\item The Stackelberg game is an appealing solution in SDN-based VNs to construct the hierarchical interaction between the SDN controller and the lower-layer nodes. In the game, the SDN controller acts as a leader for global information collection and message announcement. 
\end{itemize}

\section{Open Issues and Future Directions}
\label{sec_challenge}

\begin{figure*}[!hbt] 
	\centering
	\includegraphics[width=7in]{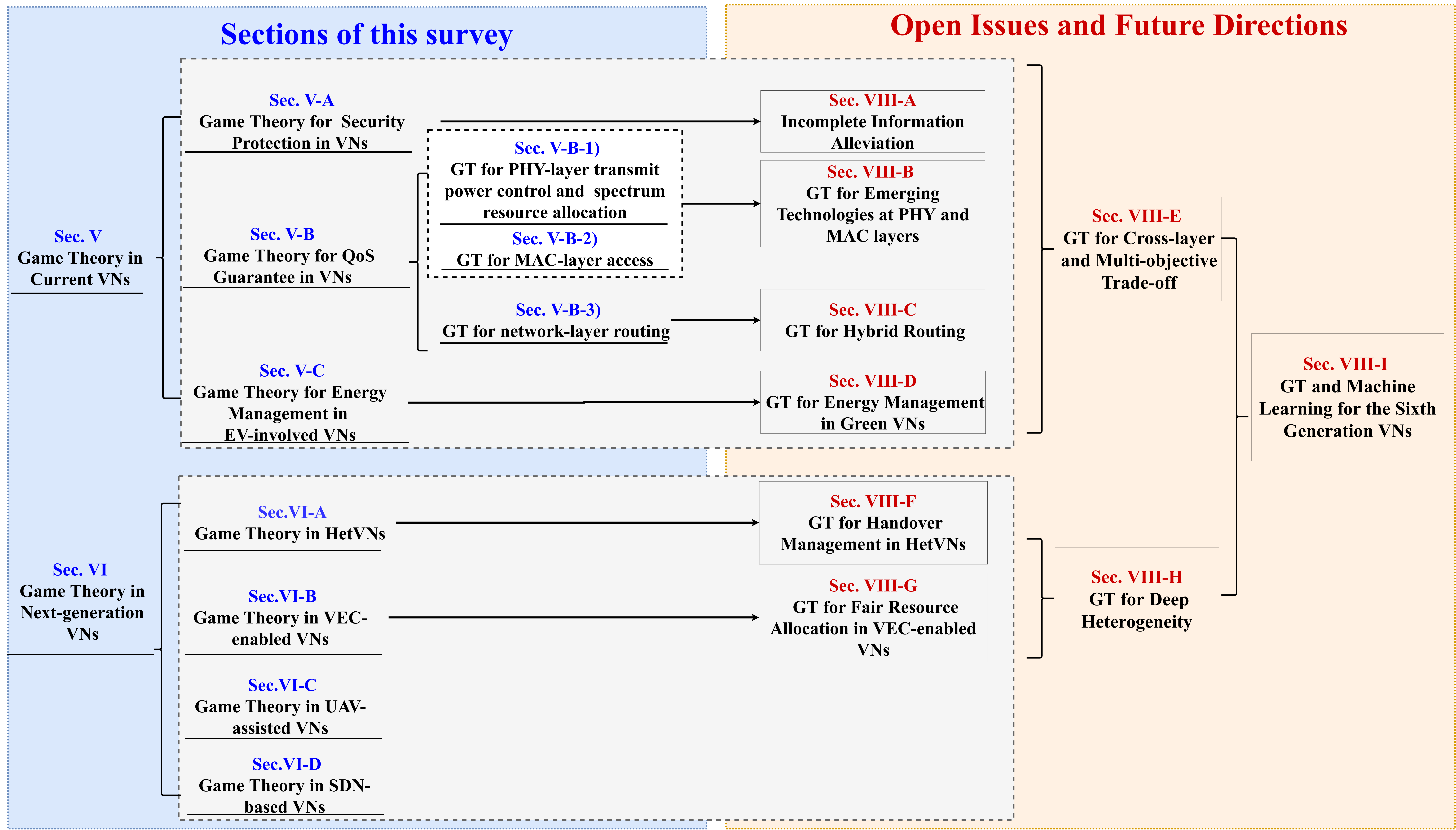}
	\caption{\textcolor{color1}{Relationship between the open issues and the sections in this survey.}}
	\label{fig_future}
\end{figure*}

The open issues and possible research directions of the GT application in VNs are discussed in this section. The visual relationship of the content in this section and the sections in this survey is illustrated in Fig. \ref{fig_future}.

\subsection{Incomplete Information Alleviation}
As discussed in Section \ref{sec_security}, GT approaches are powerful to model attackers, maliciousness, or selfishness in VNs. However, most existing studies assume complete information or finite strategy spaces, which could lead to several issues as follows.

\begin{itemize}[itemsep= 3 pt,topsep = 3 pt]
	\item The main challenge is the difficulty for legitimate vehicles to obtain complete information about the utilities and strategies of the attackers. Although the Bayesian game is able to model the incomplete-information vehicular situations, it is only based on the beliefs or prior knowledge on the types of the other nodes. 
	\item The aspects ``security level", ``privacy level", and ``trustworthiness level" are investigated independently. The trade-off among these aspects requires further study due to the limited resource in VNs. 

\end{itemize}

The emerging SDN technology is expected to be integrated into the game to alleviate the incomplete information limitation by gathering global information.

\subsection{GT for Emerging Technologies at PHY and MAC layers}

  With the ever-increasing demands of massive connectivity and transmitted data, the conventional VNs could suffer from severe collision, congestion, and overloads. The emerging technologies of nonorthogonal multiple access (NOMA) and multi-input and multi-output (MIMO) are foreseen as promising candidates to tackle the challenges at PHY and MAC layers. Constructing GT models for NOMA or MIMO-based VNs is still challenging for the future research.

  \begin{itemize}[itemsep= 3 pt,topsep = 3 pt]
 	\item NOMA provides massive connectivity and ubiquitous communications for the future VNs using the non-orthogonal access. However, the non-orthogonal nature brings new challenges for the NOMA-enabled VNs due to the severer interference, including 1) the cross-interference caused by the overlapping communication range of multiple transmitters, and 2) the co-channel interference caused by the sub-band sharing among multiple vehicles. Therefore, designing game strategies for joint subchannel and power allocation is expected to be a possible future research direction for efficient NOMA-enabled VNs.

 	\item  MIMO provides high data rate, large communication range, and high spectral efficiency for future VNs using multiple transmit and receive antennas. Although GT is expected to provide promising solutions for antenna selection, the multiple-element channel matrix between transmitters and receivers leads to large strategy space. Therefore, GT solutions for joint transmit and receive antenna selection is worth further exploration in the future research.
 \end{itemize}

\subsection{GT for Hybrid Routing}

A lot of GT solutions for routing in VNs have been proposed recently including relay-based routing, broadcast-based routing, and cluster-based routing, but there is still some work remains open.

\begin{itemize}[itemsep= 3 pt,topsep = 3 pt]
	
	\item GT for hybrid routing scheme is expected to be a future research direction by fully utilizing the advantages of the three types of routing approaches. However, it is still under explored since the utility models of hybrid routing are expected to be more complex  that requires new routing metrics.

	\item The trajectory prediction is important to decide the optimal routing strategies, but is considered by very little work. The challenge lies in the fact that vehicles lack complete information on the highly dynamic VNs. 
\end{itemize}

\subsection{GT for Energy Management in Green VNs}

 Future green vehicular communications become more heterogeneous with the integration of various battery-enabled energy facilities, including solar-powered and wind-powered RSUs and SGs. The highly dynamic VNs 
and intermittent renewable energy supplies bring new challenges for making efficient decisions on energy management in green VNs.

  \begin{itemize}[itemsep= 3 pt,topsep = 3 pt]
	\item With the development of the wireless power transfer technology, future EVs are envisioned to be charged on-the-move. How to guarantee seamless handover among  heterogeneous energy facilities is challenging due to the highly dynamic of vehicles. GT approaches for interactions among various charging entities need further investigation
	 

	\item The RSUs or SGs could be overloaded by increasing charging demands of EVs. GT models of interactions among EVs for efficient energy swapping could be a promising solution to relieve the load of the energy supply facilities.

\end{itemize}

\subsection{ GT for Cross-layer and Multi-objective Trade-off}

 GT has been widely employed in solving the single layer-targeted and single objective-specific problems, which however is insufficient to meet the stringent and diversified requirements of the future VNs. Designing flexible GT approaches for cross-layer and multi-objective trade-off in VNs can be reckoned as an open issue for the future research direction.


  \begin{itemize}[itemsep= 3 pt,topsep = 3 pt]

	\item Currently, there is no standardized cross-layer mechanisms for VNs. Supported by different VN protocols, the communication components at different VN layers are heterogeneous and operated independently. GT has the potential to construct interaction and coordination among different layers, but is challenged by the heterogeneity among different VN layers. To overcome the heterogeneity, the upper-layer strategies should be decided based on the feedback from the lower layers. Therefore, more complete information should be acquired and shared among different layers.

	\item Most of the exiting GT approaches mainly target on the single-objective optimization from the aspects of QoS, security, and energy. Several studies explore the trade-off between QoS and security  \cite{sedjelmaci2018generic,brahmi2019cyber,sun2017non}, or between QoS and energy \cite{hua2017game,tian2017self}.  With the development of 5G technology and the emergence of various applications and services, QoE is emerging as an important requirement of individual users. GT approaches need further exploration to support various specific requirements and balance all of these requirement efficiently.
\end{itemize}

\subsection{GT for Handover Management in HetVNs}
 
 There is a gap in the current studies that the application of GT for handover management in HetVNs has not been explored. The coexistence of multiple heterogeneous access technologies and the highly dynamic of vehicles result in frequent and massive handovers in VNs. Reliable collaboration among access points is necessary to fulfill the seamless and flexible handover in a short latency. Given the strong ability of GT in managing the cooperation among nodes, it is recommended to design delicate GT cooperation among different access points. However, there are still several issues that need to be addressed when implementing efficient cooperative strategies.

   \begin{itemize}[itemsep= 3 pt,topsep = 3 pt]
 	\item Hybrid cooperation strategies seem necessary for both homogeneous cooperation and heterogeneous cooperation. Researchers should be mindful about the additional overheads such as computational complexity and possible long latency caused by the hybrid strategies.
 	
 	\item How to mitigate or eliminate interference from the neighboring access points still remains a challenging problem for efficient cooperative strategy design.

 	\item Sufficient prior knowledge about vehicles, especially the trajectory information, seems crucial for effective cooperation so that the content can be cached in advance along with the vehicle. For this reason, the SDN technology is envisioned as a promising candidate for acquiring more global information.
 \end{itemize}
 

 
\subsection{GT for Fair Resource Allocation in VEC-enabled VNs}
 
GT has been extensively investigated for task offloading and resource allocation problems in VEC-based VNs. However, with the ever-increasing number of tasks, ultra-low latency requirement, and the inherent highly dynamic of vehicles, it is still challenging to design efficient GT mechanisms for task offloading and resource allocation in VEC-assisted VNs.

 \begin{itemize}[itemsep= 3 pt,topsep = 3 pt]
	
\item Although some efforts have been dedicated on the joint task offloading and resource allocation, they mainly use GT to construct the vertical coordination between vehicles and VEC servers. GT for horizontal cooperation is critical and still requires more investigation in the future work. We may recommend GT application for the following horizontal coordination: 1) horizontal coordination among vehicles for task uploading and task offloading decisions, and 2) horizontal coordination among VEC servers for task migration and resource allocation decisions.

\item Although cooperative game offers solutions to fairness, it is still challenging to ensure both efficiency and fairness when the resource requesters have heterogeneous demands.
\end{itemize}

\subsection{GT for Deep Heterogeneity}

Future VNs could be structured by deep heterogeneous space-air-ground-underwater vehicles. The following characteristics of the future VNs bring new challenges for the GT model design.
\begin{itemize}[itemsep= 3pt,topsep = 3 pt]

\item The multiple-layer heterogeneous devices such as satellites, UAVs, vehicles, and maritime surveillance \cite{zhang2021udarmf} incur large number of heterogeneous players in a game.

\item The coexistence of heterogeneous network technologies and communication types incur large strategy spaces for the players in a game.

\item The huge volume of heterogeneous big data emerges with the ever-increasing ubiquitous connectivity and various applications. Although storing more valuable data can increase its re-utilization, processing high-volume data with higher speed and greater heterogeneity leads to higher storage costs and longer delay. Therefore, how to make a trade-off between data processing cost and the big data value emerges as an interesting and open issue in modeling the utility functions of the players.
\end{itemize}

Mean-field game \cite{huang2010nce} provides an advanced  multi-agent model to analyze the massive interacting populations. It could be a potential candidate to model the interactions among the large number of nodes exchanging massive big data in the future deep heterogeneous VNs.

\subsection{GT and Machine Learning for the Sixth Generation VNs}

From the perspective of GT approaches, there are four main challenges of applying GT approaches in the VNs.

\begin{itemize}[itemsep= 3 pt,topsep = 3 pt]

\item The non-cooperative game is usually less efficient due to the lack of complete information, the absence of cooperation among selfish nodes, and the limited rationality of the nodes.

\item Although the Bayesian game has shown its potential to model the vehicular scenarios with limited information, the complex probabilistic model is required to build accurate
beliefs on the other players. This may cause large overheads and delay for the real-time communication.

\item Although the cooperative game is able to model cooperative behaviors among competitive players, the main challenge is that it is difficult to construct the externally-enforced agreement among the players.

\item Although the evolutionary game enables nodes with limited rationality to adapt to the uncertain and dynamic VNs by learning from the historic environment, designers should be aware of the convergence time and possibly overwhelmed selfishness.
\end{itemize}

With the 5G is deploying around the world, the artificial intelligence (AI)-enabled sixth generation (6G) \cite{yang2020artificial} technology will pave the way for the AI-enabled VNs in 6G. Machine learning (ML) is emerging as a promising AI technology that provides intelligent and efficient solutions for the application of communication networks. Therefore, a coherent integration of advantages of GT and ML is expected for the future 6G VNs.

\section{Conclusion}
\label{sec_conclusion}

This paper provides a survey of recent developments of GT applied to VNs, discussing the applications of GT in the existing VNs and future vehicular communications integrated with 5G mobile network technologies. First, an initial background of VNs and the unique features that may result in the challenges in VNs is presented.  Furthermore, a comprehensive overview of GT models used in VNs is presented, giving the basics and classifications from several aspects, proposing a taxonomy on the models applied in VNs, and discussing each model's key characteristics alone with their advantages to deal with the specific challenges in VNs. After discussing the requirements of VNs and the motivation of using GT, comprehensive literature reviews on GT approaches in dealing with the challenges of current VNs next-generation VNs are respectively provided.  Finally, the lessons learned and the remaining open issues along with corresponding future research directions are discussed.

The conclusion is that game-theoretical mechanisms are promising to deal with the challenges resulting from the limited resource. It can be used to model the competitive, cooperative, malicious, and defensive behaviors of the intelligent nodes in VNs. The key to an efficient GT model is to stimulate cooperation (or mitigate selfishness) and alleviate uncertainty for intelligent nodes to achieve reciprocal resource or information sharing in the highly dynamic and uncertain VNs. Among the game models introduced in Section \ref{sec_BgGT}, the cooperative, evolutionary, and games with interactive functions are powerful to construct the efficient models for VNs. Cooperative and evolutionary games are more appropriate for homogeneous players. Regarding the cooperative game, it provides powerful tools for cooperation modeling and fair resource allocation, which is much suited to the clustering-based VN issues. The key and difficulty of cooperation are designing the externally-enforced contract or agreement. Furthermore, the evolutionary game can overcome limited rationality and adapt to the uncertain and unexpected vehicular environment by learning from history. Designers should prove the convergence time to avoid the evolutionary game's possible long convergence time. Besides, the Stackelberg, bargaining, auction, and matching games are promising for heterogeneous vehicular scenarios due to their capability to stimulate interaction. The Stackelberg game is suitable for hierarchical VN structures such as VEC-enabled VNs, UAV-assisted VNs, or the future space-air-ground VNs. Bargaining, auction, and matching games are appropriate to model the one-to-one, one-to-many, and many-to-many interactions in VNs, especially for the problems of task offloading or resource allocation in VEC-enabled VNs.  Using one of the models alone is insufficient for the complex VNs with selfish nodes and limited resources. The incentive mechanisms is essential to be incorporated in games to mitigate selfish behaviors in VNs. The learning and matching schemes are promising because they address the challenges in VNs, such as randomness, uncertainty, and limited resources. Reputation or credit methods can also be considered into the game for the security protection in VNs, such as trustworthiness evaluation. In terms of the uncertainty of VNs, the Bayesian game can be incorporated to model the incomplete information in VNs. SDN is a powerful structure to gather global information and execute the punishment or reward mechanisms.

 \section*{Acknowledgment}
 This study is supported in part by the National Natural Science Foundation of China (62172186, 62002133, 61872158, 61806083), in part by the National Key Research and Development Program of China  (2018YFC0831706), in part by the Science and Technology Development Plan Project of Jilin Province (20190701019GH, 20190701002GH, 20190103051JH, 20200201166JC), and in part by the Central government funds for guiding local scientific and Technological Development (2021Szvup047).
 
 The authors would like to thank the editors and the anonymous reviewers for their valuable comments and insights which helped to improve the quality of this study.

\bibliographystyle{IEEEtran}
\bibliography{references.bib}
	
	\begin{IEEEbiography}[{\includegraphics[width=1in,height=1.25in,clip,keepaspectratio]{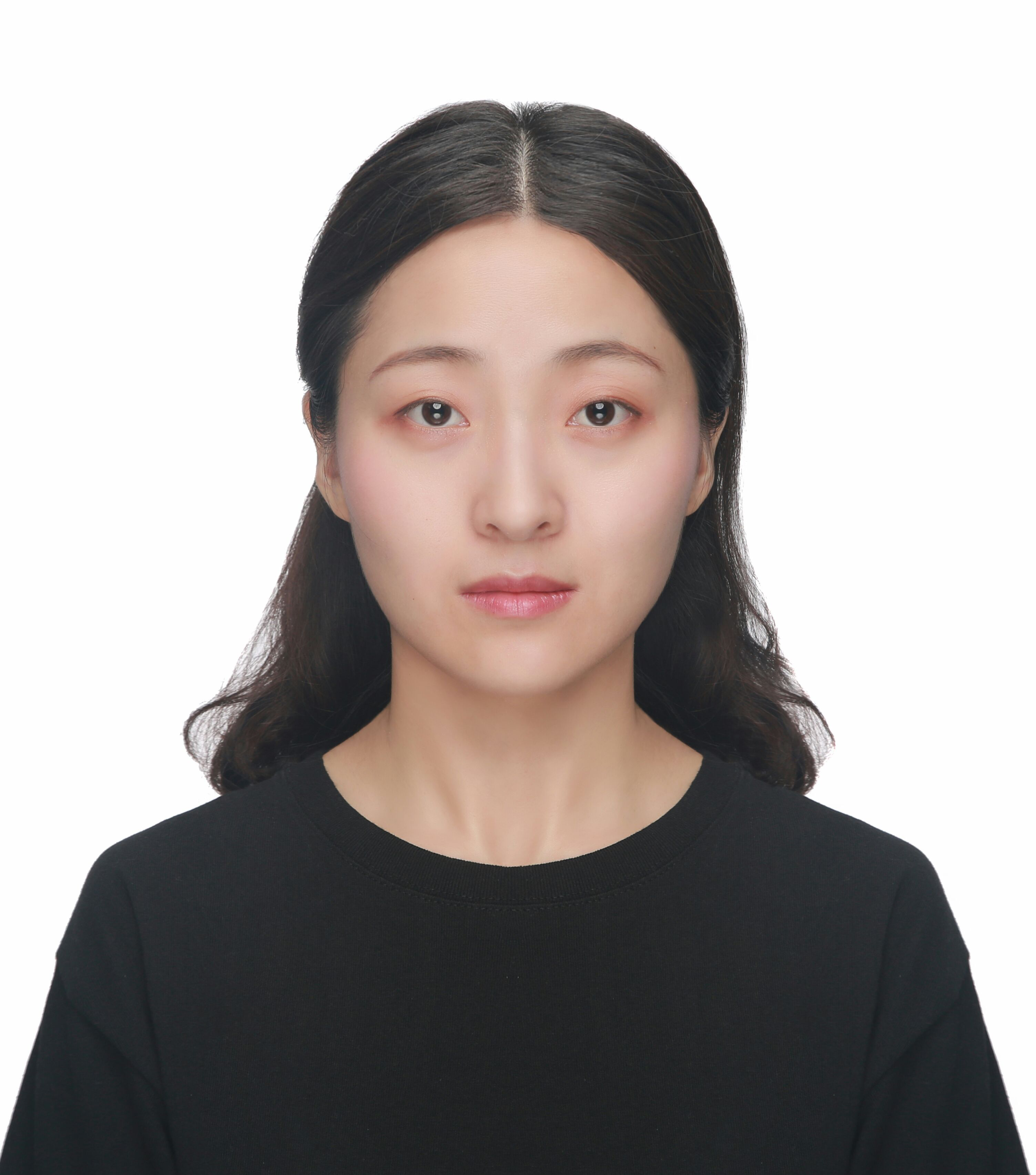}}]{Zemin Sun} received the M.Sc in computer science from Jilin University, Changchun, China, in 2018. Her research interests include communication quality and security in vehicular networks and game theory. She is currently seeking her Ph.D. degree in Jilin University.
	\end{IEEEbiography}

\end{document}